\documentclass[11pt]{article}
\pdfoutput=1

\usepackage[T1]{fontenc}
\usepackage[utf8]{inputenc}
\usepackage[english]{babel}

\usepackage{setspace}
\usepackage{simplewick}
\usepackage[dvipsnames]{xcolor}
\usepackage[normalem]{ulem}
\usepackage{cite}
\usepackage{dsfont}
\usepackage{marginnote}
\usepackage{placeins}
\usepackage{xcolor}
\usepackage{caption}
\usepackage{subcaption}
\usepackage{bbm}
\usepackage{booktabs}

\usepackage{tikz}
\usepackage{pgfplots}
\pgfplotsset{compat=1.18}
\usetikzlibrary{decorations.pathmorphing}
\tikzset{snake it/.style={decorate, decoration=snake}}
\usetikzlibrary{calc}
\usetikzlibrary{math}

\def\centerarc[#1](#2)(#3:#4:#5)
{ \draw[#1] ($(#2)+({#5*cos(#3)},{#5*sin(#3)})$) arc (#3:#4:#5); }

\usetikzlibrary{intersections}
\usetikzlibrary{patterns, arrows.meta, scopes, fadings, intersections, pgfplots.fillbetween}
\usepackage{blindtext}
\usepackage{geometry}
\usepackage{verbatim}
\usepackage{prettyref}
\usepackage{amsmath}
\usepackage[normalem]{ulem}
\usepackage{amssymb}
\usepackage{graphicx}
\usepackage{setspace}
\usepackage{esint}
\usepackage{verbatim}
\usepackage{physics}
\usepackage{listings}

\renewcommand{\Im}{\mathsf{Im }}
\renewcommand{\Re}{\mathsf{Re }}
\newcommand{\de}{\text{d}}

\newcommand{\xdownarrow}[1]{%
  {\left\downarrow\vbox to #1{}\right.\kern-\nulldelimiterspace}
}
\newcommand{\ve}{\varepsilon}

\usepackage{color}
\definecolor{darkgreen}{rgb}{0,0.5,0}
\definecolor{darkblue}{rgb}{0,0,0.6}
\definecolor{purple}{rgb}{0.4,.2,0.7}
\definecolor{orange}{rgb}{0.95, 0.5, 0.3}

\usepackage[colorlinks=true,citecolor=darkgreen,linkcolor=darkblue,urlcolor=blue]{hyperref}

\numberwithin{equation}{section}
\numberwithin{table}{section}

\def\cH{{\cal H}}

\def\a{\alpha}
\def\b{\beta}
\def\g{\gamma}
\def\d{\delta}

\def\be{\begin{equation}}
\def\ee{\end{equation}}
\def\bea{\begin{eqnarray}}
\def\eea{\end{eqnarray}}
\def\ba{\begin{align}}
\def\ea{\end{align}}

\def\L{{\mathcal L}}

\def\cO{{\cal O}}
\def\de{\text{d}}

\def\op{\operatorname}
\def\SFF{{\rm SFF}}

\definecolor{codegreen}{rgb}{0,0.6,0}
\definecolor{codegray}{rgb}{0.5,0.5,0.5}
\definecolor{codepurple}{rgb}{0.58,0,0.82}
\definecolor{backcolour}{rgb}{0.95,0.95,0.92}

\lstdefinestyle{mystyle}{
    backgroundcolor=\color{backcolour},   
    commentstyle=\color{codegreen},
    keywordstyle=\color{magenta},
    numberstyle=\tiny\color{codegray},
    stringstyle=\color{codepurple},
    basicstyle=\ttfamily\footnotesize,
    breakatwhitespace=false,         
    breaklines=true,                 
    captionpos=b,                    
    keepspaces=true,                 
    numbers=left,                    
    numbersep=5pt,                  
    showspaces=false,                
    showstringspaces=false,
    showtabs=false,                  
    tabsize=2
}

\makeatletter
\newcommand{\vast}{\bBigg@{4}}
\newcommand{\Vast}{\bBigg@{5}}
\makeatother

\begin{document}
\begin{spacing}{1.1}
  \setlength{\fboxsep}{3.5\fboxsep}

~
\vskip5mm

\begingroup
\renewcommand{\thefootnote}{\fnsymbol{footnote}}

\begin{center} 

{\Huge \textsc \textbf{Quantum simulation using Trotterized disorder Hamiltonians in a single--mode optical cavity}}

\vskip10mm

Rahel Lea Baumgartner$^{1}$\footnotemark[1], Pietro Pelliconi$^{1,2}$\footnotemark[1],  Soumik Bandyopadhyay$^{3,4}$, Francesca Orsi$^{5}$, Philipp Hauke$^{3,4}$, Jean-Philippe Brantut$^{5}$ \& Julian Sonner$^{1}$\\

\vskip1em
{\small
{\it 1) Department of Theoretical Physics, University of Geneva, 24 quai Ernest-Ansermet, 1211 Gen\`eve 4, Suisse}\\
\it 2) Department of Physics, Princeton University, Princeton NJ 08544, USA \\
\it 3) Pitaevskii BEC Center, CNR-INO and Dipartimento di Fisica, Universit\`a di Trento,  Via Sommarive 14, I-38123 Trento, Italy\\
\it 4) INFN-TIFPA, Trento Institute for Fundamental Physics and Applications, Via Sommarive 14, I-38123 Povo, Trento, Italy\\
\it 5) Institute of Physics and Center for Quantum Science and Engineering, École Polytechnique Fédérale de Lausanne (EPFL), CH-1015 Lausanne, Switzerland 
}
\vskip5mm

\end{center}

\vskip10mm

\begin{abstract}

All--to--all interacting and disordered many--body systems are notoriously hard to simulate on quantum platforms, as interactions are commonly mediated by auxiliary degrees of freedom that lower the amount of disorder, introducing undesired correlations. In this work, we show how a Trotterization scheme can be effectively utilized to densify the disorder of the model. In particular, we study the statistical properties of the resulting model, as well as Trotterization errors in the simulation that affect the time evolution and dynamical observables. As a concrete example, we propose an implementation via a single--mode cavity QED platform of the complex Sachdev--Ye--Kitaev model. We analyze several features of the effective model, such as the distribution of the effective couplings, the number of interacting sites, state preparation, and the behavior of quantum chaos probes. We conclude this work with a detailed investigation of the robustness of our findings against dissipation, both analytically and numerically.

\end{abstract}
\vfill
\footnotetext[1]{These two authors contributed equally.}
\footnotetext[0]{\tt{rahel.baumgartner@unige.ch}}
\footnotetext[0]{\tt{pelliconi@princeton.edu}}
\footnotetext[0]{\tt{julian.sonner@unige.ch }}
\endgroup

\pagebreak
\pagestyle{plain}

\setcounter{tocdepth}{2}
{}
\vfill
\tableofcontents

\newpage
\section{Introduction}
Quantum simulations in ultra-cold atomic gas platforms have opened a new avenue for exploring strongly correlated many-body quantum systems and their nonequilibrium dynamics. Ultra-cold atomic gases in optical lattices have already enabled controlled realizations of iconic condensed matter Hamiltonians, such as the Bose--Hubbard model (which exhibits the superfluid--Mott insulator transition \cite{greiner2002quantum}) as well as the Fermi--Hubbard model (where a Mott insulating phase and magnetic ordering emerge in a fermionic system \cite{schneider2008metallic, jordens2008mott}). A natural next step is to extend these approaches to quantum simulations of disordered, all--to--all interacting many-body models---systems that differ from the above in one critical aspect: their interactions are random (drawn from a suitable distribution) rather than spatially structured. Furthermore, we are interested in long--range, mainly all--to--all interactions, of the kind relevant to spin glasses and spin liquids, including systems believed to host holographic phases of matter. A particularly prominent example of this class are the Sachdev--Ye--Kitaev (SYK) models, involving $N$ fermionic modes (Majorana or complex) with all--to--all couplings among random subsets of $q$ fermions. The SYK model has emerged as a paradigm of maximal quantum chaos, non--Fermi--liquid behavior and holographic duality, giving it special importance in modern many-body physics. Despite the central role of SYK and related disordered models, realizing them experimentally remains highly non-trivial, especially because engineering disordered, all--to--all interactions in a controlled experimental platform is extremely challenging. 

In this work, we focus on the theoretical framework to tackle this challenge. Our ultimate motivation is to enable the quantum simulation of such disordered all--to--all many--body systems using ultracold atoms coupled to high--finesse optical cavities, specifically focusing on a single--mode cavity setup. As a concrete illustration, we propose an implementation of the complex SYK model with four-fermion interactions (cSYK$_4$) in a single-mode cavity QED platform (see Section \ref{sec:Experimental_implementation}), following our previous proposal \cite{Baumgartner:2024ysk} (see also Ref.~\cite{Uhrich:2023ddx}). Our approach relies on a simple but powerful idea: using Trotterization to `densify' the disorder in a controllable way. In essence, we propose to simulate the full all--to--all random Hamiltonian by sequentially switching between `sparse' versions of the model that the experimental platform can natively implement.\footnote{Here sparse means that, at a given time, only a reduced number of independent random couplings are realized. Importantly, this notion of sparsity differs from that of sparse SYK models studied in the literature \cite{Xu:2020shn, Garcia_Garcia_2021}. Later on we will be more precise about our definitions of `sparse', `dense' and `full' disorder.} Suppose the available experimental platform can only realize a sparsified version of the target model at any given time---meaning that at each realization, only a small fraction of all possible couplings are active, or equivalently that the Hamiltonian has a parametrically reduced number of independent random couplings (as happens in low--rank SYK constructions \cite{Kim:2019lwh}). Let us denote such a sparse system as $H_{\alpha}$ for $(\alpha = 1,2,\ldots R)$, representing $R$ distinct random realizations. Crucially, by combining a sufficiently large number of different realizations, the physics of the aggregate approaches that of the fully disordered target Hamiltonian, where the disorder no longer is sparse, in a sense that we will elaborate below. In fact, as $R$ becomes large, the effective Hamiltonian 
\begin{equation}
H_{\rm eff} = \sum_{\alpha=1}^R H_{\alpha}  \approx  H_{\rm target}
\end{equation}
converges (in distribution) to the dense, all--to--all model we aim to simulate. The idea of `summing over disorder realizations' in this manner follows the strategy of low--rank SYK models studied in \cite{Marcus-Vandooren_2018,Kim:2019lwh}. Here, we propose to implement this sum in time: the cavity QED platform can realize each $H_{\alpha}$ one after the other, Trotterizing the evolution so that over one full cycle of $R$ steps, the system effectively experiences the combined Hamiltonian $H_{\text{eff}}$. In other words, if $U_{\alpha}(\delta t) = e^{-iH_{\alpha}\delta t}$ is the short--time evolution under the $\alpha$-th sparse Hamiltonian, then 
\begin{equation}
U_{\rm eff}(t) = \left(\prod_{\alpha=1}^R U_{\alpha}(\delta t)\right)^n + {\cal O}(\varepsilon)\,,
\end{equation}
with $n$ cycles of duration $R\delta t= t/n$, will approximate $e^{-i H_{\text{target}}t}$ up to a chosen accuracy threshold $\varepsilon$. This digital summation of random interaction leverages existing capabilities for `single-shot' analog simulation of sparse-coupling Hamiltonians, avoiding the need for complex multi-qubit gates that a fully digital algorithm would require. Importantly, our analysis shows that the Trotter errors introduced by this procedure can be made negligible: the Trotter error does not grow with system size (a favorable result consistent with results on localized Trotter errors \cite{Hauke_trotter_Error,Hauke_Chaos_Trotter,Kargi2021,Chinni2022}), and the Trotterized time evolution converges uniformly to that of the fully disordered model as the timestep $\delta t \to 0$.

\begin{figure}[t]
\centering
\includegraphics[width=0.85\textwidth]{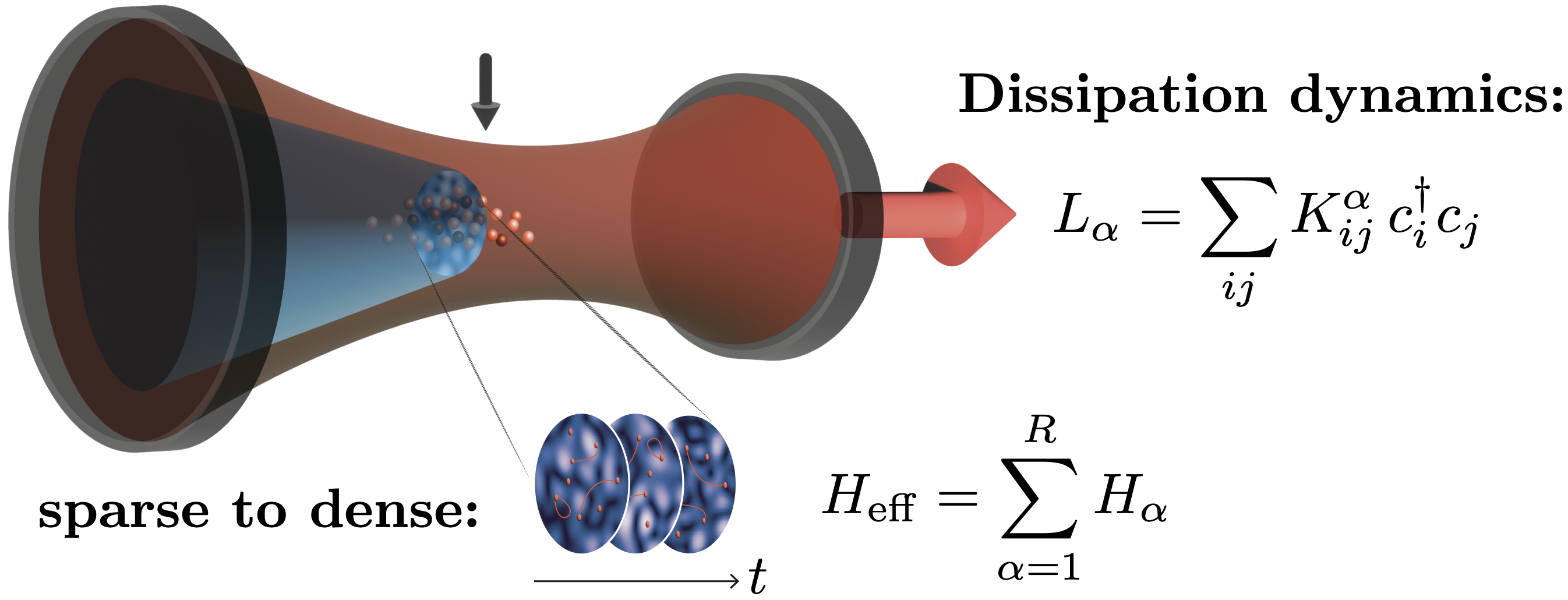}
\caption{A sparse or low-rank random disorder model, $H_{\alpha}$, is prepared in a single-mode optical cavity via a randomization of the light-matter couplings between the atoms and the cavity mode, obtained by projecting a light-shifting beam with a speckled intensity pattern. In the center of the cavity, a quantum degenerate Fermi gas scatters photons from a side pump  (black arrow) into the cavity mode, which mediates the all-to-all interactions between atoms. This low-rank model is effectively multiplexed by switching between different speckle patterns in time, approximating a full-rank or dense model, $H_{\rm eff}$, with an accuracy that follows from a usual Trotter-type analysis. The total evolution includes coherent quantum evolution of the random disorder model and dissipative losses out of the cavity, which in the dispersive regime convert into fermionic jump operators and which we model by Lindbladian operators $L_{\alpha}$. As for the coherent couplings synthesized from lower-rank tensors, the densification of random terms via the Trotter scheme also renders the dissipative and coherent aspects of the model statistically independent up to $1/R$ corrections, where $R$ is the number of speckle patterns per Trotter step.}
\label{fig:TrotterOverview}
\end{figure}

Finally, since any realistic optical cavity implementation is inevitably subject to sources of dissipation, we study in particular the role of photon leakage through the cavity mirrors. This process, quantified by the leakage rate $\kappa$, together with spontaneous photon scattering at rate $\Gamma$, gives rise to Lindblad jump operators whose structure inherits the disordered character of the engineered couplings. We analyze how such dissipative channels interplay with the coherent Trotterized dynamics, extracting numerical timescales by studying the coherent fidelity. Also, we demonstrate that cavity photon loss contributes at the same order as the unitary process in the large-$N$ limit. This implies directly that the cooperativity of the cavity plays a crucial role in being able to observe signatures of SYK physics.

\subsubsection*{Overview of the paper}

In this paper, we focus on various aspects of quantum simulating disordered systems, with the aim of proposing and studying a protocol that is able to reproduce the theoretical predictions of a fully disordered model. Our protocol involves summing over different realizations of the system with `sparse' disorder, effectively densifying the latter. 

In Section \ref{sec:ClassOfModels}, we study different aspects of this protocol, both at the level of implementing the quantum simulation, and in benchmarking the target model. In particular, in Section \ref{sec:SparseToFullSYK4} we show that the sum over different realizations can be efficiently implemented via Trotterization. We analytically find that the Trotter error on the simulation does not scale with system size, and we numerically show how the Trotterized time evolution converges to the one of the fully disordered model when the Trotter step is sufficiently small. In Section \ref{sec:Statistical_analysis_couplings}, we study the statistical convergence of the couplings from sparse disorder to independent Gaussians, while in Section \ref{sec:Relative_entropy} we study such convergence through the lens of information theoretic quantities. In particular, we find the the Shannon entropy (and the Kullback--Leibler (KL) divergence) saturate faster at the Gaussian prediction than naively expected for a certain number of realizations. In Section \ref{sec:Concrete_example}, we then focus on a particular example of such systems (dubbed low-rank SYK \cite{Kim:2019lwh}) which is closely related to our subsequent cavity implementation, studying how the spectrum changes varying $R$.

Section \ref{sec:Experimental_implementation} moves from theory to experimental implementation. Building on the ideas of Section \ref{sec:Concrete_example}, we propose a concrete cavity-QED (cQED) scheme to realize the cSYK$_4$ model. After a brief review of the required cQED ingredients in Section \ref{sec:cQED_basics} and a derivation of the effective cavity Hamiltonian in Section \ref{sec:Effective_cavity_Hamiltonian}, we present numerical evidence in Section \ref{sec.benchmarking_otoc_sff} that quantum chaos indicators in our proposed setup agree with those of an ideal SYK system. In particular, we compute out-of-time-ordered correlators (OTOCs) and the spectral form factor (SFF) for the proposed experimental implementation and find excellent agreement with theoretical SYK predictions. We further show how the effective number of interacting sites $N$ in the cavity can be determined dynamically in Section \ref{sec:determination_of_N}. Finally, Section \ref{sec:product_states} discusses state preparation (\textit{e.g.,} using product states as initial states).

In Section \ref{sec:Dissipation}, we study the impact of dissipation in the cavity platform. We focus on the two main dissipative channels affecting the coherent evolution in a high-finesse cavity, namely photon loss (leakage through the mirrors) and photon scattering (random photon emission). Section \ref{sec:Lindblad_spectrum} reviews general theoretical expectations and properties of the Lindblad spectrum of the system's open dynamics. In Section \ref{sec:Lindbald_spectrum_numerical}, we numerically simulate the spectrum including photon loss and examine the spectral signatures and fidelity decay caused by this dissipation, comparing them directly to the unitary SYK evolution. Further, in Section \ref{sec:largeN} we present an analytical argument showing that these dissipative effects persist in the large-$N$ limit. In particular, we show that photon loss contributes at the same order as the SYK Hamiltonian's unitary dynamics when $N$ is large. 

We conclude in Section \ref{sec:discussion} with a summary and an outlook of future directions. Several technical details and extensions are provided in the Appendices. 

\section{A class of randomized spin and fermion models}\label{sec:ClassOfModels}

All--to--all interacting disordered models have been extensively studied in various branches of theoretical physics, ranging from statistical mechanics all the way to quantum gravity via the holographic duality. The interest stems from emergent universal phases which arise at low energies, often in the form of spin glasses, or in theories with conformal symmetry. Examples of these models include the Sherrington--Kirkpatrick model \cite{sherrington1975solvable, Parisi_1979}, the $p$--spin model \cite{Crisanti_1992}, the Sachdev--Ye model \cite{Sachdev:1992fk} and its more modern version incarnated by the Sachdev--Ye--Kitaev model \cite{Kitaev_2015, Sachdev:2015efa, Maldacena:2016hyu}. Their Hamiltonian is generically written in the form
\begin{equation}
    H = \sum_{I} J_I \hat \cO_I \ ,
    \label{eq:all_to-all_random_interaction_generic_model}
\end{equation}
where $I = \{i_1, \dots, i_q\}$ is a set of $q$ indices labeling the constituents participating in the interaction associated with the random coupling $J_I$, and $\hat \cO_{I}$ groups the corresponding operators (typically, spin or fermionic operators) acting on those sites. The distinctive all--to--all type of interaction is theoretically implemented running $I$ over all possible $q$-tuples of indices. This specific feature allows for a rich variety of phenomena, but it is particularly hard to engineer for experimental quantum simulations, especially in cases where a many--body interaction is needed (as is the case for $q > 2$). An ingenious idea to overcome this difficulty is to mediate the all--to--all nature of the interaction via an auxiliary (bosonic) particle, detuned by an energy $\Delta$ from the systems degrees of freedom and coupling to them schematically as 
\begin{equation}
    H = \Delta \, a^\dagger a + \sum_{I} J_I \Big( \hat \cO_I  a^\dagger + \hat \cO_I^\dagger  a \Big ) \ .
\end{equation}
In situations where the mediating particle can be integrated out (or, as we will refer to it later, be {\it adiabatically eliminated}), and where the bosonic mode remains in its vacuum sector such that only virtual excitations occur, the resulting Hamiltonian has the form 
\begin{equation}
    H = -\frac{1}{\Delta}\sum_{I,K} J_I J_K \,  \hat \cO_I^\dagger  \hat \cO_K \ .
    \label{eq:low--rank_all_to_all}
\end{equation}
To give a concrete example, suppose that we are given a system with single--body ($q = 2$) all--to--all random interactions of the form \eqref{eq:all_to-all_random_interaction_generic_model}, and we want to engineer a two--body\footnote{Sometimes these are called {\it four--body} terms, counting the number of interacting fermionic modes ($q$) instead of the number of particles they are acting on, as we do here.}  interacting model. Using the method just described via coupling the system to an auxiliary mediator, we can experimentally implement the interaction \eqref{eq:low--rank_all_to_all}, which is an all--to--all interacting model with $q = 4$. A major advantage of this scheme is its direct applicability to experimental quantum simulation schemes, for example in cQED \cite{Uhrich:2023ddx,Baumgartner:2024ysk}. The only remaining difference between the models \eqref{eq:all_to-all_random_interaction_generic_model} and \eqref{eq:low--rank_all_to_all} is in the disordered couplings, where a rank--$q$ tensor in \eqref{eq:all_to-all_random_interaction_generic_model} is written as the outer product of a rank--$q/2$ tensor with itself in \eqref{eq:low--rank_all_to_all}. As we will later see, this difference is not purely mathematical, but has deep physical consequences, first and foremost in reducing the amount of disorder in \eqref{eq:low--rank_all_to_all} with respect to \eqref{eq:all_to-all_random_interaction_generic_model}, introducing spurious correlations between the various disordered couplings. For this reason, models of the form \eqref{eq:low--rank_all_to_all} are called {\it low--rank} models \cite{Kim:2019lwh}, as opposed to the full-rank nature of \eqref{eq:all_to-all_random_interaction_generic_model}. 

The main aim of this Section is to study a protocol to enhance the disorder of a low--rank model, in order to obtain a full--rank one. The physical idea behind this enhancement is to sum different realizations of the low--rank disorder through a Trotterization approach, schematically defining an effective Hamiltonian of the form
\begin{equation}
    H_{\rm eff} = \sum_{\alpha = 1}^R H_\alpha = \frac{1}{\Delta}\sum_{\alpha = 1}^R \sum_{I,K} J_I^\alpha J_K^\alpha \,  \hat \cO_I  \hat \cO_K^\dagger \quad \to \quad \sum_{I,K} \mathcal J_{I;K} \,  \hat \cO_I  \hat \cO_K^\dagger \ .
    \label{eq:Effective_Hamiltonian_intro_sec_2}
\end{equation}
We will then study several aspects of this proposal, from a detailed analysis of the Trotterization scheme to a study of the statistical properties of the couplings at different values of $R$.

\subsection{From low-- to full--rank by taking it Trotter step by step}
\label{sec:SparseToFullSYK4}

With the aim of performing quantum simulations of dense disordered models, we are now tasked with finding a protocol to realize the dynamics of Eq.~\eqref{eq:Effective_Hamiltonian_intro_sec_2}. 
A standard way to realize the time evolution of a sum of Hamiltonians is through the Trotter--Suzuki formula \cite{Trotter_1959, Suzuki_1976}, which states that 
\begin{equation}
    U_{\rm eff}(T) \equiv e^{- i H_{\rm eff} t} = \Bigg( \prod_{\alpha = 1}^{R} e^{-i H_\alpha t/n} \Bigg)^n + \frac{t^2}{2n} \sum_{\a < \b}^R [H_\a,H_\b] \, + \, \dots  \ . 
    \label{eq:U_T_steps}
\end{equation}
Beyond its mathematical validity, the main feature of this formula is to give an operational way to perform a quantum simulation of \eqref{eq:Effective_Hamiltonian_intro_sec_2}. In practice, equation \eqref{eq:U_T_steps} says that to simulate \eqref{eq:Effective_Hamiltonian_intro_sec_2}, one may as well cycle through each individual component $n$ times, for a time $t/n$, up to a small error, whose leading contribution is $\frac{t^2}{2n} \sum_{\a < \b}^R [H_\a,H_\b]$. In particular, fixing the total time of the simulation $t$, the higher the number of cycles, the smaller the error. 

While the Trotter--Suzuki formula \eqref{eq:U_T_steps} is standard practice in quantum simulations, there are several aspects related to the present context that require closer scrutiny.
\begin{enumerate}

    \item As we are interested in large--$N$ many--body systems, it could happen that the `small' simulation error scales unfavorably with $N$, growing larger and larger with system size. Even though the scaling is at most polynomially in $N$ (through the commutators $[H_\a,H_\b]$), for large $N$ this would in practice either drastically limit the total simulation time attainable or require an enormous number of cycles to obtain a reasonable tolerance.

    \item Even in the case when the aforementioned error is under control, we might be worried that the disorder prevents the simulation from accurately following the evolution of the effective model, in particular at the level of observables that depend on the dynamics.  This might happen, \textit{e.g.}, when the value of the observable becomes parametrically low in $N$, which for large $N$ may become much smaller than the tolerance allowed. 

\end{enumerate}
In this Section, we will tackle all these issues to show that a Trotterized evolution of sparse models is able to precisely reproduce both the time evolution of a dense model and the evolution of dynamical observables. The main arguments of this Section have been presented in Ref.~\cite{Baumgartner:2024ysk}, which we we corroborate here with additional details. We will mainly focus on the complex SYK model, for which
\begin{equation}
    H_{\rm eff} = \sum_{\alpha = 1}^R H_{\alpha} = \sum_{i_1 i_2, k_1 k_2} \left( \sum_{\alpha=1}^R J_{i_1 k_1}^{\alpha} J_{i_2 k_2}^{\alpha} \right) \, c_{i_1}^{\dagger} c_{k_1} c_{i_2}^{\dagger} c_{k_2} \ ,
    \label{eq:H_alpha_simple_model}
\end{equation}
where $c_i$ ($c_i^\dagger$) are annihilation (creation) operators of complex spinless fermions. While this specification might seem restrictive, it is foreseeable that the lessons obtained in this Section are generalizable to other dense disordered models, see for instance \cite{Hauke_Chaos_Trotter, Hauke_trotter_Error}. Certainly an analogous scheme also works for the Majorana SYK model; our reason for working with the complex spinless fermion version of the model is the more direct connection of the complex model to cavity--QED experiments (see Section \ref{sec:cQED_basics} and Refs.~\cite{Uhrich:2023ddx,Baumgartner:2024ysk}).

Before proceeding, let us highlight two noteworthy features of our Trotter approach. First, our scheme differs from those commonly considered in the literature, where each four--body interaction is evolved independently in the Trotterization \cite{Garcia_Garcia_2021, Xu:2020shn, Granet2025}. To make the simulation more efficient, such works propose to retain only $k N$ interaction vertices, with $k = \mathcal O(1)$. As first argued in Ref.~\cite{Kim:2019lwh}, the Trotterized approach also necessitates $R \sim N$ steps, but is able to generate all interaction vertices at once. Interestingly, while sparse models require $k > 1$ to be chaotic, our Trotterization approach is chaotic for any $\cO(1)$ ratio $R/N$ at low temperatures. In our case, the sole purpose of the Trotterization is to eliminate the statistical correlations of the sparse disorder, and implies a notable improvement in efficiency. Additional details are given in subsequent Sections. The second feature we wish to highlight is that, by cycling multiple times over a fixed set of Hamiltonian realizations, our scheme effectively reproduces static disorder. To this aim, the periodic cycling over the same realizations is crucial. This should not be confused with Brownian disorder that randomly changes over time, such as in Brownian SYK \cite{Saad:2018bqo}, which would be realized by selecting a different realization of the Hamiltonian at every instance in time. The associated time-evolution operator $ \prod_{\alpha = 1}^{R} e^{-i H_\alpha t/n} $ would be similar to the one in Eq.~\eqref{eq:U_T_steps} but without the $n$-fold reproduction of the same disordered Hamiltonians. Although this situation is also of high relevance, we will not be interested in analyzing this case here.


\subsubsection*{Simulation Error}

To address our concerns 1.\ and 2.\ in the previous subsection, we have to understand how `close' the Trotterized time evolution is to the effective time evolution. For a large number of steps $n$, the perturbative expansion \eqref{eq:U_T_steps} holds, and this issue amounts to understanding the `size' of the error. As the error term in Eq.~\eqref{eq:U_T_steps} is given by an operator, its value depends on the operator norm we choose to estimate it. From a mathematical perspective, the spectral norm, defined as
\begin{equation}
    \|A\|_{\infty} = \sup_{\ket{\psi} \in \mathcal H} \frac{ \sqrt{ \bra{\psi} A^\dagger A \ket{\psi}}}{\sqrt{\braket{\psi}}} = \sqrt{ \lambda_{\rm max}(A^\dagger A) } \ .
    \label{eq:spectral_norm}
\end{equation}
is perhaps the most sensible, as it selects the direction where the two evolutions disagree the most. Indeed, as seen in the RHS of Eq.~\eqref{eq:spectral_norm} this norm is equal to the square root of the maximal eigenvalue of the Hermitian matrix $A^\dagger A$. Loosely speaking, the spectral norm gives an upper bound to the difference between the effective model and the Trotterized simulation. On the other hand, for our purposes, the norm \eqref{eq:spectral_norm} is far from ideal. On the technical side, it is very hard to give analytical estimates of \eqref{eq:spectral_norm}, as it would implicitly involve diagonalizing the matrix, which in our case would depend on the specific realizations of the couplings. Moreover, from a physical perspective, we expect the dynamics of the system to be ergodic, so that it seems more appropriate to consider a norm that takes into account a suitable average over all directions, rather than the worst case. A class of such norms is the $(p,q)$--matrix norms, defined for $D\times D$ square matrices as
\begin{equation}
    \|A\|_{p,q} = \left( \sum_{j = 1}^D \left( \sum_{i = 1}^D |a_{ij}|^p \right)^{\frac{q}{p}} \right)^{\frac{1}{q}} \ , \qquad \text{where} \qquad [A]_{ij} = a_{ij} \ .
    \label{eq:p_q_matrix_norms}
\end{equation}
This class of norms considers a `weighted' average over the various entries of the matrix of interest. These norms are easier to handle for our purposes, as they work at the level of the matrix entries, rather than the matrix eigenvalues. In particular, the matrix entries are directly related to the disordered couplings, a fact that greatly simplifies the disorder average. 

In the following, we will focus on arguably the simplest one: the $(2,2)$--matrix norm, also called Frobenius norm, which for a square matrix can be written as
\begin{equation}
    \|A\|_{2,2} = \frac{1}{D} \left( \sum_{i,j = 1}^D |a_{ij}|^2  \right)^{\frac{1}{2}} = \frac{1}{D} \sqrt{\Tr[A^\dagger A]} \ .
    \label{eq:Frob_norm_definition_sec_3}
\end{equation}
To simplify the notation, we will omit the $(2,2)$ subscript from now onward. Notice that we have changed the normalization in Eq.~\eqref{eq:Frob_norm_definition_sec_3} with respect to Eq.~\eqref{eq:p_q_matrix_norms}, which is a particularly convenient choice since in this normalization unitary matrices are unit norm vectors, and thus Eq.~\eqref{eq:Frob_norm_definition_sec_3} allows us to compute relative differences in unitary evolutions\footnote{We thank Adrián Sánchez-Garrido for suggesting this normalization.}. 

\begin{figure}[t]
\centering
\includegraphics[width=0.96\textwidth]{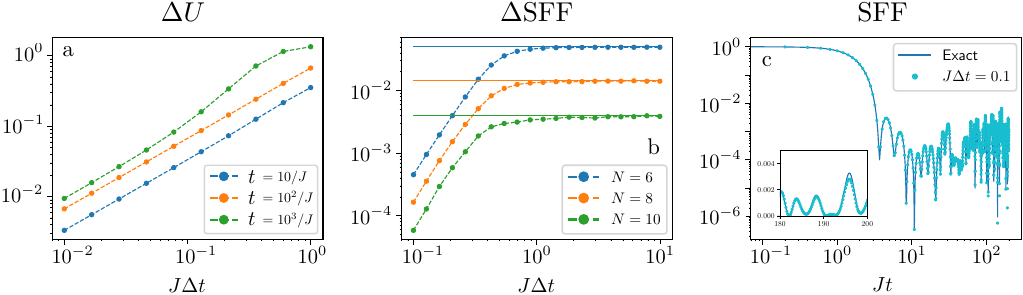}
\caption{ ({\bf a}) Numerical simulations for the average error \eqref{eq:DU} for $N = 8$ and different simulation times, as a function of $J \Delta t$. ({\bf b}) Average time difference between the exact SFF and the Trotterized SFF, as defined in Eq.~\eqref{eq:DSFF}, for different values of $J \Delta t$. We notice a power law convergence in the limit $J \Delta t \to 0$, with a saturation for $J \Delta t \gtrsim 1$ to the plateau value of $1/D$ (solid lines). For each evolution, $n_{\rm max} = 2000$. ({\bf c}) Exact and Trotterized simulations of the SFF, for $J \Delta t = 0.1$ and $N = 10$. We zoom in on a time window at late times to show that the Trotterized SFF still follows the exact SFF to a good accuracy. All three plots were taken from Ref.~\cite{Baumgartner:2024ysk} with the consent of the Authors.}
\label{SFF_trotter_and_Delta_SFF_Simulation}
\end{figure}

Our main goal now is to estimate the error
\begin{equation}
    \ve  = \frac{t^2}{2n} \sqrt{\overline{\Big \|\sum_{\a < \b} [H_\a,H_\b] \Big \|^2}} \ ,
\end{equation}
which quantifies the difference to leading order between the two time evolutions. The calculations are a bit unwieldy, and we relegate them to Appendix \ref{App:second_order}. What is most interesting is the final result, for which we find that
\begin{equation}
    \overline{\Big \|\sum_{\alpha < \beta} [H_{\alpha}, H_{\beta}] \Big \|^2} \lesssim 2 \times 10^{2} \, \frac{J^4 R^2}{N^2} + \mathcal O(R^2/N^3) \ ,
    \label{eq:error_trotter_in_SYK}
\end{equation}
which was first reported in Ref.~\cite{Baumgartner:2024ysk}. While the numerical coefficient in the RHS of Eq.~\eqref{eq:error_trotter_in_SYK} can considerably overestimate the actual error, the remarkable feature of Eq.~\eqref{eq:error_trotter_in_SYK} is that, for $R \sim N$, the result is independent of $N$. This means that, even when $N$ is large, we do expect the accuracy of the evolution to depend solely on the number of cycles $n$. In particular, one needs
\begin{equation}
    n \gtrsim \frac{M J^2 t^2 R}{\ve N}
\end{equation}
Trotter steps for the error in Eq.~\eqref{eq:U_T_steps} to be suppressed to any desired value $\ve$. Here, $M$ is an order one number, which can simply be derived from Eq.~\eqref{eq:error_trotter_in_SYK} for an analytical estimate, even though numerically one can check that it is much smaller. In order to confirm these expectations, we can numerically study how, on average, the two evolutions differ for different $n$'s or $\Delta t = t / n_{\rm max}$. In particular, we numerically realize the effective model and the Trotterized simulation, and fixing a total evolution time $t$, we consider the `global' probe
\begin{equation}
    \Delta U = \frac{1}{n_{\rm max}} \sum_{n = 1}^{n_{\rm max}} \overline{\Bigg \| U_{\rm eff}(t_n) - \Bigg( \prod_{\alpha = 1}^{R} e^{-i H_\alpha \Delta t} \Bigg)^n \Bigg\|} \ .
    \label{eq:DU}
\end{equation}
The additional sum over the various $n$'s gives the average error of the entire evolution, which helps in obtaining a smooth prediction.  
Arguably, $\Delta t$ is experimentally more convenient than $n_{\rm max}$, as it is the (inverse) frequency at which various realizations have to be cycled within the Trotterization. The prediction of this Section is that, for $\Delta t$ sufficiently small, $\Delta U$ increases proportionally with $\Delta t$. This is confirmed by the numerical results, which are shown in panel (\textbf{a}) of Figure \ref{SFF_trotter_and_Delta_SFF_Simulation}. It is rather interesting to notice that, when the simulation time becomes large ({\it e.g.}, for $t \gtrsim 10^3 /J$), the error no longer scales linearly with $t$, but the numerical results start showing hints of the quantum localization for the Trotter error of digital quantum simulations of chaotic systems, first found in Ref.~\cite{Hauke_trotter_Error}. 

As a side remark, we end this Section noticing that the chaotic nature of the evolution is also manifested in the fact that when $\Delta t$ becomes too large, the two unitary matrices become essentially random vectors, signaling an ergodic regime. In this case, the probe $\Delta U$ saturates to (approximately) $\sqrt{2}$, which is the expected length of the difference between two unit random vectors in a large Hilbert space\footnote{The two vectors are on average almost orthogonal, and thus the length of the difference is $\sqrt{2}$.}. 
The curves in Figure \ref{SFF_trotter_and_Delta_SFF_Simulation}a would saturate to this value if they would be continued to larger $\Delta t$ (not shown).

\subsubsection*{Bounds on observables}

Now that we have established that the Trotterization allows for a uniform approximation of the dynamics, we can approach the dynamical evolution of observables. In particular, we want to verify that observables that depend on the dynamics can be uniformly approximated by the Trotterized time evolution. This issue will typically depend strongly on the specific observable we choose. 

As a concrete example, we focus on the Spectral Form Factor (SFF), a quantity ubiquitously studied in many--body quantum chaos due to its universal properties. Recalling the definition of the canonical partition function $Z(\beta) = \Tr[e^{- \beta H}]$, the SFF is defined as
\begin{equation}
    \SFF (\beta, t) = \frac{|Z(\beta + i t)|^2}{Z^2(\beta)} = \frac{1}{Z^2(\beta)} \sum_{n,m} e^{- \beta (E_n + E_m) - i t (E_n - E_m)} \ ,
    \label{SFF_definition}
\end{equation}
where $E_n$ are the eigenenergies of $H$. 
For chaotic systems, the Spectral Form Factor shows a universal dip--ramp--plateau behavior when its characteristic fluctuations are smoothed out by coarse--graining in time, or by averaging over ensembles in disordered systems \cite{Haake_book}. Moreover, the connection with time evolution becomes transparent at infinite temperature ($\beta = 0$), since 
\begin{equation}
    \SFF (0, t) =  \frac{\big | \! \Tr[U(t)] \big|^2}{D^2} \equiv \SFF (t) \ .
\end{equation}
It is then apparent how the fluctuations in the SFF are a direct consequence of the erratic motion of the time evolution of the model within the space of unitary matrices, and we can ask the question whether the Trotterized time evolution can resolve such fluctuations for every single realization. 

Following the same route we adopted for the time evolution, we define the average distance between the two infinite-temperature SFF's
\begin{equation}
    \Delta \SFF  = \frac{1}{n_{\rm max}} \sum_{n = 1}^{n_{\rm max}} \overline{\big| \SFF_{\rm eff}(t_n) - \SFF_{\mathsf{T}}(t_n) \big|} \ ,
    \label{eq:DSFF}
\end{equation}
where $\SFF_{\rm eff}(t_n)$ is the SFF for the effective model and $\SFF_{\mathsf{T}}(t_n)$ is the Trotterized one. In particular, we wish to understand if the Trotterized approximation bounds uniformly the effective evolution of the SFF or not, when the Trotter step length $\Delta t$ vanishes. 

To tackle this problem, we may start from the fact that the SFF is not only a continuous function of the time evolution, but also everywhere differentiable. If the derivative is everywhere bounded, this implies that a small deviation in the time evolution operator $U$ would in turn also bound the error in the SFF. Because of this, one expects also a general bound on $\Delta U$ and $\Delta$SFF (as defined above) of the form
\begin{equation}
    \Delta \SFF \leq \kappa_{\rm L} \, \Delta U \ .
    \label{SFF_Lipschitz_relation_U}
\end{equation}
Functions that respect this relation are called Lipschitz continuous, and it can indeed be shown that the SFF belongs to this class. We devote the first part of Appendix~\ref{sec:lipscitz_bound} to showing that the SFF is Lipschitz continuous with constant
\begin{equation}
     1 \leq \kappa_{\rm L} \lesssim \frac{2 (D-1)}{D} \leq 2 \ ,
    \label{Lip_const_estimate}
\end{equation}
which implies that the error on the SFF is at most proportional to the one on the time evolution, with a constant of proportionality that does not scale with $N$ (and $D$). If we compare this result with the late time value of the plateau of the SFF, which is 
\begin{equation}
    \SFF(0, t) \to \frac{1}{D} \qquad \text{for} \qquad t \to \infty \ ,
\end{equation}
it is apparent that the bound \eqref{Lip_const_estimate} is not very constraining, as at late times the Trotterized SFF and the effective SFF could be very distant, at least relative to their value. On the other hand, a numerical evaluation of the SFF for both the effective model and the Trotterized time evolution shows that this, in general, does not happen. This is presented in panel (\textbf{c}) of Figure \ref{SFF_trotter_and_Delta_SFF_Simulation}, where we have computed one realization of the effective model for $N = 10$ in the half--filling sector. Here, we see that the Trotterized SFF follows the effective SFF not only at the beginning of the time evolution, but also when it reaches the plateau value of $1/D \approx 0.004$. The intuition behind this behavior is that the points in the space of unitary matrices where $\kappa_{\rm L} \simeq \cO(1)$ are very few, compared to a `bulk' where $\kappa_{\rm L} \lesssim \cO(D^{-1})$. In the second part of Appendix \ref{sec:lipscitz_bound}, we numerically check this fact by sampling $\kappa_{\rm L}$ from $10^{5}$ random pairs of unitary matrices with a small difference $\Delta U$, for each $4 \leq D \leq 40$. We find that $\kappa_{\rm L}$ fits well with
\begin{equation}
    \kappa_{\rm L} \approx \frac{1}{D} \ .
\end{equation}
This finding is also confirmed numerically by checking $\Delta$SFF for several values of $J  \Delta t$, keeping the total number of Trotter steps fixed at $n_{\rm max} = 2000$, as shown in panel (\textbf{b}) of Figure \ref{SFF_trotter_and_Delta_SFF_Simulation}. In this case, we notice that when $J\Delta t \sim 0.1$, the Trotterized SFF can follow precisely the effective SFF, where $\Delta U \approx 10^{-1}$ and $\Delta$SFF is $\approx 10^{-4}$, despite \eqref{SFF_Lipschitz_relation_U} ({\it i.e.} despite a possible $\cO(1)$ constant of proportionality between the two). On the other hand, when $J \Delta t \sim 1$, the two SFF's become effectively independent, since $\Delta$SFF saturates at the value of $1/D$.

The above analysis concludes our investigation on the differences between the target and Trotterized time evolution. We have shown that both at the level of unitary matrices, and at the level of observables that depend on the dynamics, there is a uniform convergence between the two below a certain threshold, which depends on the number of Trotter steps we want to consider (or, equivalently, the timescale we want to probe). With these results, we conclude that a Trotterized experimental simulation will closely approximate the exact dense effective model, with controlled errors as per our estimates.


\subsection{Statistical analysis of couplings} \label{sec:Statistical_analysis_couplings}

Motivated by the results of the previous Section, we now focus on the statistical distribution of couplings of our many--body disordered model, which amounts to studying the statistical distribution of
\begin{equation}
    \mathcal J_{I;K} \equiv \sum_{\alpha = 1}^R J^\alpha_{I} J^\alpha_{K} \ ,
    \label{eq:couplings_definition_sec_2}
\end{equation}
where, as stated above, $I$ and $K$ are two sets of indices, and $\alpha$ is a superscript that enumerates the various realizations. To make analytical progress, we must make some assumptions about the properties of $J_{I}^\alpha$. In particular, we assume all $J_{I}^\alpha$'s to be independent and identically distributed for all sets of indices $I$ and for each realization $\alpha$. A rather convenient choice is to take a Gaussian distribution, so that
\begin{equation}
    P \big(J_{I}^\alpha \big) \, \de J_{I}^\alpha = \frac{1}{\sqrt{2\pi \sigma^2}} \, e^{- \frac{1}{2\sigma^2} \big( J_{I}^\alpha \big)^2 } \, \de J_{I}^\alpha \ ,
\end{equation}
where $\sigma$ is the variance. 

We now initiate a study of the statistical properties of the effective model couplings \eqref{eq:couplings_definition_sec_2}, mostly focusing on their convergence to i.i.d. Gaussian random variables.
For details on the results presented in this Section, we refer the Reader to Appendix~\ref{App:Statistical_analysis_couplings_extended}.

\subsubsection*{Single--coupling convergence}

The first distribution we study is the distribution of a single coupling $\mathcal J_{I;K}$. To find it, we proceed in two steps. First, we find the distribution of a product of two independent Gaussian random variables, from which we derive the distribution after summing $R$ independent realizations. Both can be found explicitly. Taking two random variables $X_{1,2} \sim \mathcal N(0, \sigma^2)$, we can consider the product $Y = X_1 X_2$, and it is not hard to show that it is distributed as a Bessel function, namely
\begin{equation}
    P(Y) \, \de Y = \frac{1}{\pi \sigma^2} \, K_0\left(\frac{|Y|}{\sigma^2 } \right) \, \de Y \ .
    \label{eq:single_coup_marg}
\end{equation}
This is the distribution of the product of two Gaussian random variables. A simple calculation shows that the mean of this distribution vanishes, as expected, and the variance is $\mathbb E[Y^2] = \sigma^4$, thus the product of the two variances. This is not physically surprising, as the units of $Y$ are the products of the units of $X_{1,2}$. Now, to find the sum over different realizations, we can consider the {\it characteristic function}
\begin{equation}
    \varphi_Y(s) = \mathbb E \Big[ e^{i s Y} \Big] = \frac{1}{\sqrt{1 + \sigma^4 s^2}}  \ ,\label{eq:characteristic_function_single_coupling}
\end{equation}
which is just the Fourier transform of the probability distribution. A well--known statistical fact is that the characteristic function of the sum of two independent random variables is the product of the single characteristic functions, namely $\varphi_{Y_1+Y_2}(s) = \varphi_{Y_1}(s) \varphi_{Y_2}(s)$, which can be shown exploiting the independence of $Y_1$ and $Y_2$. The generalization for a sum of $R$ variables is the product of the $R$ characteristic functions. We use this fact to find the distribution of the sum of $R$ variables distributed as \eqref{eq:single_coup_marg}. We consider the variable
\begin{equation}
    \mathcal Y = \frac{1}{\sqrt{R}} \sum_{\alpha = 1}^R Y_\alpha \ .
\end{equation}
Notice that, differently from \eqref{eq:couplings_definition_sec_2}, here we are also adding a $R^{- \frac{1}{2}}$ prefactor which keeps the variance of $\mathcal Y$ equal to that of the $Y_{\alpha}$'s. This allows a cleaner comparison, but it is not very natural for experimental implementations, so we will only do it in this Section. Finding the characteristic function of $\mathcal Y$ (and performing an inverse Fourier transform) gives then the distribution of couplings for the sum of $R$ realizations, which is 
\begin{equation}
    P(\mathcal Y) \, \de \mathcal Y = \frac{\sqrt{R}}{\sqrt{\pi} \, \sigma^2 \, \Gamma(R/2)} \left( \frac{\sqrt{R} \, |\mathcal Y|}{2 \sigma^2} \right)^{\frac{R-1}{2}} K_{\frac{1-R}{2}} \left( \frac{\sqrt{R}  \, |\mathcal Y|}{\sigma^2} \right) \, \de \mathcal Y \ .
    \label{P_Sigma_exact}
\end{equation} 
This is the distribution of a single coupling $\mathcal J_{I;K}$ after integrating out all the other random variables. It is not hard to show that it has vanishing mean and variance $\sigma^4$, as we found before. In the limit of $R$ large, it approaches a Gaussian distribution, even though it is not immediate from the expression \eqref{P_Sigma_exact}. To see it, the simplest way is to expand the characteristic function for large $R$ in a power series in $1/R$, and then take the inverse Fourier transform\footnote{Notice that \eqref{P_Sigma_expansion_second_order} is not a probability distribution if we truncate the expansion at a finite order in $1/R$, as it becomes negative in some regions. However, it shows the Gaussian convergence in a clean way.}. We find that
\begin{multline}
    P(\mathcal Y) = \frac{e^{-\frac{\mathcal Y^2}{2 \sigma ^4}}}{\sqrt{2 \pi \sigma^4}} 
    + \frac{1}{ R } \left( \frac{3}{4} - \frac{3 \mathcal Y^2}{2\sigma ^{4}} + \frac{\mathcal Y^4}{4\sigma^8} \right) \frac{e^{-\frac{\mathcal Y^2}{2 \sigma ^4}}}{\sqrt{2 \pi \sigma^4}} \\
    + \frac{1}{ R^2 } \left(\frac{25}{32} - \frac{45 \mathcal Y^2}{8 \sigma^{4}} + \frac{65 \mathcal Y^4}{16\sigma^8} - \frac{17 \mathcal Y^6}{24\sigma^{12}} + \frac{\mathcal Y^8}{32\sigma^{16} } \right) \frac{e^{-\frac{\mathcal Y^2}{2 \sigma ^4}}}{\sqrt{2 \pi \sigma^4}} + \mathcal O(R^{-3})  \ ,
    \label{P_Sigma_expansion_second_order} 
\end{multline}
showing the convergence to a Gaussian distribution for large values of $R$.

\subsubsection*{Pairwise convergence and independence}

A similar strategy can be employed to study the joint probability distributions of two couplings $\mathcal J_{I;K}$ and $\mathcal J_{J;L}$, where at least one between $I$, $J$, $K$, and $L$ is different from the others\footnote{If they are all equal to each other, the two couplings coincide, and the situation is trivial.}. If the two sets of indices are such that $(I \cup K) \cap (J \cup L) = \varnothing$, the situation reduces to the previous Section: the two couplings are completely independent, and their joint probability distribution factorizes into a product of the form \eqref{P_Sigma_exact}. In contrast, when two pairs of indices coincide and two differ, a non--trivial joint probability distribution appears. To find it, we use the same strategy employed above. Assume $X_{1,2,3}$ to be i.i.d. random variables normally distributed as $\mathcal N(0, \sigma^2)$, and consider $Y_1 = X_1 X_3$ and $Y_2 = X_2 X_3$. The joint probability distribution can easily found to be (we refer to Appendix \ref{App:Statistical_analysis_couplings_extended} for details) 
\begin{equation}
    P(Y_1 , Y_2) = \frac{e^{- \sqrt{Y_1^2 + Y_2^2}/\sigma^2}}{2 \pi \sigma^2 \sqrt{Y_1^2 + Y_2^2} } \ ,
    \label{eq:two_couplings_probability_distribution}
\end{equation}
which shows, as expected, that $Y_1$ and $Y_2$ are not independent, as the joint probability distribution does not factorize\footnote{An interesting point to make here is that, on the other hand, $Y_1$ and $Y_2$ are uncorrelated, since
\begin{equation*}
    \mathbb E_{Y_1, Y_2} [Y_1 Y_2] = \mathbb E_{X_1, X_2, X_3} [X_1 X_2 X_3^2] = 0 \ .
\end{equation*}}. On the other hand, as before it can be shown that sums over different realizations does make them independent. To prove it, we employ the same strategy presented above. We compute the characteristic function, which is 
\begin{equation}
    \varphi_{Y_1, Y_2}(s_1, s_2) = \mathbb E_{Y_1, Y_2} \Big[ e^{i s_1 Y_1 + i s_2 Y_2} \Big] = \frac{1}{\sqrt{1 + \sigma^4 \big(s_1^2 + s_2^2 \big)}}  \ ,
\end{equation}
clearly reminiscent of \eqref{eq:characteristic_function_single_coupling}. As before, we use this Fourier transform to find the joint probability distribution of sums of independent random variables distributed as \eqref{eq:two_couplings_probability_distribution}, appropriately normalized, such as 
\begin{equation}
    \mathcal Y_1 = \frac{1}{\sqrt{R}} \sum_{\alpha = 1}^R Y_{1,\alpha} \ , \qquad \text{and} \qquad \mathcal Y_2 = \frac{1}{\sqrt{R}} \sum_{\alpha = 1}^R Y_{2,\alpha} \ .
\end{equation}
The joint probability distribution is then found taking the inverse Fourier transform of the characteristic function for $\mathcal Y_1$ and $\mathcal Y_2$. Unfortunately, , in contrast to \eqref{P_Sigma_exact}, we have not been able to perform analytically this inverse transformation. Nonetheless, we can still find the large--$R$ convergence, which is 
\begin{equation}\label{eq.convergenceTrotterizedModel}
    P(\mathcal Y_1, \mathcal Y_2) = \frac{e^{-\frac{\mathcal Y_1^2 + \mathcal Y_2^2}{2 \sigma^4}}}{2 \pi \sigma^4} + \frac{1}{R} \left(2 -  \frac{2( \mathcal Y_1^2 + \mathcal Y_2^2)}{\sigma^{4}} + \frac{(\mathcal Y_1^2 + \mathcal Y_2^2)^2}{4\sigma^{8}} \right) \frac{e^{-\frac{\mathcal Y_1^2 + \mathcal Y_2^2}{2 \sigma^4}}}{2 \pi \sigma^4} + \mathcal O(R^{-2}) \ .
\end{equation}
Again, we see that in the limit of large--$R$, the joint distribution factorizes, making the two couplings effectively independent, up to $1/R$ corrections.

\subsubsection*{Effective vs local independence}

Finally, we can also study the joint probability distribution between the local couplings $J_{I}^\alpha$ and the effective coupling $\mathcal J_{I;K}$. Such a probability distribution is interesting as it quantifies the independence between the effective model and any spurious effect that couples to the experimental apparatus, in particular to the single--body coupling $J_{I}$. 
We employ the same strategy used for the two previous examples. Assume two random variables $X_{1,2} \sim \mathcal N(0, \sigma^2)$, and consider again the product $Y = X_1 X_2$. Using a simple change of coordinates, we find that their joint probability distribution is 
\begin{equation}
    P(X_1,Y)  = \frac{1}{2 \pi \sigma^2 |X_1|} \, e^{-\frac{Y^2}{2 \sigma^2 X_1^2} - \frac{X_1^2}{2 \sigma^2}}  \ .
\end{equation}
As before, in order to consider a sum over different realizations, it is convenient to compute the characteristic function. For simplicity, we omit its expression here but it can be found in \eqref{eq:characteristic_function_single_vs_double}. Defining now
\begin{equation}
    \mathcal X = \frac{1}{\sqrt{R}} \sum_{\alpha = 1}^R X_{1, \alpha} \  \qquad \text{and} \qquad \mathcal Y = \frac{1}{\sqrt{R}} \sum_{\alpha = 1}^R Y_{\alpha} \ ,
\end{equation}
an inverse Fourier transform of the characteristic function for $\mathcal X$ and $\mathcal Y$ gives the joint probability distribution, but it is not possible to express it in terms of elementary functions. We employ a large--$R$ expansion to find
\begin{equation}
    P(\mathcal X, \mathcal Y) = \frac{e^{-\frac{\mathcal X^2}{2 \sigma^2} -\frac{\mathcal Y^2}{2 \sigma^4}}}{2 \pi \sigma^3} + \frac{1}{R}\left(\frac{5}{4} - \frac{\mathcal X^2}{2\sigma^2} - \frac{2 \mathcal Y^2}{\sigma^4} + \frac{\mathcal X^2 \mathcal Y^2}{2\sigma^6} + \frac{\mathcal Y^4 }{4\sigma^8 } \right) \frac{e^{-\frac{\mathcal X^2}{2 \sigma^2} -\frac{\mathcal Y^2}{2 \sigma^4}}}{2 \pi \sigma^3} + \mathcal O (R^{-2}) \ ,
    \label{eq:joint_distr_local_vs_eff}
\end{equation}
which mimics the behavior found previously, namely independence between the various couplings at large--$R$. Notice that, as expected by dimensional analysis, the variances of $\mathcal X$ and $\mathcal Y$ are different, as it is shown in the exponents of \eqref{eq:joint_distr_local_vs_eff}.

\subsubsection*{Summary of the results}

The results obtained in this Section establish statistical independence between all random variables at large enough $R$. At first sight, this analysis might look rather redundant, following the general statistical lore that {\it distributions converge to independent Gaussian variables}, formalized in the Central Limit Theorem (CLT). This is certainly accurate, and the statistical independence does indeed arise because of the CLT. However, on a deeper look, it is a rather striking fact. For instance, considering the case of the pairwise convergence studied above in \eqref{eq.convergenceTrotterizedModel}, at large--$R$ the two variables $\mathcal Y_1$ and $\mathcal Y_2$ are constructed sharing $R$ identical and $R$ independent random variables, making the statistical independence a non--trivial phenomenon to occur. 

Unfortunately, as pointed out above, a unified treatment of the statistical convergence of all couplings at generic $R$ and $N$ is not possible. In particular, finding a convenient set of independent random variables at generic $R$ is rather difficult, and the corresponding probability distribution functions are even harder to obtain. We end this study on the statistical properties of the couplings of such models computing some information--theoretic quantities, such as the Shannon entropy and the relative entropy, to further benchmark the convergence to a fully disordered model.


\subsection{Entropic measures} \label{sec:Relative_entropy}

We now quantify the difference between the low-- and full--rank versions of the models above by calculating the Shannon entropy and the relative entropy of the coupling distributions found above.

\subsubsection*{Shannon entropy}

The Shannon entropy \cite{Shannon_1948} is an information--theoretic quantity that quantifies the amount of uncertainty (or equivalently of information) contained in a random variable $X$. It is defined as 
\begin{equation}
    H[X] = - \int_{\mathsf X} P(X) \log \left( P(X) \right) \ ,
\end{equation}
where $P(X)$ is the probability density function of the random variable $X$, and $\mathsf X$ is the set of all possible values of $X$. In physical terms, the Shannon entropy measures the amount of randomness contained in a variable. Furthermore, among all distributions that have a fixed variance, it is maximized by the Gaussian distribution. 

We can compute the Shannon entropy for the single coupling distribution \eqref{P_Sigma_exact}. In order to find analytical expressions, we will only consider the expansion over large values of $R$, already presented in \eqref{P_Sigma_expansion_second_order}. Using this result, the Shannon entropy is obtained with a straightforward calculation, yielding
\begin{equation}
    H[\mathcal Y]  = \frac{1}{2} \Big(1 + \log(2 \pi \sigma^4) \Big) - \frac{3}{4 R^2} + \mathcal O(R^{-3}) 
    \label{Shannon_entropy_K_distr}
\end{equation}
up to second order in $1/R$. 
The first term in the RHS is the result for a Gaussian with variance $\sigma^4$, which was expected given the Gaussian convergence at infinite $R$, as shown in \eqref{P_Sigma_exact}. The second term is a $\sigma$--independent correction which interestingly is of second order in $1/R$ since the term proportional to $1/R$ vanishes. Moreover, we notice that the negative sign was to be expected, since the Shannon entropy at fixed variance is maximized by the Gaussian distribution. 

Interestingly, the result \eqref{Shannon_entropy_K_distr} tells us that in our specific case, while the convergence of the distribution scales as $1/R$, the entropy saturates faster, in particular as $1/R^2$. This is a very specific feature of the large--$R$ expansion \eqref{P_Sigma_expansion_second_order}, since in principle one could have only guessed a $1/R$ convergence. For instance, it can be generically shown that for any small perturbation of the form
\begin{equation}
    P(x) = \mathcal N_{\sigma^4}(X) + \varepsilon \, F(X) + \mathcal O(\varepsilon^2) \ , \qquad \text{with} \qquad \int_{\mathsf X} F(X) = 0 \ ,
\end{equation}
where $\mathcal N_{\sigma^4}(X)$ is a Gaussian of variance $\sigma^4$ and the second condition ensures a proper normalization, the Shannon entropy of $P(X)$ is easily found to be 
\begin{equation}
    H[X]  = \frac{1}{2} \Big(1 + \log(2 \pi \sigma^4) \Big) + \varepsilon \int_{\mathsf X}F(X) \log(\mathcal N_{\sigma^4}(X)) + \mathcal O(\varepsilon^2) \ ,
    \label{Shannon_entropy_generic_perturbation}
\end{equation}
and thus in general shows a convergence that scales linearly in $\varepsilon$. The fact that the Shannon entropy shows a quadratic perturbative behavior implies that integral on the RHS of \eqref{Shannon_entropy_generic_perturbation} vanishes, which requires some fine tuning as the integrand is an even function.

\subsubsection*{Relative entropy, or KL divergence}

Another entropic quantity we can measure is the relative entropy, also called the Kullback–Leibler (KL) divergence, between the target Gaussian distribution and the single coupling distribution~\eqref{P_Sigma_exact}. The KL divergence is a statistical distance between two probability distributions, and quantifies how much they differ. It is defined by the formula
\begin{equation}
    D_{\rm KL}\left(Q || P \right) = \int_{\mathsf X} Q(X) \log \left( \frac{Q(X)}{P(X)}\right) \ .
    \label{eq:relative_entropy}
\end{equation}
We are interested in this quantity when $Q(X)$ is the target Gaussian distribution, and $P(X)$ is the low--rank single coupling distribution. Before going to any specific computations, the Berry--Esseen equality allows us to bound the convergence to zero with $R$ of $D_{\rm KL}\left(Q || P \right)$ quite generally, stating that there exists a constant $K$, such that \cite{bobkov2014berry}
\begin{equation}
    D_{\rm KL}\left(Q || P \right) \le \frac{K}{R}\,,
\end{equation}
where $K$ depends on the moments of the individual distributions. Again, in the case of interest to us, the rate of convergence is actually faster, since it is of the order $1/R^2$. This can be shown simply using the result \eqref{P_Sigma_expansion_second_order}. In particular, another tedious but simple computation shows that
\begin{equation}
    D_{\rm KL}\left(\mathcal N_{\sigma^4} || P \right) = \frac{3}{4 R^2} + \mathcal O(R^{-3}) \ .
    \label{K_dis_relative_entropy}
\end{equation}
This result is not surprising, given what we have found in \eqref{Shannon_entropy_K_distr} and the discussion below, but it is nonetheless interesting to remark that the first correction term is completely independent of the variance of the individual distributions. 

To summarize, in this Section we have shown that entropic quantities that can be computed from the low--rank distributions found earlier in Section \ref{sec:Statistical_analysis_couplings} show a fast $1/R^2$ convergence to the Gaussian value, as opposed to the most generic behavior of $1/R$. At present, we do not have any clear physical interpretation of this result, and at this level of analysis we can regard it as a fortunate coincidence that increasing $R$, the amount of disorder grows faster than expected. On the other hand, it is tempting to speculate on a possible relation with Ref.~\cite{Kim:2019lwh}, which showed that the SYK physics is reached when $R \sim N$, where the rank of the tensor of couplings $\mathcal J_{i_1i_2k_1k_2}$ is well below saturation (which is the case $R \sim N^2$). It is therefore possible that the results \eqref{Shannon_entropy_generic_perturbation} and \eqref{K_dis_relative_entropy} should be read as avatars of the fact that physical quantities converge faster to the fully disordered physics than probability distributions. Unfortunately, to corroborate this idea it would be best to analyze entropic quantities for the total joint distribution of the couplings at generic $N$ and $R$, which as pointed out before is highly non--trivial.


\subsection{A concrete example: low--rank \texorpdfstring{cSYK${}_4$}{}}
\label{sec:Concrete_example}

In the previous Section, we showed how it is possible to generate fully disordered many--body Hamiltonians from fewer--body interactions, summing over different realizations of the couplings. In particular, we studied how the ensemble of the couplings approaches a set of independent Gaussian random variables. To give a concrete example of the utility of this study, we come back to the low--entropy model
\begin{equation}
    H_{\alpha} = \sum_{i_1 i_2, k_1 k_2} J^{\alpha}_{i_1 k_1} J^{\alpha}_{i_2 k_2} \, c_{i_1}^{\dagger} c_{k_1} c_{i_2}^{\dagger} c_{k_2} \ , \qquad \qquad \overline{J_{i_1 k_1}^\alpha J_{i_2 k_2}^\alpha} =  \frac{\sqrt{2} J}{N^{\frac32}} \, \delta_{i_1 k_2} \delta_{i_2 k_1} \ ,
    \label{eq:H_sparse}
\end{equation}
which resembles a complex SYK (cSYK) model. The model in \eqref{eq:H_sparse} is quite interesting, as it appears when fermionic sites exchange a disordered interaction through a mediator which is integrated out, effectively creating a two--body all--to--all disordered model. Variations of this idea have been considered in \cite{Marcus-Vandooren_2018, Kim:2019lwh, Esterlis_2019, Wang_2020, Wang_2020_2, Choi_2022, Davis_2023, Solis:2025clm}, and we devote Section \ref{sec:Experimental_implementation} to a particular cQED implementation. The identification with the cSYK model is not quite exact, for two reasons. 
\begin{enumerate}

    \item The random couplings $\mathcal J_{i_1 i_2 k_1 k_2} = J_{i_1 k_1} J_{i_2 k_2}$ are not all independent, effectively reducing the full amount of disorder. Taking inspiration from our previous Section, summing over different realizations of the couplings, we obtain the effective model
    \begin{equation}
        H_{\rm eff} = \frac{1}{\sqrt{R}} \sum_{\alpha = 1}^R H_{\alpha} = \sum_{i_1 i_2, k_1 k_2} \left( \frac{1}{\sqrt{R}} \sum_{\alpha=1}^R J_{i_1 k_1}^{\alpha} J_{i_2 k_2}^{\alpha} \right) \, c_{i_1}^{\dagger} c_{k_1} c_{i_2}^{\dagger} c_{k_2} \ .
        \label{eq:H_alpha_simple_model_2}
    \end{equation}
    where each $H_{\alpha}$ is of the form \eqref{eq:H_sparse} for a different realization $J^\alpha_{ik}$ of the couplings. The factor of $R^{-1/2}$ in front of the sum allows for the meaningful comparison of different numbers of realizations, as it keeps the variance of the couplings fixed. Our statistical study shows that the couplings 
    \begin{equation}
        \mathcal J_{i_1 i_2 k_1 k_2} \equiv \frac{1}{\sqrt{R}} \sum_{\alpha=1}^R J_{i_1 k_1}^{\alpha} J_{i_2 k_2}^{\alpha} \ , \qquad \qquad \overline{\mathcal J_{i_1 i_2 k_1 k_2}^2} = \frac{2 J^2}{N^3}\ ,
        \label{eq:Low_rank_SYK_definition}
    \end{equation}
    are independent Gaussian random variables, provided we take $R$ sufficiently large. How large strictly depends on the physics we are interested in, but to be concrete, Refs.~\cite{Marcus-Vandooren_2018, Kim:2019lwh} showed that when $R \sim N$, at large--$N$ (a Majorana version of) this model becomes maximally scrambling. Throughout this work, we will mainly leave $R$ unspecified, but in most cases from now on we should think of it as $R \sim N$. Additionally, we remark that not all couplings $\mathcal J_{i_1 i_2 k_1 k_2}$ have zero mean, in particular the ones of the form $\mathcal J_{i_1 i_2 i_2 i_1}$, but such means are all equal so that they can be taken out of `diagonal' couplings $\mathcal J_{i_1 i_2 i_2 i_1}$ at the expense of adding an operator of the form $f(\hat N)$ (where $\hat N$ is the number operator), which is a constant shift in the Hamiltonian at fixed filling. 

    \item One might be worried that the fermionic ladder operators in \eqref{eq:H_sparse} are not normal ordered, and when putting them in this form, additional one--body terms appear, `contaminating' the pure cSYK$_4$ physics. The two--body terms are (neglecting a proportionality factor of $R^{-1/2}$)
    \begin{equation}
        \sum_{i_1 k_2, i_2} \left( \sum_{\alpha=1}^R J_{i_1 i_2}^{\alpha} J_{i_2 k_2}^{\alpha} \right) \, c_{i_1}^{\dagger} c_{k_2} = \sum_{i_1, i_2} \left( \sum_{\alpha=1}^R J_{i_1 i_2}^{\alpha} J_{i_2 i_1}^{\alpha} \right) \, c_{i_1}^{\dagger} c_{i_1} + \sum_{i_1 \neq k_2, i_2} \left( \sum_{\alpha=1}^R J_{i_1 i_2}^{\alpha} J_{i_2 k_2}^{\alpha} \right) \, c_{i_1}^{\dagger} c_{k_2}  \ .
    \end{equation}
    On the RHS, we have divided the various terms into a `diagonal' piece and an `off--diagonal' piece, respectively. The former has a non--zero mean, which is again proportional to the number operator, namely
    \begin{equation}
        \sum_{i_1, i_2} \left( \frac{1}{\sqrt{R}} \sum_{\alpha=1}^R \overline{J_{i_1 i_2}^{\alpha} J_{i_2 i_1}^{\alpha}} \right) \, c_{i_1}^{\dagger} c_{i_1} = \frac{\sqrt 2 J}{\sqrt{N}}\sum_{i_1} \, c_{i_1}^{\dagger} c_{i_1} = \sqrt 2 J \, \frac{\hat N}{\sqrt{N}} \ .
    \end{equation}
    As before, this is just a constant shift in the Hamiltonian, which we can neglect. The remaining pieces contribute to an SYK$_{2}$ with variance
    \begin{equation}
        \frac{1}{R}\sum_{i_2, j_2}\sum_{\alpha, \beta =1}^R \overline{J_{i_1 i_2}^{\alpha} J_{i_2 k_2}^{\alpha} J_{i_1 j_2}^{\beta} J_{j_2 k_2}^{\beta}} = \frac{2 J^2}{N^2} \ .
    \end{equation}
    This variance has a $N^{-2}$ scaling, parametrically smaller than the $N^{-1}$ scaling of SYK$_2$, meaning that this contribution of the Hamiltonian will vanish at large $N$.
\end{enumerate}
The outcome of the above discussion is that taking $R \sim N$ and $N$ sufficiently large, the effective model of equation \eqref{eq:H_alpha_simple_model_2} reproduces the physics of cSYK$_4$.

\subsubsection*{Numerical simulations of the effective model}

\begin{figure}[t]
\centering
\includegraphics[width=0.960\textwidth]{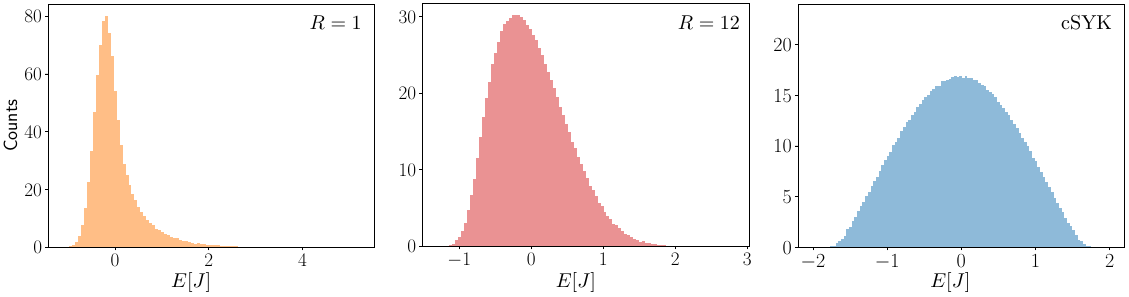}
\caption{Spectral densities of the effective model \eqref{eq:H_alpha_simple_model} for $R = 1$ and $R = N$, choosing $N = 12$, $J = 1$ and diagonalizing the Hamiltonian in the half--filled sector. We also show the spectral density of cSYK. While for low $R$'s the density is rather skewed, it slowly converges to the density of cSYK. }
\label{fig:probability_density_spectrum}
\end{figure}

To put in practice the ideas presented in the previous Section, we want to study an observable of the effective model \eqref{eq:H_alpha_simple_model} for different values of $R$ in order to benchmark the convergence to SYK. As previously mentioned, similar studies for Majorana SYK were performed in Refs.~\cite{Marcus-Vandooren_2018, Kim:2019lwh}, showing how taking $R = \gamma N$ converges to (Majorana) SYK for different observables, either analytically or by solving the Schwinger--Dyson equations numerically. Keeping in mind our goal of implementing a similar model in an optical cavity, where possible `contaminations' can appear in coupling distributions (more on it in Section \ref{sec:Experimental_implementation}), we decide to analyze the model through exact diagonalization. This method is clearly limited by the relatively small system attainable ($N \lesssim 14$), but it can be applied to any variation of SYK, making it a suitable tool to compare with experimental predictions. 

For the present example, we focus on the spectral density of the model \eqref{eq:H_sparse} (where we have taken out the two--body term), diagonalizing several realizations of the Hamiltonian in the half--filling sector. We have realized the system 100 times, combining the resulting spectra. Unfortunately, the statistical nature of the system does not allow for clean numerical comparison with the results of the previous Section at large $R$. However, on a qualitative basis, it is clear from Figure \ref{fig:probability_density_spectrum} that the spectrum slowly converges to that of cSYK.

To summarize the results up to here, we have shown that summing over different realizations of a Hamiltonian with correlated disorder, it is possible to increase the entropy of the couplings, effectively making them independent by summing over different realizations of the Hamiltonian. While on one hand, this is a simple consequence of the Central Limit Theorem, we have shown how the convergence rate also affects observables of the model. In the next Section, we will focus on proposing an experimental scheme to perform dynamical quantum simulations of such effective models.


\section{Experimental implementation of cSYK}
\label{sec:Experimental_implementation}
We now turn our attention to the applicability of our previous theoretical results in experimental platforms, specifically in the setup of single--mode optical cavities. Our ultimate goal is to engineer the dense disordered interactions in an experimental platform, and we will thus briefly present the main experimental ingredients to engineer and control our desired disordered all--to--all interactions. 

After that, we will show in detail how the cavity Hamiltonian effectively reduces to that of cSYK$_4$ in a suitable range of the parameters. We will then check the chaotic nature of our effective model utilizing prominent benchmarks of chaotic dynamics (SFF and OTOCs), finishing with some remarks on state preparation and the effective number of interacting sites.


\subsection{Cavity QED basics} 
\label{sec:cQED_basics}
 
The platform on which we aim to simulate cSYK is a quantum degenerate Fermi gas in a high-finesse optical cavity. Thanks to numerous recent advances in this field, optical cavities have emerged as a highly promising playground for the realization of disordered many--body systems. For a pedagogical review, see, {\it e.g.}, Ref.~\cite{Mivehvar_2021}. Particularly noteworthy is the fact that cavity QED systems naturally realize long--range one--body interactions \cite{PhysRevLett.84.4068, PhysRevLett.91.203001, Esslinger_2010, PhysRevX.8.011002, science.aar3102, science.abd4385, Periwal2021, RevModPhys.95.035002, RevModPhys.85.553, Brantut_2023}, which is extremely advantageous for simulating models like SYK.\footnote{This is due to the electromagnetic field mediating collective atom--photon interactions over long distances, in contrast to, \textit{e.g.}, the typically short-range Coulomb interactions between charged particles.}
Moreover, recent seminal work has demonstrated that such interactions can be experimentally controlled with high precision, including the introduction of tunable disorder in the atom--light coupling \cite{Brantut_rand_spin_model_2023}. These capabilities have already been exploited in previous proposals for implementing the SYK model, such as in Ref.~\cite{Uhrich:2023ddx}. In that work, to construct the desired two--body couplings of the SYK model from the naturally occurring one--body couplings $J_{ij}$, it was proposed to utilize the multi--mode structure of an optical cavity. By coupling to a large number of cavity modes, one can enhance the connectivity of the system, effectively promoting a sparsely interacting model to one with a denser interaction structure characteristic of SYK.

In contrast, our proposal (first presented in Ref.~\cite{Baumgartner:2024ysk}) differs primarily in the choice of the cavity mode structure. Specifically, we consider an experimentally simpler setup, requiring coupling to only one single cavity mode. However, this simplification comes at the cost of requiring multiple realizations of the interactions via repeated Trotter steps. To better understand the technical differences, we first provide a brief overview of optical cavity QED systems, followed by a discussion of the atom--light interactions responsible for generating the disordered one--body couplings $J_{ij}$.

\subsubsection*{Light--Matter interactions}

We consider a cloud of $N_{\rm at}$ trapped ${}^6$Li atoms placed inside an optical resonator formed by two high--quality mirrors, which create a standing electromagnetic wave. The spatial profile of this cavity mode is determined by the cavity geometry, {\it i.e.}, the spacing and curvature of the mirrors. 
The atomic cloud, modeled as an ensemble of two--level systems, consists of a ground state $\ket{g}$ and an excited state $\ket{e}$, separated by an atomic transition frequency $\omega_{\rm a}$, see Figure \ref{fig:disorder_detunings}. When this frequency is close to the cavity photon frequency $\omega_{\rm c}$, interactions between atoms are enabled through the exchange of cavity photons (denoted by the operator $a$ and with intensity $\Omega_{\rm c}$). The high reflectivity of the cavity mirrors enhances these photon--mediated couplings, allowing photon exchange to occur on timescales faster than photon loss. While the system operates in the strong--coupling regime of cavity QED, the cooperativity is finite, and some photon leakage inevitably occurs---this will be discussed in more detail in Section \ref{sec:Dissipation}.

\begin{figure}[t]
\centering
\begin{tikzpicture}
    \node[anchor=south west, inner sep=0] (img1) at (-4,0) {\includegraphics[width=0.4\textwidth]{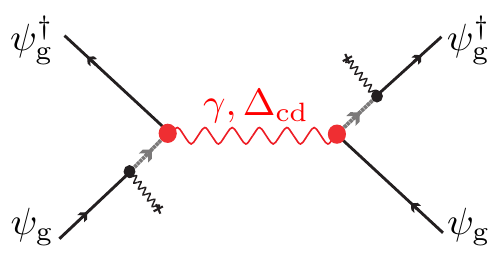}};
    
    \node[anchor=south west, inner sep=0] (img2) at (3.2,0.4) {\includegraphics[width=0.1\textwidth]{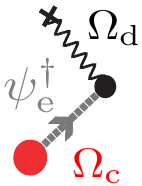}};

    \draw[gray] (0.65,2.2) ellipse [x radius=0.5, y radius = 0.65];
    \draw[gray] (4.1,1.45) ellipse [x radius=1.0, y radius = 1.3];
    
    \draw[gray] (0.68, {2.2 + 0.65}) to[out=-20, in=-160] (3.7, {1.45 + 1.2});
    \draw[gray] (0.65, {2.2 - 0.65}) to[out=-5, in=150] (3.7, {1.45 - 1.2});
    
        
    \draw[->, thick] (5.8,1.8) -- (6.8,1.8);
    
    \node[anchor=south west, inner sep=0] (img3) at (6.8,0) {\includegraphics[width=0.29\textwidth]{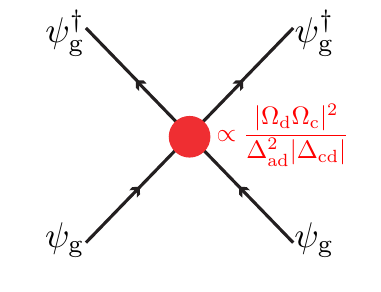}};
\end{tikzpicture}
\caption{\textbf{Left}: Diagram of the interaction process involving ground--state atoms $\psi_{\rm g}$, excited states $\psi_{\rm e}$, and the cavity photons $\gamma$. The atoms are driven by an external drive field with Rabi frequency $\Omega_{\rm d}$, detuned from the excited state by $\Delta_{\rm ad}$, while the cavity detuning is $\Delta_{\rm cd}$. \textbf{Right}: Effective four--fermion interaction generated after adiabatic elimination of the excited state and cavity photons, with coupling strength scaling as $\tfrac{|\Omega_{\rm d}\Omega_{\rm c}|^2}{ \Delta^2_{\rm ad}|\Delta_{\rm cd}|}$.}
\label{fig:cavity_interactions}
\end{figure}
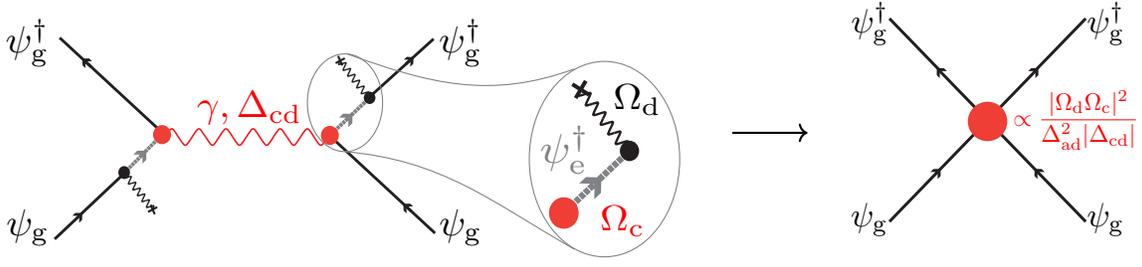

To induce coherent dynamics between the atomic states and the cavity mode, we introduce a classical transverse pump laser, often referred to as the drive beam. This laser is detuned from the atomic resonance by a large amount $\Delta_{\rm ad} = \omega_{\rm a}-\omega_{\rm d}$, where  $\omega_{\rm d}$ is the drive frequency and $\Omega_{\rm d}$ its intensity. The purpose of this large detuning is to prevent real population of the excited state $\ket{e}$ and thereby enable, in a later step, its adiabatic elimination from the dynamics. However, the drive beam facilitates virtual transitions through $\ket{e}$, which---when combined with the cavity--mediated coupling---lead to effective interactions between atoms remaining in the ground state $\ket{g}$. Figure \ref{fig:cavity_interactions} shows the Feynman diagrams corresponding to the cavity interactions before and after the adiabatic elimination. 

The effective interactions are determined by the spatial overlap of the atomic cloud with the drive and cavity mode profiles. By tailoring the intensity and spatial dependence of the drive field, one can exert experimental control over both the strength and spatial structure of the induced atom--atom couplings. This capability provides a platform for engineering programmable disordered interactions. Indeed, recent advances \cite{Brantut_rand_spin_model_2023, Baghdad2023, Lei2023} have demonstrated methods for arbitrary control of disorder in the cavity QED context. 
In the following, we outline how such platforms can, in turn, be used to realize cSYK${}_4$ interactions between atoms in the cavity.

\subsubsection*{Disorder}

\begin{figure}[t]
\centering
\includegraphics[width=0.75\textwidth]{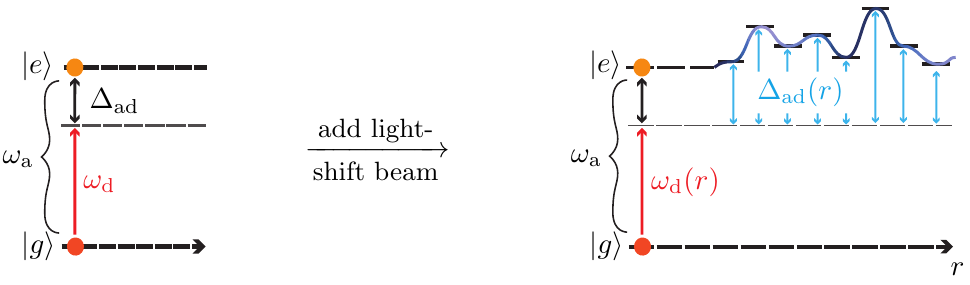}
\caption{Comparison of the atomic two--level structure without and with disordered detunings, induced by an additional light--shift beam.}
\label{fig:disorder_detunings}
\end{figure}

We consider a scenario where the atomic cloud is shaped like a pancake---thin along the cavity axis and extended in the two--dimensional transverse plane. In this geometry, we again assume ${}^6$Li atoms to be two--level systems, and a classical drive beam couples their internal states. The corresponding level structure is depicted on the left side of Figure \ref{fig:disorder_detunings}.
To introduce controlled disorder, we add a second laser field referred to as the light--shift beam, with frequency $\omega_{\rm b}$. This beam is engineered to have a spatially random intensity profile, such as a speckle pattern, making its Rabi frequency $\Omega_{\rm b}(r)$ position--dependent.
As such, this light--shift beam causes an AC--Stark shift of the energy of the excited state $\ket{e}$, which in turn induces a shift of the detuning frequency $\Delta_{\rm ad}$.

Experimentally, such a disordered light field can for example be created using an optical speckle pattern---for instance, by shaping the beam with a spatial light modulator (SLM) and focusing it onto the atomic cloud \cite{orsi:2024ab}. 

This generates an effective sparse model, as the rank of the interaction is low. To densify the disorder using only a single--mode optical cavity, we additionally cycle through different speckle patterns in time, which in turn make the atom--drive detuning $\Delta_{\rm ad}$ both spatial and time dependent. However, as shown in Section \ref{sec:SparseToFullSYK4}, by quickly cycling over a fixed amount of detuning profiles such time dependence disappears in the effective model, increasing the rank of the effective couplings.


\subsection{Effective cavity Hamiltonian} \label{sec:Effective_cavity_Hamiltonian}

Having explored the basic ingredients of the cavity QED platform, we are now ready to analyze a concrete setup, deriving an effective Hamiltonian that realizes SYK${}_{4}$ long--range interactions. This setup was first proposed in Ref.~\cite{Uhrich:2023ddx} for a multi--mode cavity, and later adapted for a single--mode cavity in Ref.~\cite{Baumgartner:2024ysk}. In the following, we will focus on the latter. 

We continue to assume the pancake--shaped geometry for the atomic cloud, which is thin along the cavity axis and extended in the transverse directions, neglecting any movement in the direction of the cavity axis. The total Hamiltonian can schematically be written as
\begin{equation}
    H = H_{\rm kt} + H_{\rm c} +  H_{\rm a} + H_{\rm ac} + H_{\rm ad} \ .
\end{equation} 
We now introduce each term in detail. The first contribution $H_{\rm {kt}}$ describes the kinetic and potential energy associated with the center--of--mass motion of atoms in both ground and excited states, confined by a harmonic trap. It is given by, 
\begin{equation}
    H_{\rm kt} = \sum_{\rm s \in \{ \rm e,g\}} \int \de^2r \, \psi_{\rm s}^{\dagger}(r) \left( \frac{p^2}{2 m_{\rm at}} + V_{\rm t}(r) \right) \psi_{\rm s}(r) \ .
\end{equation}
Here, $\psi_{\rm s}(r)$ denotes the field operator for an atom in state $\rm s\in \{\rm g, e\}$, $m_{\rm at}$ is the atomic mass (for \textsuperscript{6}Li), and $V_{\rm t}(r)$ is the external trapping potential in the transverse plane of the atomic cloud. The second term $H_{\rm c}$ accounts for the energy of a single-mode optical cavity,
\begin{equation}
    H_{\rm c} = \omega_{\rm c}\, a^\dagger a \ ,
\end{equation}
where $\omega_{\rm c}$ is the cavity frequency, and $a, a^{\dagger}$ are the photon creation and annihilation operators for the cavity mode. Unlike the multi--mode setup discussed in \cite{Uhrich:2023ddx}, here we restrict ourselves to a single--mode cavity, which leads to important differences in the resulting effective interactions.
The third contribution $H_{\rm a}$ captures the internal energy of atoms in the excited state
\begin{equation}
    H_{\rm a} =  \int \de^2 r \;  \omega_{\rm a}(r) \, \psi_{\rm e}^{\dagger}(r) \,\psi_{\rm e}(r) \ .
\end{equation}
The local transition frequency $\omega_{\rm a}(r)$ acquires a spatial dependence due to an externally applied light--shift beam, as explained in the previous subsection. More precisely, the light shift beam introduces a position dependency on the detuning $\Delta_{\rm ad}(r)$ which therefore changes $\omega_{\rm a}$ to $\omega_{\rm a}(r) = \Delta_{\rm ad}(r) + \omega_{\rm d}$ (see Figure \ref{fig:disorder_detunings}). 

The final two terms describe the atom--light interactions $H_{\rm ac}$ accounts for the coupling between atoms and the cavity mode, while $ H_{\rm ad}$ describes the coupling to the drive field:
\begin{align}
    H_{\rm ac} \, = & \,  \frac{\Omega_{\rm c}}{2} \int \de^2 r \, \left( g_{\rm c}(r) \psi_{\rm e}^{\dagger}(r)\psi_{\rm g}(r) a + \rm h.c. \right) \ ,\label{eq.Hac} \\ 
    H_{\rm ad} \, = & \, \Omega_{\rm d} \int \de^2 r \, \left(  g_{\rm d}(r) \, e^{-i\omega_{\rm d}t}\psi_{\rm e}^{\dagger}(r)\psi_{\rm g}(r)  + \rm h.c. \right) \, .\label{eq.Had_oscillating} 
\end{align}
Here, $\Omega_{\rm c}$ and $\Omega_{\rm d}$ are the respective coupling strengths, and $g_{\rm c}(r)$ and $g_{\rm d}(r)$ describe the spatial profiles of the cavity and drive fields over the atomic cloud. Attentive readers will recognize equations $\eqref{eq.Hac}$ and $\eqref{eq.Had_oscillating}$ as the QED interaction vertices between charged particles and light, suitably divided between the interaction with the classical field (drive) and the quantum field (photons from cavity mode). In our two--dimensional cloud, the cavity and drive mode profiles $g_{\rm c/d}(r)$ are described by Hermite--Gauss modes. However, we take the experimentally relevant limit where the waist $w_{\rm c}$ of the cavity mode is much larger than the atomic cloud's spatial extent $x_{0}$, so that $g_{\rm c}(r)\approx 1$. Similarly, assuming the drive beam is spatially homogeneous over the cloud, we set $g_{\rm d}(r)=1$ in a long--wave approximation, which corresponds to pumping the cavity on the axis, or with a small angle. \footnote{This differs from the multimode proposal of \cite{Uhrich:2023ddx}, where the atomic cloud and cavity mode profiles have comparable spatial structure.} 

In the next step we eliminate the explicit time dependence introduced by the drive beam in Eq.~\eqref{eq.Had_oscillating} (which oscillates at frequency $\omega_{\rm d}$), by transforming into a \textit{rotating frame} generated by 
\begin{equation}
    H_{\rm RF} = \omega_{\rm d}\int \de^2 r \, \psi_{\rm e}^{\dagger}(r)\psi_{\rm e}(r) + \omega_{\rm d} a^{\dagger}a \,.
\end{equation}
Applying this transformation to the full Hamiltonian, we get
\begin{equation}
    H = H_{\rm kt} +  \Delta_{\rm cd}\, a^\dagger a + \int \de^2 r \,  \Delta_{\rm ad}(r) \, \psi_{\rm e}^{\dagger}(r)\psi_{\rm e}(r) +  \int \de^2 r \, \left( \Phi \, \psi_{\rm e}^{\dagger}(r)\psi_{\rm g}(r)  + \Phi^{\dagger} \, \psi_{\rm g}^{\dagger}(r)\psi_{\rm e}(r) \right) \ ,
\label{eq.Htotal_RF}
\end{equation}
where we have used the approximations $g_{\rm c}(r)\approx g_{\rm d}(r) \approx 1$ and introduced a new parameter 
$\Phi = \Omega_{\rm d} + \frac{1}{2}\Omega_{\rm c} \, a$. Notice also that the kinetic term remains unaltered by the transformation to the rotating frame, $[H_{\rm kt}, H_{\rm RF}]=0$. The detailed computation to show this explicitly is given in the Supplementary Material of \cite{Uhrich:2023ddx}.  
 
\subsubsection*{Adiabatic elimination}
In systems where atoms are driven far off--resonantly from an excited state, the population of that excited state remains negligible, even though it plays an important virtual role in mediating interactions. This regime allows for an adiabatic elimination, a simplification technique where the excited state is removed from the description, leaving an effective model that captures its influence indirectly. This is particularly powerful in cavity QED, where it allows one to derive effective atom--photon and atom--atom interactions that are mediated by virtual excitations, without tracking fast excited--state dynamics explicitly. Formally, it means that we can set Heisenberg evolution of the operator to zero $-\rm i\tfrac{\rm d}{\rm dt}$$\psi_{\rm e}(r) = 0$, so that its equation of motion becomes
\begin{equation}
    \left( \tfrac{p^2}{2 m_{\rm at}} + V_{\rm t}(r) - \Delta_{\rm ad}(r) \right) \psi_{\rm e}(r) - \Phi \, \psi_{\rm g}(r) = 0 \qquad \to\qquad \psi_{\rm e} = -\frac{\Phi \, \psi_{\rm g}(r)}{\Delta_{\rm ad}(r)} +  \mathcal{O}(\Delta^{-2}_{\rm ad}) \ ,
    \label{eq.adiabatic_elim}
\end{equation}
where we have assumed $|\Delta_{\rm ad}(r)| \gg |\tfrac{p^2}{2 m_{\rm at}} + V_{\rm t}(r)|$ for the expansion. Additionally, we need $|\Delta_{\rm ad}(r)| \gg |\Phi|$ to ensure the scarcely populated excited states. Complementarily, $\Delta_{\rm ad}$ also sets a lower limit on the duration of the Trotter time steps one can choose, such that the adiabatic elimination holds. In particular, we are limited by $\Delta_{\rm ad}\Delta t \gg 1$. 

Substituting expression \eqref{eq.adiabatic_elim} into the Hamiltonian~\eqref{eq.Htotal_RF} yields an effective model involving only ground-state atoms and photons. We truncate the Hamiltonian at second order in the inverse detuning, $\mathcal{O}(\Delta_{\rm ad}^{-2})$, and remain with
\begin{equation}
    H =  \int \de^2r \, \psi_{\rm g}^{\dagger}(r) \left( \frac{p^2}{2 m_{\rm at}} + V_{\rm t}(r) \right) \psi_{\rm g}(r) +  \Delta_{\rm cd}\, a^\dagger a + \int \de^2 r \,  \frac{|\Phi|^2}{\Delta_{\rm ad}(r)} \, \psi_{\rm g}^{\dagger}(r)\psi_{\rm g}(r) \ .
\label{eq.H_eliminated_e_states}
\end{equation}
Now let us turn our attention to the cavity photons. We have already assumed that the most dominant energy scale is the detuning between atomic and drive frequency, see Eq.~\eqref{eq.adiabatic_elim}. In a similar fashion, we can assume the cavity to follow adiabatically the atoms, which is achieved for large enough $\Delta_{\rm cd}$ or large enough cavity linewidth \cite{Mivehvar_2021}. 
We thus employ $-\rm{ i\tfrac{\rm d}{\rm dt}}$$a = 0$, which yields the equation
\begin{equation}
a \left( \Delta_{\rm cd} + \frac{\Omega_{\rm c}^2}{4} \int \de^2 r \,  \frac{1}{\Delta_{\rm ad}(r)} \psi_{\rm g}^\dagger(r) \psi_{\rm g}(r) \right) +  \frac{\Omega_{\rm d}\Omega_{\rm c}}{2} \int \de^2 r \,  \frac{1}{\Delta_{\rm ad}(r)} \psi_{\rm g}^\dagger(r) \psi_{\rm g}(r) = 0  \ ,
\end{equation}
This second adiabatic elimination is justified for the regime 
\begin{equation}
    \left| \tfrac{\Omega_{\rm c}^2}{\Delta_{\rm cd}\Delta_{\rm ad}} \right| \ll\left| \tfrac{\Omega_{\rm d}\Omega_{\rm c}}{\Delta_{\rm cd}\Delta_{\rm ad}} \right| \ll 1 \ , 
\end{equation}
together with the hierarchy $|\Delta_{\rm ad}| \gg |\Delta_{\rm cd}|$, and $ \Delta_{\rm cd} \Delta t\gg 1$. All these assumptions conspire to give the solution
\begin{equation}
a = -\frac{\Omega_{\rm d}\Omega_{\rm c}}{2\Delta_{\rm cd}\Delta_{\rm ad}} \int \de^2 r \,  \frac{\Delta_{\rm ad}}{\Delta_{\rm ad}(r)} \, \psi_{\rm g}^\dagger(r) \psi_{\rm g}(r) + \mathcal{O}(\Delta^{-2}_{\rm ad}) \, ,
\end{equation}
where we expanded again for large $\Delta_{\rm ad}$ and large $\Delta_{\rm cd}$. We can now plug this solution back into the original Hamiltonian to obtain the effective model, which is
\begin{multline}
    H = H_{\rm kt, g} +  \frac{\Omega^2_{\rm d} }{\Delta_{\rm ad}}\int \de^2 r \frac{\Delta_{\rm ad}}{\Delta_{\rm ad}(r)} \, \psi^\dagger_{\rm g}(r) \psi_{\rm g}(r) \\
    - \frac{\Omega^2_{\rm d}\Omega^2_{\rm c}}{4\Delta_{\rm cd}\Delta^2_{\rm ad}} \int \de^2 r \, \de^2 r' \frac{\Delta_{\rm ad}^2}{\Delta_{\rm ad}(r) \Delta_{\rm ad}(r')} \, \psi^\dagger_{\rm g}(r) \psi_{\rm g}(r) \psi^\dagger_{\rm g}(r') \psi_{\rm g}(r') \ .
    \label{eq:final_micro_hamiltonian_in_psi_g}
\end{multline}
The above Hamiltonian comprises three terms. The first one on the RHS is the kinetic term for the atoms in their ground state $\ket{g}$, which occupy levels of the harmonic ladder. 
The second one is a one--body term that arises solely from the effect of driving the system, while the third is a two--body term, which we will rewrite as a sparse SYK$_{4}$ of the form \eqref{eq:H_sparse} momentarily. While in our approach the space dependence of the two--body term is inherited from that of the atom--drive detuning, a similar effect could be produced by direct shaping of the pump beam profile, a capability foreseen in the next generation of cavity experiments \cite{Bolognini:25}. 

It is now convenient to expand the ground state field operator in the basis of the Harmonic trap, namely
\begin{equation}
    \psi_{\rm g} (r) = \sum_{i} \phi_i(r) \, c_i \qquad \text{with} \qquad  \left( \frac{p^2}{2 m_{\rm at}} + V_{\rm t}(r) \right) \phi_i(r) = E_i \, \phi_i(r) \ ,
\end{equation}
where the $\phi_i(r)$'s are the Hermite--Gauss functions. Physically, the fermionic ladder operators $c_i$ (and $c_i^\dagger$) create (and annihilate) an atom in the ground state $\ket{g}$ in the $i$-th level of the Harmonic trap. Those levels represent delocalized energy levels that can be occupied or empty, and play the role of sites in the traditional formulation of the SYK model, in the spirit of synthetic dimensions in ultracold gases \cite{PhysRevLett.112.043001, PhysRevA.95.023607}. 
We comment on the degeneracy of the levels of the two--dimensional harmonic trap in Appendix \ref{sec:App_on_speckles}.

The energy scale of the harmonic trap is also much smaller than the remaining two terms, and we can safely neglect it. Moreover, the one--body term is a random SYK${}_2$ interaction that can be compensated by an auxiliary dipole interaction. We will comment on it in a subsequent subsection shortly after. We are then left with the two--body term. Defining now the couplings
\begin{equation}
    J_{ik} = \frac{1}{2}\int \de^2 r \,  \frac{\Delta_{\rm ad}}{\Delta_{\rm ad}(r)} \, \phi^*_{i}(r)\phi_{j}(r)
    \label{eq:experimental_couplings_theory_expression}
\end{equation}
the two--body interaction can be conveniently rewritten in terms of the fermionic ladder operators of the harmonic trap, giving
\begin{equation}
    H_{\rm eff} = - \frac{\Omega^2_{\rm d}\Omega^2_{\rm c}}{\Delta_{\rm cd}\Delta^2_{\rm ad}}\,\sum_{i_1i_2, k_1 k_2}\, J_{i_1 k_1} J_{i_2k_2} \, c_{i_1}^{\dagger} c_{k_1}c_{i_2}^{\dagger} c_{k_2} \ .
    \label{eq:H_sparse_cQED}
\end{equation}
We have used the same subscript `eff' as in \eqref{eq:H_sparse} due to the evident similarities between the two Hamiltonians. In \eqref{eq:H_sparse_cQED} we have highlighted the energy scale 
\begin{equation}
    \mathcal E = \frac{\Omega^2_{\rm d}\Omega^2_{\rm c}}{\Delta_{\rm cd}\Delta^2_{\rm ad}} \ ,
\end{equation}
leaving the couplings in \eqref{eq:H_sparse_cQED} as pure numbers.  Moreover, due to the analysis around \eqref{eq:H_sparse}, we can safely place the fermionic operators in \eqref{eq:H_sparse_cQED} as normal ordered, without affecting the large--$N$ physics. We proceed by checking the distributions of various couplings, in order to better understand the differences and similarities between \eqref{eq:H_sparse} and \eqref{eq:H_sparse_cQED}.

\subsubsection*{Distribution of experimental couplings}

\begin{figure}[t]
\centering
\begin{tabular}{lll}
\includegraphics[width=0.30\textwidth]{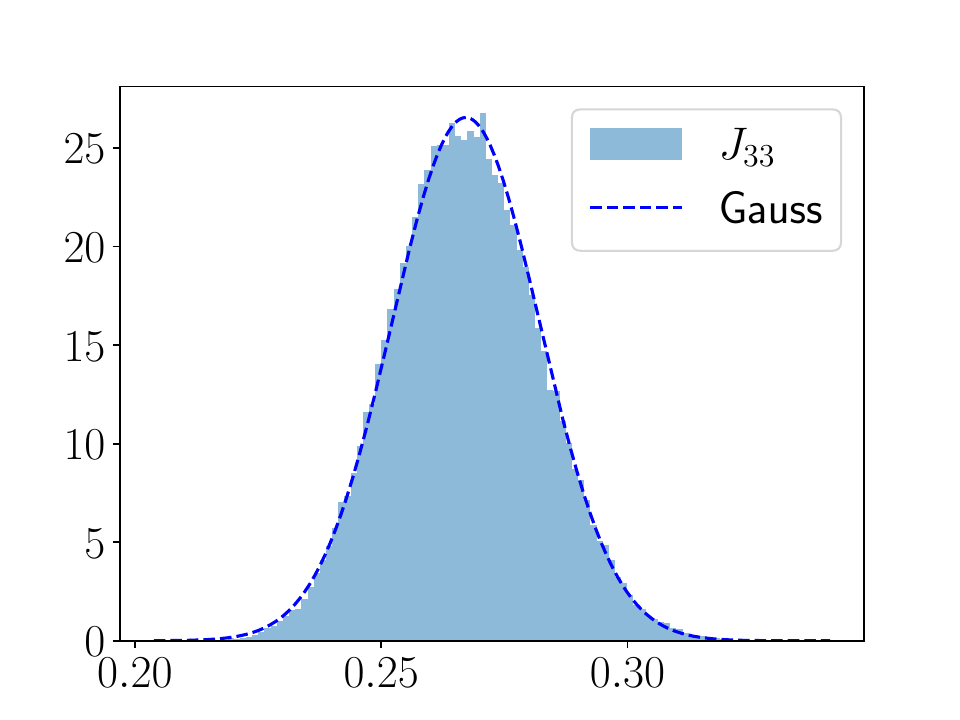}
&
\includegraphics[width=0.30\textwidth]{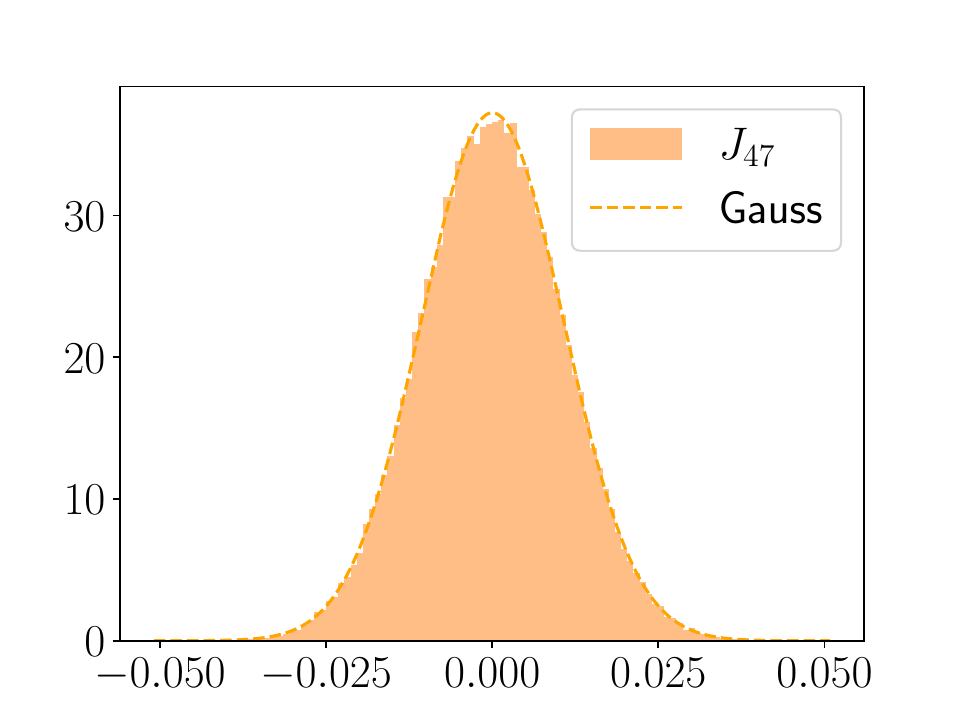}
&
\includegraphics[width=0.30\textwidth]{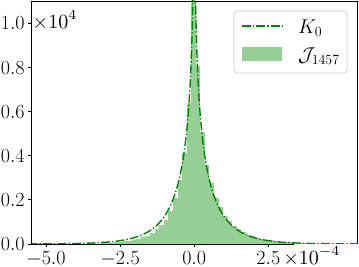}
\end{tabular}
\caption{Probability distribution for one--body couplings $J_{ik}$, fitted by a Gaussian distribution. Both couplings are realized $10^5$ times, and plotted in histograms with 100 bins. {\bf Left}: Distribution for $J_{33}$, which resembles a Gaussian with mean $\mu = 0.26$ and standard deviation $\sigma = 0.015$. {\bf Center}: Distribution for $J_{47}$, which resembles a Gaussian with zero mean ($\approx 10^{-6}$) and standard deviation $\sigma = 0.010$. {\bf Right}: Distribution for $\mathcal J_{14;57} = J_{15}J_{47}$, which is well approximated by a Bessel $K$-distribution of the form \eqref{P_Sigma_exact} with $R = 1$.}
\label{fig:probability_density_two_body_couplings}
\end{figure}

The first consideration we can do about the couplings in Eq.~\eqref{eq:experimental_couplings_theory_expression} is that, while most of them have zero mean, not all do. This is apparent when performing an average over various realizations. The disorder only affects the detuning $\Delta_{\rm ad}(r)$, and assuming $\overline{\Delta_{\rm ad}(r)} \approx 2 \Delta_{\rm ad}$ (see Appendix \ref{sec:App_on_speckles}, Equation \eqref{eq:average_of_detuning_appendix}), crucially independent of position, we have
\begin{equation}
    \overline{J_{ik}} \approx \frac{1}{2}\int \de^2 r \, \phi^*_{i}(r)\phi_{k}(r) = \frac{1}{4} \, \delta_{ik} \ ,
    \label{eq:experimental_couplings_two_body_avg}
\end{equation}
where in the last passage, we have used the orthonormality of the Hermite--Gauss functions. We can check this fact numerically, realizing many times a coupling between two modes $i$ and $k$ for different speckle patterns, to compute its probability distribution, which in turn gives the mean and the standard deviation. This is shown on the left panel of Figure~\ref{fig:probability_density_two_body_couplings}, where we see that the prediction \eqref{eq:experimental_couplings_two_body_avg} is met. Moreover, the distribution of both `diagonal' and `off--diagonal' couplings fits well with a Gaussian, with approximately equal standard deviation $\sigma$, and likewise the distribution of effective couplings $\mathcal J_{i_1 i_2; k_1 k_2}$ agrees with the prediction of \eqref{P_Sigma_exact}. This is shown on the left, center, and right panels of Figure \ref{fig:probability_density_two_body_couplings}.

In order to finally identify the two models \eqref{eq:H_sparse} and \eqref{eq:H_sparse_cQED}, we can shift the `diagonal' couplings by their means, namely
\begin{equation}
    J_{ik} \to J_{ik} - \frac{1}{4} \, \delta_{ik} \ .
    \label{eq:coupling_shift}
\end{equation}
Doing so, the `new' set of couplings is Gaussianly distributed with zero mean. On the other hand, the effect of the shift \eqref{eq:coupling_shift} is to introduce a one--body interaction, since
\begin{multline}
    H_{\rm eff} = - \frac{\Omega^2_{\rm d}\Omega^2_{\rm c}}{\Delta_{\rm cd}\Delta^2_{\rm ad}}\,\sum_{i_1i_2, k_1 k_2} \! \! \Big(J_{i_1 k_1} - \frac{1}{4} \delta_{i_1 k_1} \Big) \Big(J_{i_2 k_2} - \frac{1}{4} \delta_{i_2 k_2} \Big) \, c_{i_1}^{\dagger} c_{k_1}c_{i_2}^{\dagger} c_{k_2} \\
    = - \frac{\Omega^2_{\rm d}\Omega^2_{\rm c}}{\Delta_{\rm cd}\Delta^2_{\rm ad}}\,\sum_{i_1i_2, k_1 k_2}\, J_{i_1 k_1} J_{i_1 k_1} \, c_{i_1}^{\dagger} c_{k_1}c_{i_2}^{\dagger} c_{k_2}  + \frac{\Omega^2_{\rm d}\Omega^2_{\rm c}}{2 \Delta_{\rm cd}\Delta^2_{\rm ad}}\, N_{\rm at} \sum_{i k}\, J_{i k}  \, c_{i}^{\dagger} c_{k}\ ,
    \label{eq:H_sparse_cQED_shifted}
\end{multline}
where, in the last passage, we have omitted a constant shift in the Hamiltonian. In the second line of \eqref{eq:H_sparse_cQED_shifted}, the first term is exactly the same as \eqref{eq:H_sparse}, where now all couplings have zero mean. On the other hand, the second term can be rewritten as 
\begin{equation}
    \frac{\Omega^2_{\rm d}\Omega^2_{\rm c}}{2 \Delta_{\rm cd}\Delta^2_{\rm ad}}\, N_{\rm at} \sum_{i k}\, J_{i k}  \, c_{i}^{\dagger} c_{k} = \frac{\Omega^2_{\rm d}\Omega^2_{\rm c}}{2 \Delta_{\rm cd}\Delta^2_{\rm ad}}\, N_{\rm at}  \int \de^2 r \, \frac{\Delta_{\rm ad}}{\Delta_{\rm ad}(r)} \, \psi_{\rm g}^\dagger(r) \psi_{\rm g}(r) \ ,
\end{equation}
where $N_{\rm at}$ is the number of atoms present in the cavity, a factor coming from the evaluation of a number operator. This term is very similar to the one--body term  in Eq.~\eqref{eq:final_micro_hamiltonian_in_psi_g} we have yet to discuss. Let us address this issue now.

\subsubsection*{Dipole compensation} 

The effective Hamiltonian \eqref{eq:final_micro_hamiltonian_in_psi_g} is affected by the one--body term that would hide the SYK${}_4$ physics. We now discuss, in the same spirit as \cite{Uhrich:2023ddx}, how to cancel such a term via dipole compensation. The total one--body term we have to cancel is 
\begin{equation}
    \frac{\Omega_{\rm d}^2}{\Delta_{\rm ad}} \left(1 + \frac{\Omega^2_{\rm c}}{2 \Delta_{\rm cd}\Delta_{\rm ad}}\, N_{\rm at} \right)  \int \de^2 r \, \frac{\Delta_{\rm ad}}{\Delta_{\rm ad}(r)} \, \psi_{\rm g}^\dagger(r) \psi_{\rm g}(r) \ ,
    \label{eq:two_body_term_total}
\end{equation}
and the ingenious idea presented in \cite{Uhrich:2023ddx} stems from the fact that a term of this sort could arise from the adiabatic elimination of excited states, when the atoms are coupled to an additional auxiliary state. Driving the transition with a laser of intensity $\Omega_{d'}$, the Hamiltonian would acquire a term of the form
\begin{equation}
    \int \de^2 r \, \frac{\Omega_{\rm d'}^2}{\Delta_{{\sf aux},{\rm d'}}(r)} \, \psi_{\rm g}^\dagger(r) \psi_{\rm g}(r) \ .
\end{equation}
Choosing $\Delta_{{\sf aux},{\rm d'}}(r)$ to be exactly anticorrelated with $\Delta_{{\rm ad}}(r)$ ({\it i.e.} $\Delta_{{\sf aux},{\rm d'}}(r) = - \Delta_{{\rm ad}}(r)$) and $\Omega_{\rm d'}$ to match the prefactor in \eqref{eq:two_body_term_total}, we have an exact cancellation of the one--body term. Experimentally, the space dependence can be precisely copied from the light--shift beam to the compensation beam using light propagating through the same optics and using a tuneout wavelength for the excited state. For the case of $^6$Li, we expect such a tuneout wavelength to lie only a few hundreds of MHz away from resonance, owing to the small fine structure of excited states. This small difference ensures the absence of chromatic shifts between the light--shifting and compensation beams. Using experimentally sensible parameters (see Appendix \ref{App:exp_parameters}), and assuming $\Omega_{\rm d'}$ has a high accuracy, we can estimate the error in the detunings we are allowed to have in order to see the SYK${}_4$ interaction to be
\begin{equation}
    \frac{\delta \Delta_{{\sf aux},{\rm d'}}}{\Delta_{\rm ad}}  \sim 10^{-4} \ .
\end{equation}


\subsection{Benchmarking experimental predictions with \texorpdfstring{cSYK${}_4$}{}}
\label{sec.benchmarking_otoc_sff}

In what follows, we present numerical simulations of the out--of--time--ordered correlation function (OTOC) and the spectral form factor (SFF) of the effective Hamiltonian derived in Eq.~\eqref{eq:H_sparse_cQED_shifted}, compared to those of the complex SYK$_4$ model. These two observables are widely used diagnostics of quantum chaos and serve as benchmarks to assess how closely our effective model reproduces the physics of the target SYK system.\footnote{Throughout this manuscript, we set $\hbar=1$ for convenience. This allows us to express energies in terms of frequency units.}

\subsubsection*{OTOCs}

\begin{figure}[t]
\centering
\begin{tabular}{ll}
\includegraphics[width=0.49\textwidth]{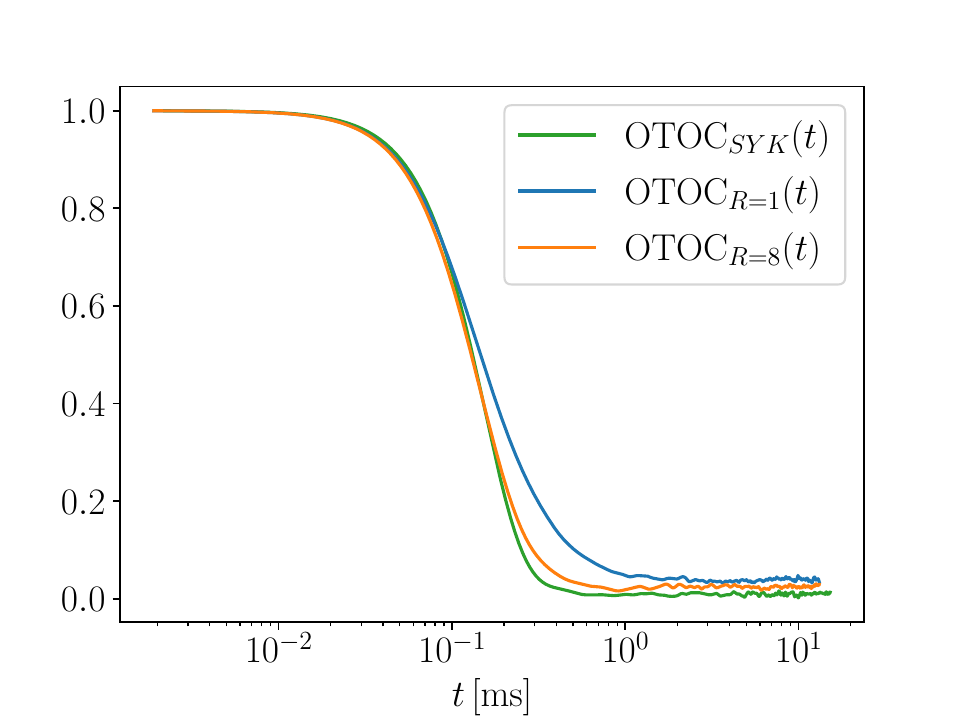}
&
\includegraphics[width=0.49\textwidth]{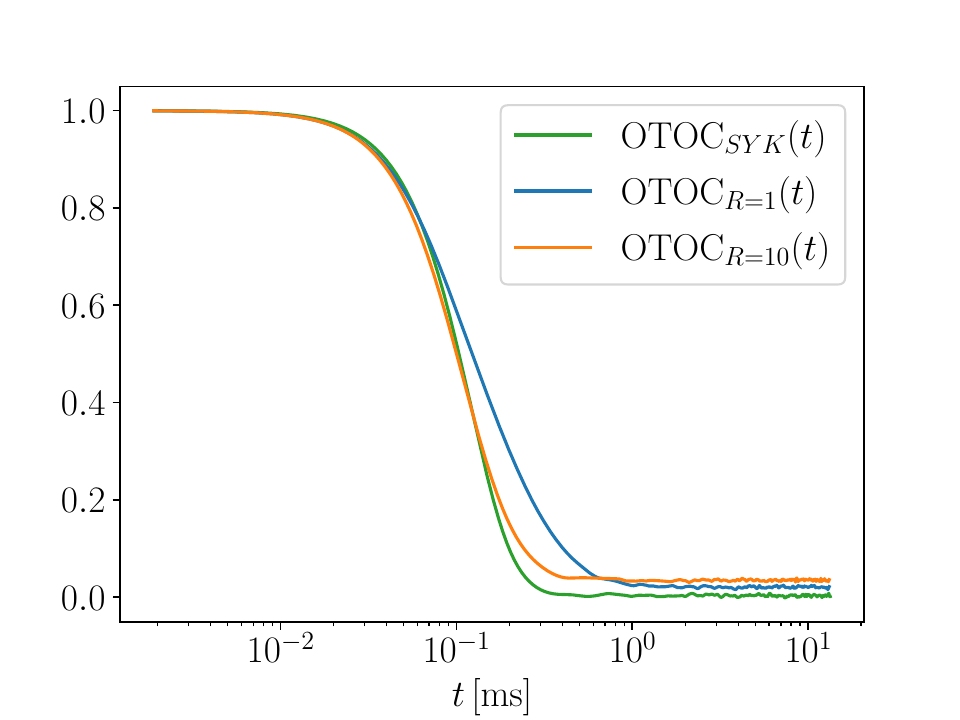}
\end{tabular}
\caption{Out--of--time order correlation functions at infinite temperature ($\beta =0$) for $N=8$, averaged over $50$ realizations on the left, respectively $N=10$ and $10$ realizations on the right. We compare the effective model with two different numbers of Trotter steps $R\in \{1,N\}$ to the OTOC of complex SYK$_4$.} 
\label{fig:OTOC_effective_model}
\end{figure}

We compute the out--of--time--ordered correlator (OTOC) at infinite temperature (\(\beta = 0\)) for both the SYK model and its Trotterized analogs. The OTOC we are using is
\begin{equation}
    \mathrm{OTOC}(t) = \frac{\mathrm{Tr} \left[ W(t) V W(t) V + V(t) W V(t) W \right]}{2\, \mathrm{Tr} \left[ W^2 V^2 \right]} \ ,
    \label{eq:OTOC_definition}
\end{equation}
where $W$ and $V$ are local fermionic hopping operators acting on different site pairs and $W(t), V(t)$ their Heisenberg--evolved analogues, {\it i.e.}, $W(t) = e^{i H t} W e^{-i H t}$. The denominator in \eqref{eq:OTOC_definition} is a convenient choice of normalization as it ensures that $\mathrm{OTOC}(0) = 1$, as the operators commute at equal times. In chaotic systems like SYK, the OTOC decays exponentially from its initial value at early times and saturates to a small constant at late times. As shown in Figure~\ref{fig:OTOC_effective_model}, the Trotterized model reproduces the qualitative features of the SYK dynamics. Increasing the number of Trotter steps $R$ enhances this agreement, particularly in the early--time decay as well as decay to the saturation plateau. 
In our implementation, the Trotterized interaction tensor is constructed by summing over $R$ independent coupling realizations and normalizing by $1/\sqrt{R}$. While this normalization may not correspond to a physical process in an actual experiment, it is essential in simulations to maintain a constant interaction strength and enable a meaningful comparison with the SYK model and different Trotterizations. To ensure alignment of energy scales, we also rescale the time axis of the SYK data by a small constant factor ($\sim 0.7$), which compensates for minor numerical discrepancies in the coupling statistics introduced during the mean subtraction procedure. Altogether, the OTOC comparison confirms that the effective Trotterized model faithfully captures the key signatures of quantum chaos characteristic of SYK--type dynamics.

\subsubsection*{Spectral form factor}
\begin{figure}[t]
\centering
\begin{tabular}{ll}
\includegraphics[width=0.49\textwidth]{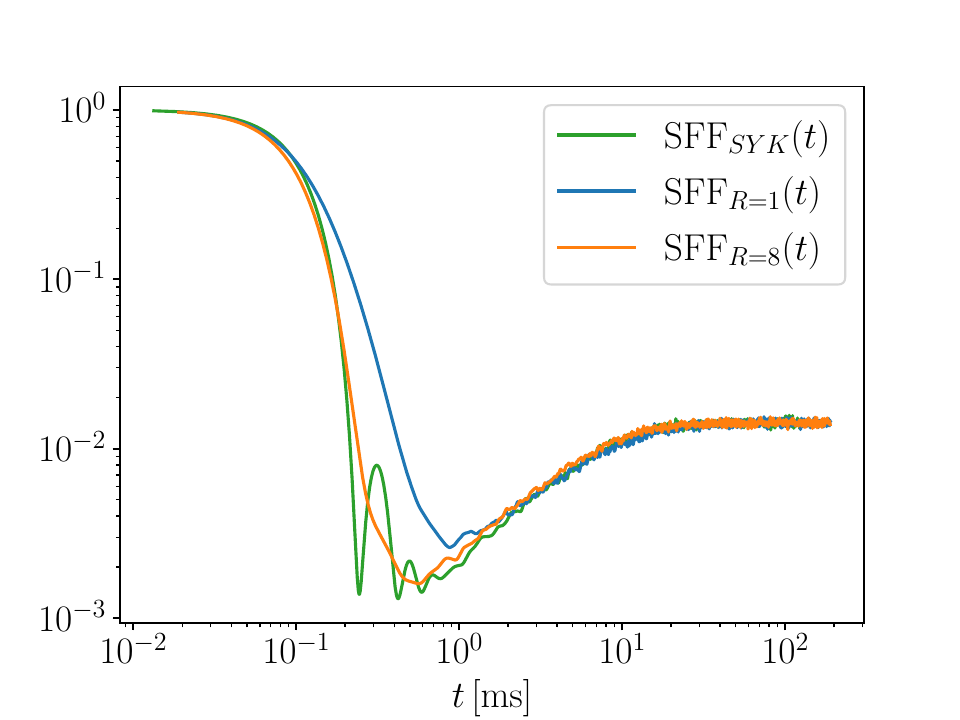}
&
\includegraphics[width=0.49\textwidth]{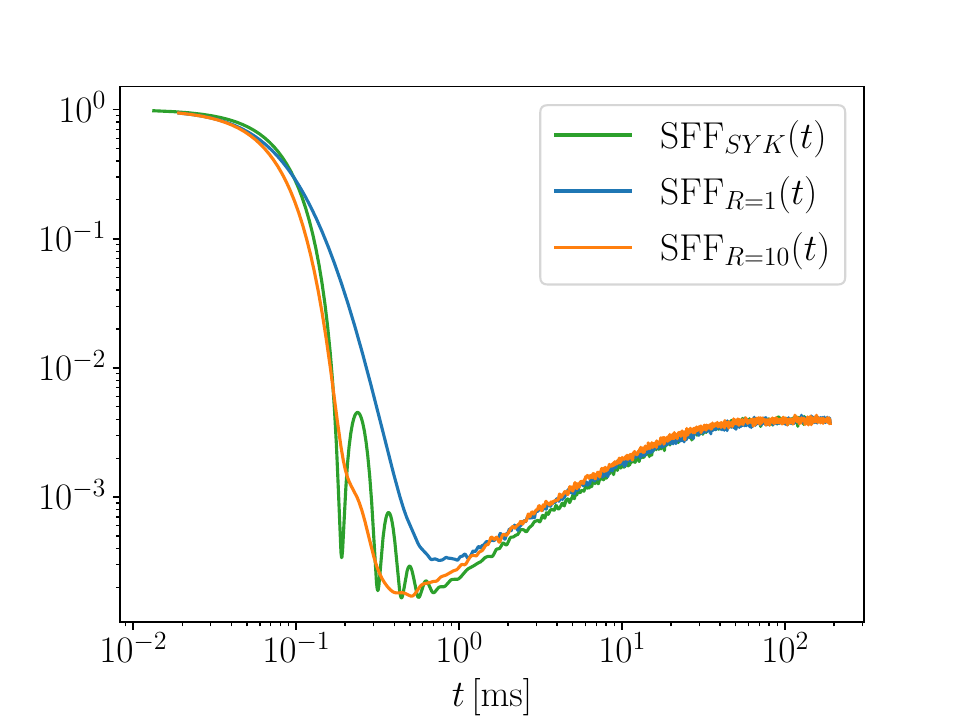}
\end{tabular}
\caption{Spectral form factor for $N=8$ (left) and $N=10$ (right), each for $10^4$ realizations. We compare the effective model with two different numbers of Trotter steps $R \in \{1,N\}$ to the SFF of complex SYK$_4$.} 
\label{fig:SFF_effective_model}
\end{figure}

As a second probe of chaotic dynamics, we compute the spectral form factor (SFF), which is known to display a characteristic dip--ramp--plateau structure in systems such as the SYK model. This time--resolved quantity reflects the discreteness and correlations in the energy spectrum. The SFF at infinite temperature is
\begin{equation}
\mathrm{SFF}(t) = \frac{1}{D^2} \left| \mathrm{Tr}\left(e^{-i H t} \right) \right|^2,
\end{equation}
where $D$ is the Hilbert space dimension. For each model, we obtain the spectrum via exact diagonalization and evaluate the trace of the time--evolution operator $e^{-iHt}$ at logarithmically spaced time points. 
Figure~\ref{fig:SFF_effective_model} shows the result for the complex SYK$_4$ model and the effective Hamiltonians with $R=1$ and $R=N$ Trotter steps. As in the case of the OTOC, we find that increasing the number of Trotter steps improves the agreement between the effective model and the SYK target model. In particular, the long--time saturation value and the shape of the intermediate ramp are well reproduced by the fully Trotterized model with $R=N$, but more interestingly, even the overall early--time behavior where the SFF is highly sensitive to the detailed structure of the spectrum is well captured. We checked also, that increasing the number of Trotter steps to $R\sim N^2$, the effective model is also able to resolve additionally the individual oscillations of the decay of the SFF, pointing to an exact agreement between effective and target model in the large $R$ limit. 

The effective Hamiltonians are constructed by averaging over $R$ independent coupling realizations and normalizing by $1/\sqrt{R}$, which ensures consistency in the interaction strength across different Trotter depths. As before, we apply a global rescaling of the time axis (by a factor of approximately $\sim 0.7$) to align the energy scales and correct for small numerical deviations due to mean subtraction in the coupling statistics. Each curve is averaged over $10^4$ realizations to suppress statistical noise and clearly reveal universal features. Altogether, the agreement in both early-- and late--time regimes reinforces the validity of the Trotterized model as an effective simulator of complex SYK dynamics.

\subsection{Dynamical determination of \texorpdfstring{$N$}{}} \label{sec:determination_of_N}

\begin{figure}[t]
\centering
\includegraphics[width=0.96\textwidth]{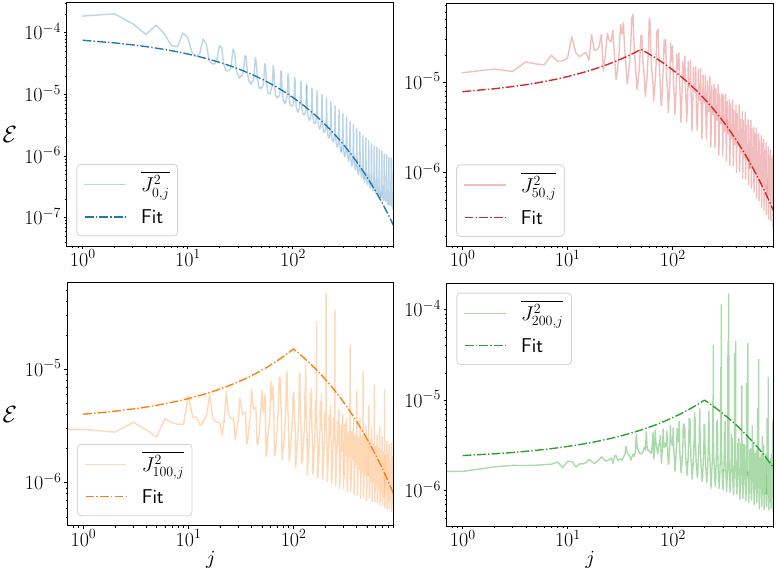}
\caption{Average of $J_{ij}^2$ in units of $\mathcal E$, fixing $i = \{0, 50, 100, 200 \}$ and varying $j \in (0, 10^3)$. The width of the speckle mask is $r = 6$ pixels (see Appendix \ref{sec:App_on_speckles} for details).}
\label{fig:SYK_N_Sites_determination}
\end{figure}

We now propose an operational method to determine $N$, the number of fermionic sites of the effective model. This is crucially different from the initial number of atoms present in the cavity $N_{\rm at}$, as that number just determines the number of `filled' sites of the effective model at the beginning of the evolution. To determine how many sites interact, one has to determine which levels of the trap are connected through a momentum kick given by emitting or absorbing a virtual photon. This is essentially given by $J_{ij}$, which is (the square root of) the probability of two sites interacting through such a virtual photon. We then proceed to study the distribution of $J_{ij}^2$ for the experimental couplings realized in the cavity. The physical idea behind this is that an excitation (an atom in the ground state) at site $i$ will interact more strongly with the `neighboring sites' at a similar energy, since in order to interact with site $j$ much further away (in energy) a more energetic virtual photon has to be exchanged. In practice, fixing various values of $i$ (to $\{ 0, 50, 100, 200 \}$ in Figure \ref{fig:SYK_N_Sites_determination}), we can scan all $j$'s within a large window (in particular $j \in (0, 10^3)$ for Figure \ref{fig:SYK_N_Sites_determination}) and see how the average of $J_{ij}^2$ behaves. The result qualitatively follows the expectation, peaking around $j = i$, and slowly decaying. However, the behavior is quite noisy due to the presence of `resonances', which one can picture as arising due to the fact that each level of the trap is degenerate and it just so happens that some of these interact more strongly with the site $i$. Because of this, we can only perform a rather qualitative analysis. In particular, plotting the result on a log--log plot, it is clear that the general behavior is super--polynomial and sub--exponential. An educated guess, which captures the behavior of $\langle J^2_{ij} \rangle$ within a large window of $j$'s is 
\begin{equation}
    \langle J^2_{ij} \rangle \propto \exp \left( - \left|\frac{\sqrt{i}-\sqrt{j}}{\sqrt{N}} \right| \right) \ .
    \label{eq:J_ij_distribution_for_N}
\end{equation}
We can motivate this ansatz in the following way. Every site $i$ represents an energy level of the two--dimensional harmonic trap that confines the atom in the cavity. If the site $i$ belongs to the $n$-th energy level, there are $n-1$ other sites $j$ which belong to the same energy level. By numerical evidence, \eqref{eq:J_ij_distribution_for_N} seems to suggest that the strength of the interactions does not decay exponentially in the number of sites, but rather exponentially in the level of the harmonic trap. We believe that this effect arises because of the properties of the Hermite--Gauss functions, together with our numerical generation of speckles. We then proceed noticing that \eqref{eq:J_ij_distribution_for_N} gives an operational meaning to the number of interacting fermions, which is the parameter $N$, namely the scale at which two sites effectively stop interacting though a first--order photonic emission/absorption\footnote{Of course, in the full effective model they could still interact through the second order interaction, where the virtual photon is exchanged.}. We fit the ansatz \eqref{eq:J_ij_distribution_for_N} to the values of $\langle J_{ij}^2 \rangle$ obtained numerically, to find $N$. Before presenting the values obtained, it is interesting to notice that this parameter will depend on the size of the speckle mask used to generate the speckle pattern in the cavity. In particular, a larger mask would more finely sample the speckle pattern (due to the inverse relation between $x$ and $k$ in a Fourier transform), lowering the value of the $J_{ij}$'s and its dependence on the site difference. Thus, increasing the mask size relative to the waist of the cavity, we expect an increase of $N$. This was also confirmed in our numerical extrapolation of $N$. As mentioned previously, we performed fits for $i = \{ 0, 50, 100, 200 \}$, for relative mask sizes\footnote{See Appendix \ref{sec:App_on_speckles} for details.} of $r = 6$ pixels and $r = 15$ pixels, as on the left and right of Figure \ref{fig:SYK_N_Sites_determination}, respectively. We find to following values:

\vspace{10pt}
\begin{center}
\begin{tabular}{| c | c | c |} 
 \hline
 $i$ & $r = 6$ & $r = 15$ \\
 \hline\hline
 0 & $N_0 \approx 20$ & $N_0 \approx 120$ \\ 
 \hline
 50 & $N_{50} \approx 31$ & $N_{50} \approx 129$ \\ 
 \hline
 100 & $N_{100} \approx 46$ & $N_{100} \approx 151$ \\
 \hline
 200 & $N_{200} \approx 86$ & $N_{200} \approx 170$ \\
 \hline
\end{tabular}
\end{center}

\vspace{10pt}
\noindent From the values found, it is clear that each site $i$ interacts strongly with a certain amount $N_i$ of `neighboring' states (in energy), which depend on the site $i$ itself. For instance, for $r = 6$, the site labeled with $i = 0$ interacts strongly with sites up to $j \approx 20$, while for $i  =100$ we have a `cutoff' at around $j \approx 46$. Increasing the optical aperture $r$ of the speckle mask, the resulting speckle pattern is more fine grained, which results in lower variances of the couplings and thus a larger cutoff. 

To summarize, in this Section we have proposed an operational way to obtain an estimate for the number of interacting fermions that enter the SYK Hamiltonian.


\subsection{Product states as initial states} \label{sec:product_states}

\begin{figure}[t]
\centering
\begin{tabular}{ll}
\includegraphics[width=0.467\textwidth]{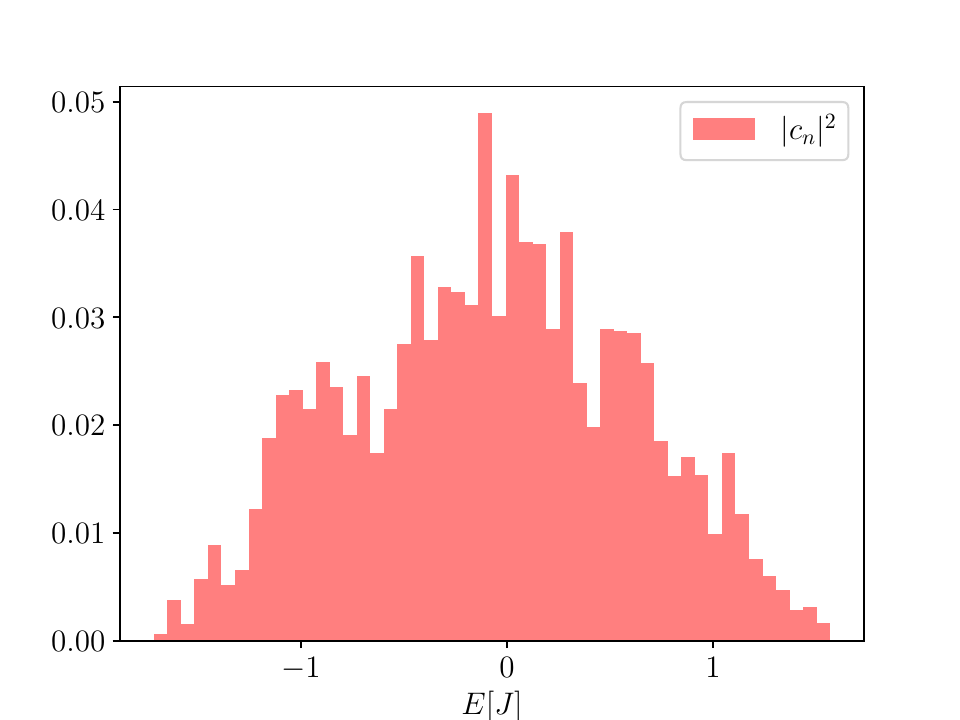}
&
\includegraphics[width=0.493\textwidth]{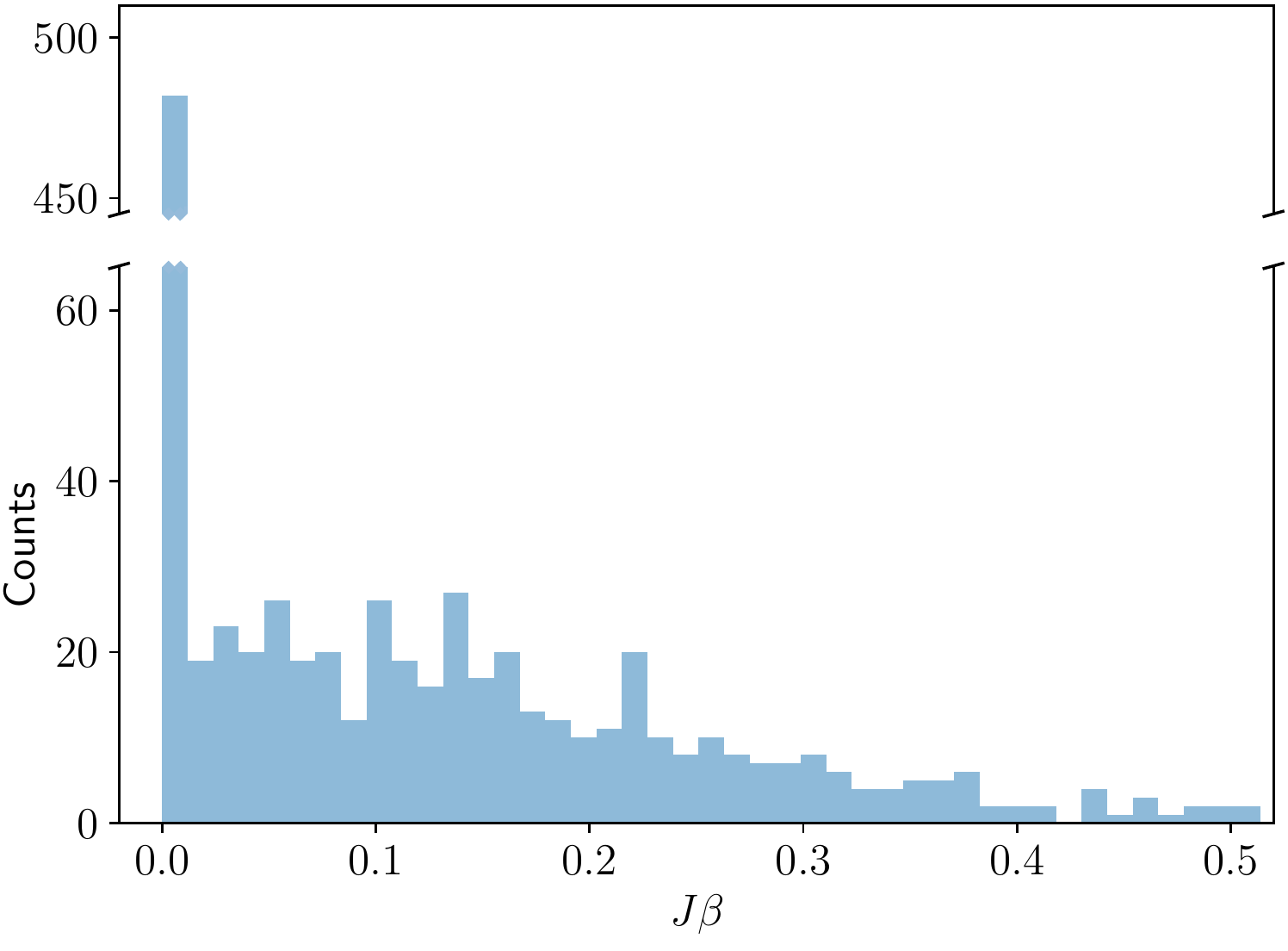}
\end{tabular}
\caption{{\bf Left}: The probability to find a product state of the form \eqref{eq:product_states} in various microcanonical energy windows of the spectrum. For each bin, the plot shows the sum of all the $|c_n|^2 = |\langle E_n | \Psi\rangle|^2$ for energy eigenvector in that window. We notice that the state is highly delocalized in the energy basis. {\bf Right}: Distribution.
For each product state \eqref{eq:product_states} at half filling, we compute the effective temperature (in units of $J^{-1}$). }
\label{fig:State_preparation}
\end{figure}

The previous discussion suggests that, at the beginning of the time evolution, the initial state is rather close to a product state in the `computational basis', thus it has the form
\begin{equation}
    \ket{\Psi} = \ket{11 \dots 1100 \dots 00 } \ .
    \label{eq:product_states}
\end{equation}
Such product states have minimal entanglement, and they just represent atoms that fill the lowest levels of the harmonic trap. These states are typically highly delocalized in the energy eigenbasis, since the list of overlaps 
\begin{equation}
    c_n = \langle E_n | \Psi \rangle
\end{equation}
is compatible with $\ket{\Psi}$ being a random vector with respect to the eigenstates $\ket{E_n}$. For the product state \eqref{eq:product_states}, we show the list of $|c_n|^2$ on the left of Figure \ref{fig:State_preparation}, where for each bin we have summed the various probabilities $|c_n|^2$ for all eigenstates belonging to that microcanonical window. It is apparent that the state is completely delocalized in the energy eigenbasis, and by typicality, we expect that any product state of this sort will yield similar results. Moreover, this extreme delocalization implies that the state will be quite energetic and have a generally high temperature. To quantify this intuition, we can define an effective temperature of the state $\ket{\Psi}$ matching the expected energy with the thermal energy, thus solving the equation
\begin{equation}
    \bra{\Psi} H \ket{\Psi} = E_{\rm cSYK}(\beta_{\rm eff})
    \label{eq:effective_beta_equation}
\end{equation}
for $\beta_{\rm eff}$. Unfortunately, this is not always possible, as the symmetry of the SYK spectrum with respect to $E = 0$ implies that the thermal energy $E(\beta)$ is always negative. To circumvent this problem, we can associate $\beta_{\rm eff} = 0$ to any state with a positive energy, which is physically motivated by the fact that, being almost random vectors, they can be thought of as states at infinite temperature. Keeping this in mind, it is numerically quite simple to solve \eqref{eq:effective_beta_equation}. We perform the procedure just outlined for any product state of the form \eqref{eq:product_states} within the half--filled sector (thus with an equal number of `0's and `1's), for $N = 12$, and the results can be found on the histogram on the right of Figure \ref{fig:State_preparation}. Out of the whole $924$ product states, we find that at least half have an effective temperature\footnote{As most of them have a positive expected energy.} $J \beta_{\rm eff} \lesssim 10^{-2}$, while the remaining are typically within the range $10^{-2} \lesssim J \beta_{\rm eff} \lesssim 0.5$. This numerically confirms the expectation that product initial states are very energetic, and that in a realistic experimental scenario one would need to find a way to cool down, from the point of view of the effective Hamiltonian, the initial product state (see \cite{Schuster:2025kwh} for recent work on the subject). Alternatively, it could also be interesting to explore the high energy physics, even though the connection with a holographic phase it is not fully clear yet. We refer the Reader to the Discussion section and to Appendix \ref{App:connections_to_gravity} for a few more details, and we leave this issue for future work.

\section{Lindbladian evolution of the cSYK model} \label{sec:Dissipation}

We conclude this work by examining the role of dissipation in the proposed experimental implementation. As with any open quantum system, cold--atom experiments are inherently subject to dissipative effects, primarily arising from cavity photon losses due to imperfect mirrors and spontaneous photon scattering into free space (see for example \cite{Mivehvar_2021} for a review). The main aim of this Section is to understand the interplay between these effects and the disordered nature of the interactions, to determine precisely the timescale of coherent dynamics, a matter of central importance for any experiment.

We employ a Markovian approach employing Lindblad superoperators to write the open system Schrödinger equation
\begin{equation}
    \mathcal{L}[\rho] \equiv \frac{\de \rho}{\de t} = - i [H_{\rm eff}, \rho] + \sum_{\alpha} \mathcal D_\alpha [\rho] \ .
    \label{eq:Lindblad_evolution_general}
\end{equation}
The first term in the RHS is the unitary evolution generated by the Hamiltonian, while the sum in the second term runs through all dissipating modes, and $\mathcal D_{\alpha}$ are commonly called {\it dissipators}. Requiring the evolution to be a completely positive and trace--preserving (CPTP) Markovian evolution requires each dissipator to be of the form
\begin{equation}
    \mathcal D_\alpha [\rho] \equiv L_\alpha \rho L_{\alpha}^\dagger - \frac{1}{2} \Big \{ L_{\alpha}^\dagger L_\alpha , \rho \Big\} \ ,
\end{equation}
where the $L_\alpha$'s are called {\it jump operators}. As previewed, the main sources of dissipation that affect a cavity simulation are photon loss through imperfect cavity mirrors and photons leaking out in free space. The rates at which these events take place are $\kappa$ and $\Gamma$, respectively. The relative strength of coherent and dissipative processes is measured by the cavity {\it cooperativity}, which is defined as $\eta =4\Omega_{\rm c}^2/\kappa \Gamma \gg 1$ and which is independent of the details of choices of detunings. It is determined by the geometry of the mirror arrangement and the reflectivity and losses of the mirrors. For optical cavities operating with fermionic atoms, this number is of order $1$ in existing systems \cite{science.abd4385,Brantut_2023}, and can reach up to $\sim 20$ in the next generation experiments \cite{Bolognini:25}. For other atomic species, values on the order of $100$ have been reported \cite{Baghdad2023,grinkemeyer:2025aa}, and emerging technologies for mirror fabrication suggests that up to $1000$ may be reachable in the near future \cite{ding2025high}. After adiabatic elimination of the photons, the jump operators describing photon loss acting in the atomic ground-state manifold are given by \cite{Uhrich:2023ddx}
\begin{equation}
    L_{\kappa} = \frac{\sqrt{\kappa} \, \Omega_{\rm d} \Omega_{\rm c}}{2(\Delta_{\rm cd} - i \kappa/2)} \int \de^2 r \,  \frac{g_{\rm d} (r) g_{0} (r)}{\Delta_{\rm ad}(r) + i \Gamma/2} \, \psi_{\rm g}^{\dagger}(r)\psi_{\rm g} (r) \ ,
    \label{eq:Lindblad_op_photon_loss}
\end{equation}
which depends of the cavity decay constant $\kappa$ and on the spontaneous decay constant $\Gamma$. It is, of course, not a coincidence that the functional form of these jump operators is reminiscent of (the `square root' of) the two--body interaction that contributes to the SYK unitary dynamics, as the two arise from the same physical effect. On the other hand, the dissipator describing photons scattering outside the cavity is \cite{Uhrich:2023ddx}
\begin{multline}
    \qquad \mathcal D_{\Gamma} [\rho] = \int \de^2 r \left( L_\Gamma(r) \rho L_{\Gamma}^\dagger(r) - \frac{1}{2} \Big \{ L_{\Gamma}^\dagger(r) L_\Gamma(r) , \rho \Big\} \right) \ , \\
    \text{with} \qquad L_{\Gamma} (r) = \frac{\sqrt{\Gamma} \, \Omega_{\rm d} \, g_{\rm d} (r)}{\Delta_{\rm ad}(r) + i \Gamma/2} \, \psi_{\rm g}^{\dagger}(r) \psi_{\rm g} (r) \ , \qquad 
    \label{eq:Lindblad_op_photon_scattering}
\end{multline}
which is induced by photons that, instead of being virtually exchanged, are spontaneously emitted in free space by excited atoms. Because of this, its functional form is reminiscent of the one--body term in the Hamiltonian that we compensate in Section \ref{sec:Effective_cavity_Hamiltonian}. 

The dissipator associated with photon losses can be interpreted as resulting from the averaging over an ensemble of non--local continuous measurements of the overlap of the atomic distribution with the light--matter coupling pattern. Importantly, these photons can in principle be collected for each experimental realization with a finite efficiency, allowing for mitigation strategies based on post--selection. In contrast, spontaneous emission effectively projects the atomic position on one particular point in space, and occurs simultaneously and independently for each atom, yielding a dephasing.  While the overall dissipation rate is determined by the cooperativity~\cite{Solis:2025clm}, the relative weight of spontaneous emission versus photon leakage can be controlled through the choice of the pump detuning with respect to the cavity.

The interplay between the Trotterization protocol laid out before and the dissipative effects is that of an effective Lindbladian that takes into account every Trotter step as a different dissipation channel. More precisely, at every step of the experiment, the dynamics is run by a Lindbladian of the form
\begin{equation}
    \mathcal L_{\alpha}[\; \cdot \; ] = - i [H_{\alpha}, \; \cdot \; ] + \mathcal D_{\alpha}[\; \cdot \; ] \ ,
\end{equation}
where $\alpha$ runs over all different speckle realizations. The main idea is that our Trotterization protocol not only reproduces the effective unitary dynamics, but also the full Lindblad generator \eqref{eq:Lindblad_evolution_general} through
\begin{equation}
    e^{t \mathcal L} = \Bigg( \prod_{\alpha = 1}^{R} e^{\frac{t}{n} \mathcal L_{\alpha}} \Bigg)^n + \mathcal O \left(\frac{t^2}{2n}\right)  \ , 
    \label{eq:Lindblad_trotterization}
\end{equation}
where the error can be shown to be generally bounded \cite{Kliesch_2011}. Employing this result, our analytical and numerical studies will work directly at the level of the effective Lindbladian, rather than the Trotterized version (with the exception of Figure \ref{fig:dissipation_threepanel}).

In the following, we will concentrate on photon loss rather than photon scattering because it is illustrative of the specific properties of the platform, and yields interesting connections with existing work on Lindbladian SYK models \cite{Ryu_2022, Prosen_2022}.


\subsection{Lindblad spectrum} \label{sec:Lindblad_spectrum}

In this Section, we study the spectrum of the Lindblad evolution \eqref{eq:Lindblad_evolution_general} with jump operators of the form \eqref{eq:Lindblad_op_photon_loss} and \eqref{eq:Lindblad_op_photon_scattering}. Before doing that, we review some generic facts about spectra of superoperators. The fact that the evolution \eqref{eq:Lindblad_evolution_general} is CPTP can be simply shown using the Kraus decomposition
\begin{equation}
    \rho(t) = \sum_{a} M_a (t) \rho(0) M_a^\dagger(t) \ , \qquad \text{with} \qquad \sum_{a} M_a (t) M_a^\dagger(t) = \mathbbm 1 \ .
\end{equation}
Then, one has
\begin{equation}
    \Tr[\rho(t)] = 1 \qquad \text{and} \qquad \bra{\Psi} \rho(t) \ket{\Psi} \geq 0 \quad \forall \ket{\Psi} \ ,
\end{equation}
where the first condition is found using the cyclic property of the trace, while the second is evident by diagonalizing $\rho(0)$. However, another useful decomposition of $\rho(t)$ is in eigenvectors of the superoperator $\mathcal L$. In this case, we view the superoperator as a linear operator acting on the vector space of Hermitian matrices, so that 
\begin{equation}
    \mathcal L : \mathcal H \otimes \mathcal H^* \to \mathcal H \otimes \mathcal H^* \ .
\end{equation}
Within the vector space $\mathcal H \otimes \mathcal H^*$, to diagonalize $\mathcal L$ means to find matrices $\tilde \rho_n$ and $c$--numbers $\lambda_n \in \mathbb C$, so that
\begin{equation}
    \mathcal L [\tilde \rho_n] = \lambda_n \, \tilde \rho_n \ .
    \label{eq:Lindblad_eigenvalues_equation}
\end{equation}
Writing then the initial density matrix in the basis of $\tilde \rho_n$, the time evolution is completely determined by the eigenvalues
\begin{equation}
    \rho(0) = \sum_{n} c_n \tilde \rho_n \ , \qquad \to \qquad \rho(t) = \sum_n c_n \, e^{\Re(\lambda_n) t + i \Im(\lambda_n)t} \tilde \rho_n \ .
    \label{eq:expansion_density_matrix_eigenvalues}
\end{equation}
In the equation above, we have divided the real and imaginary parts of $\lambda_n$ to highlight some features of the evolution. First, to avoid any instabilities, one has that $\Re(\lambda_n) \leq 0$ for every $\lambda_n$. At late times, only eigenstates with vanishing real part of the eigenvalues survive, so that
\begin{equation}
    \rho(t) \to \sum_{ n \in \mathcal S } c_n \, e^{i \Im(\lambda_n) t} \tilde \rho_n \qquad \text{where} \qquad \mathcal S  = \{ n | \Re(\lambda_n) = 0 \} \neq \varnothing \ .
\end{equation}
The condition that $\mathcal S$ is non--empty is necessary to preserve the trace. In order to characterize the imaginary part of the eigenvalues as well, taking the adjoint of \eqref{eq:Lindblad_eigenvalues_equation}, one finds that 
\begin{equation}
    \mathcal L [\tilde \rho_n^\dagger] = \lambda_n^* \, \tilde \rho_n^\dagger \ ,
    \label{eq:Lindblad_eigenvalues_equation_conjugate}
\end{equation}
where we have used that the Lindbladian satisfies $\mathcal L^\dagger = \mathcal L$, needed to ensure that the density matrix is Hermitian at every instant of the evolution\footnote{The condition $\mathcal L^\dagger = \mathcal L$ implies that $\frac{\de}{\de t} \big( \rho - \rho^\dagger \big) = 0$, thus ensuring that the density matrix stays Hermitian. Notice that the condition $\mathcal L^\dagger = \mathcal L$ with the adjoint operation induced from the Hilbert space $\cH$ does not imply that $\mathcal L$ is adjoint as a linear operator in the vector space $\cH \otimes \cH^*$. For this reason, $\mathcal L$ is allowed to have complex eigenvalues, as shown in the main text.}. Equation \eqref{eq:Lindblad_eigenvalues_equation_conjugate} implies that, for any eigenvector $\tilde \rho_n$ with a complex eigenvalue $\lambda_n$ such that $\Im(\lambda_n) \neq 0 $, it exists another eigenvector $\tilde \rho_n^\dagger$ with eigenvalue $\lambda_n^*$. The reader should not be confused by the fact that $\tilde \rho_n$ is not Hermitian, as the diagonalization of the superoperator $\mathcal L$ is not bound to contain only Hermitian eigenvectors. The only requirement is that the overall density matrix is Hermitian, which forces a schematic rewriting of the form
\begin{equation}
    \rho(t) = \frac{1}{2} \sum_n e^{\Re(\lambda_n) t}  \left( c_n \, e^{i \Im(\lambda_n) t} \, \tilde \rho_n + c_n^* \, e^{-i \Im(\lambda_n) t} \, \tilde \rho_n^\dagger \right) \ ,
\end{equation}
implying that the coefficients in the expansion \eqref{eq:expansion_density_matrix_eigenvalues} for $\tilde \rho_n$ and $\tilde \rho_n^\dagger$ are complex conjugates of each other, to preserve the Hermitian nature of the density matrix.\footnote{For the Reader who might not be familiar with these concepts, we invite them to consider the case of a purely unitary evolution, where the Lindbladian is of the form $\mathcal L[\, \cdot \, ] = - i [H, \, \cdot \,]$. Given $\{ \ket{E_n} \}$, the set of eigenvectors of the Hamiltonian with eigenvalue $E_n$, any matrix $\ketbra{E_n}{E_m}$ is an eigenvector of $\mathcal L$ with eigenvalue $i (E_m - E_n)$. Therefore, the real part of all eigenvalues is zero (consistently with a unitary evolution), with the diagonal $\ketbra{E_n}{E_m}$ being steady states. Moreover, the eigenvector $\ketbra{E_n}{E_m}^\dagger = \ketbra{E_m}{E_n}$ has complex eigenvalue $ \big[i (E_m - E_n)\big]^* = i (E_n - E_m)$. In the more general case of a non--unitary evolution, some eigenvalues have non--zero real parts.}

We conclude this paragraph on the spectrum of Lindblad operators, noticing that the infinite temperature state
\begin{equation}
    \rho_{\infty} = \frac{\mathbbm 1}{D} 
\end{equation}
is an eigenvector of the Lindblad evolutions \eqref{eq:Lindblad_op_photon_loss} and \eqref{eq:Lindblad_op_photon_scattering} with zero eigenvalue, thus belonging to the set of steady states of the evolution.\footnote{We thank Yi-Neng Zhou for pointing this out.} Therefore, because of photons leaking out of the cavity, we expect the system to heat up to infinite temperature. In practice, this means that temperature will grow up to the point where the adiabatic eliminations yielding the effective Lindbladian breaks down. In this regime, the many--body physics of the effective model is less relevant, and we expect on general grounds the system to reach the single--atom cavity--cooling temperature limit which scales as $T \sim \Delta_{\rm cd}$ \cite{RevModPhys.85.553,Piazza:2014ab}.


\subsection{Numerical simulations} \label{sec:Lindbald_spectrum_numerical}
Let us now look first at numerical results of the Lindblad spectrum before turning to the analytical large--$N$ results of the Lindbladian evolution.
\subsubsection*{Spectrum}
To analyze the dissipative dynamics of the system numerically, we compute the spectrum of the Lindbladian superoperator $\mathcal{L}$ defined in Eq.~\eqref{eq:Lindblad_evolution_general}. Analogous to section \ref{sec:largeN}, we focus for the moment still solely on the effect of photon loss. This can be motivated in two ways. Firstly, even though in standard cQED dynamics a `good cavity' is quantified by having very small photon leakage through the mirrors, we are also interested in having some output of the cavity in order to have a measurable quantity. Thus, although ultimately the cooperativity determines the ratio of dissipation rates $\Gamma$ and $\kappa$ versus interaction strength \cite{Solis:2025clm}, one may wish to operate in a regime where $\Gamma$ is reduced as far as possible while $\kappa$ is kept finite. 
Secondly, numerically it is easier, in the sense that it is less memory--demanding to compute the jump operator~\eqref{eq:Lindblad_op_photon_loss_trap_modes} than computing the jump operator associated to photon scattering $\Gamma$. The latter involves an additional integral, scaling the numerical integration from $N^2$ to $N^4$ degrees of freedom. Therefore, we focus for the time being on the numerical analysis of the jump operator $\L_{\kappa}$. 

We then construct the two parts of the Lindbladian, a unitary contribution (the Trotterized  Hamiltonian built from experimental couplings generated via the speckle patterns) and the dissipative part (built from the jump operator of the form ~\eqref{eq:Lindblad_op_photon_loss_trap_modes}). For the numerical implementation, we chose values for the various cavity parameters such that the effective dissipative couplings reproduce an experimentally viable regime. In Appendix \ref{App:exp_parameters}, we list the used values and comment briefly on such realistic experimental parameters. 

\begin{figure}[t]
    \centering
\includegraphics[width=1\linewidth]{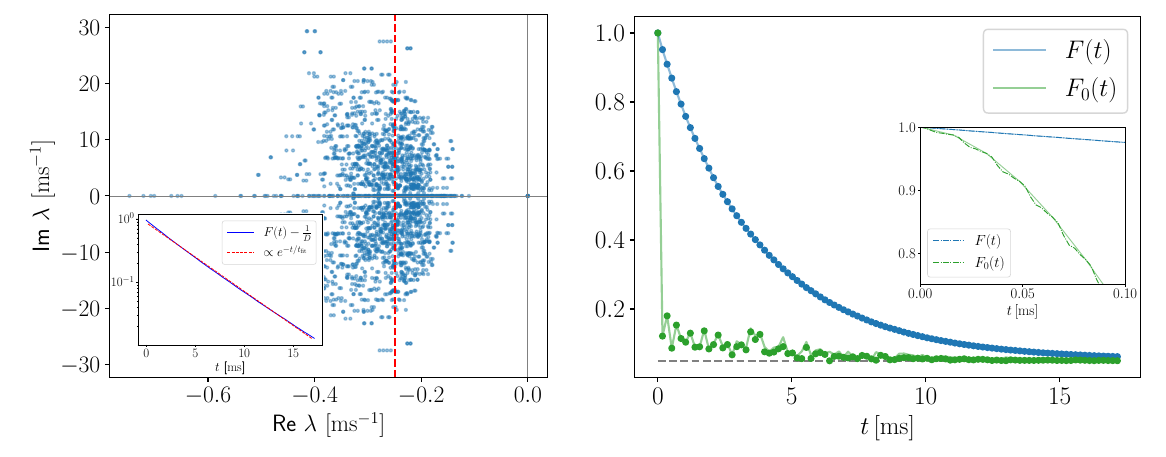}
\caption{{\bf Left}: Spectrum of the effective Lindbladian $\mathcal{L}$ for the Trotterized model with $R=6$,  $N=6$ fermions averaged over 20 realizations. The unique zero eigenvalue in the spectrum corresponds to the steady state. The red dashed line indicates the decay rate extracted from an exponential fit to the fidelity (left inset), and highlights the region of the spectrum governing the dominant relaxation processes.   {\bf Right}: Survival fidelity $F_0(t)$ and coherent fidelity $F(t)$ obtained from continuous evolution under the effective Lindbladian (solid lines). Dots show the corresponding fidelities computed from the discrete Trotterized protocol, averaged over realizations. The inset compares continuous and Trotterized evolution for a single realization, displaying all intermediate Trotter steps.}  
    \label{fig:dissipation_threepanel}
\end{figure}

Putting both contributions together into the final form of the Lindbladian, we numerically diagonalize $\mathcal{L}$ and plot first the complex spectra of the Lindbladian. We perform a total of 20 realizations, although the spectrum seen in Figure~\ref{fig:dissipation_threepanel} is averaged only over 6 realizations in order to be able to visually distinguish better between different states. Each eigenvalue $\lambda_n$ of $\mathcal{L}$ controls the decay (via $\Re(\lambda_n)$) and oscillatory (via $\Im(\lambda_n)$) behavior of the corresponding eigenmode $\tilde \rho_n$ in the time evolution of the density matrix as in Eq.~\eqref{eq:expansion_density_matrix_eigenvalues} via its real and imaginary part. As expected, the eigenvalue at the origin, seen in the small inset at the top left of Figure \ref{fig:dissipation_threepanel} corresponds to the infinite--temperature steady state $\rho_\infty = \mathbbm{1}/D$, with $D$ the Hilbert space dimension of the half--filling sector. In Figure~\ref{fig:dissipation_threepanel}, it is also visible that a spectral gap in the spectrum separates the origin from the first non--zero real part of the spectrum. The red dashed line plotted on top of the spectrum marks the inverse decay time $\tau^{-1}$ extracted independently from the exponential fit of the fidelity decay
(see lower left inset and right panel of Figure~\ref{fig:dissipation_threepanel}, more on it just below). Let us also remark that, fixing $N$ and increasing $R$, the spectral gap widens, implying a stronger dissipation. This is consistent with the findings of \cite{Ryu_2022}, and with our results of Section \ref{sec:largeN}, which we will comment on below.

\subsubsection*{Fidelity}

After computing the spectrum of the Lindbladian evolution, to better quantify the time dependence of dynamical observables we employ different notions of quantum state fidelities. Generally speaking, the fidelity between two density matrices $\rho$ and $\sigma$ is a measure of how close they are. An often-used definition is through the Uhlmann fidelity \cite{Liang_2019}
\begin{equation}
    F(\rho, \sigma) = \left( \Tr \left[ \sqrt{ \sqrt{\rho} \, \sigma \, \sqrt{\rho} } \right] \right)^2 \ .
\end{equation}
This quantity is symmetric in its arguments($F(\rho, \sigma) = F(\sigma, \rho)$), bounded between $0 \leq F(\rho, \sigma) \leq 1$, and equals 1 if and only if $\rho = \sigma$. Notably, if one of the two states is pure, say $\rho = \ketbra{\Psi}$, the fidelity simplifies to
\begin{equation}
    F(\ketbra{\Psi}, \sigma) = \bra{\Psi} \sigma \ket{\Psi} = \Tr[\sigma \ket{\Psi}\bra{\Psi} ] \ ,
\end{equation}
which is the expression we use in our analysis. In our numerical simulations, we compute two types of fidelities to analyze the dynamics under the Lindblad evolution:

\paragraph{(i) Survival Fidelity $F_0(t)$:}
We initialize the system in a random pure state $\rho_0 = \ketbra{\Psi}$ and let it evolve under the Lindbladian superoperator $\mathcal{L}$, which captures both coherent (only Hamiltonian) and dissipative dynamics. The evolved state at time $t$ is given by
\begin{equation}
    \sigma(t) = e^{t \mathcal{L}}\Big[\ketbra{\Psi}\Big] \ .
\end{equation}
We then compute the fidelity of $\sigma(t)$ with respect to the initial state,
\begin{equation}
    F_0(t) = \Tr[\sigma(t) \rho_0] = \bra{\Psi} e^{t \mathcal L} \Big[ \ketbra{\Psi} \Big] \ket{\Psi} \ ,
\end{equation}
which quantifies how much memory of the initial state is retained under the full dissipative evolution. Since dissipation typically drives the system toward a mixed steady state, $F_0(t)$ decays over time. This fidelity provides a pessimistic estimate of decoherence, as it measures loss of overlap with the exact initial state.

\paragraph{(ii) Coherent Fidelity $F(t)$:}
To isolate the effect of dissipation, we also evolve the same initial pure state $\rho_0$ under the Hamiltonian only, ignoring dissipation. This yields a unitary state:
\begin{equation}
    \rho_\text{unitary}(t) = e^{-iHt} \rho_0 e^{iHt} \ .
\end{equation}
We then compare this with the actual dissipative evolution $\sigma(t)$, and compute
\begin{equation}
    F(t) = \Tr[\sigma(t) \rho_\text{unitary}(t)] \ ,
\end{equation}
which quantifies how far the dissipative evolution goes away from the coherent one. This fidelity decays due to dissipation alone and is therefore used to extract a physical dissipation timescale. 
In the absence of dissipation, this fidelity would remain exactly equal to one, while the survival fidelity would still decay. As dissipation increases, the coherent fidelity starts decaying as well, but differently. In particular, $F_0(t)$ measures loss of overlap with the initial state (including both coherent and dissipative dynamics), while $F(t)$ isolates the irreversible decoherence effects, since both states being compared evolve from the same initial condition under different dynamics. 

In the right panel of Figure \ref{fig:dissipation_threepanel}, we show both fidelities, and see first the decay of $F_0(t)$ followed by the exponential decay of $F(t)$. Focusing on this latter decay, we can fit it to an exponential function in an appropriate time window. From the slope of $F(t) - 1/D$, we extract an associated dissipation timescale $\tau$. This timescale, the red dashed line in Figure \ref{fig:dissipation_threepanel}, is then compared to the spectrum of the Lindbladian. The fact that, increasing $R$, the gap in the spectrum widens also implies a stronger decay of the coherent fidelity, which approaches that of the survival fidelity at very large $R$. This had to be expected from the results on the spectrum, and one might be worried that, in the interesting experimental scenario where both $N$ and $R$ are large, the dissipative dynamics is too strong to keep any coherent dynamics. In the next Section, we will show that this is not the case, and that the effect of dissipation versus coherent interaction strength only depends on the ratio $\gamma = R / N$.


\subsection{Large--\texorpdfstring{$N$}{} analysis}
\label{sec:largeN}

This analysis parallels the one already presented in an Appendix of \cite{Baumgartner:2024ysk}, with some additional details.

As already mentioned, we only focus on photons leaking out of the cavity through the cavity mirrors. 
This is described by the jump operator \eqref{eq:Lindblad_op_photon_loss}, which in the basis of the trap modes becomes
\begin{equation}
    L_{\kappa} = \sum_{ij} \left(\frac{\sqrt{\kappa} \, \Omega_{\rm d} \Omega_{\rm c}}{2(\Delta_{\rm cd} - i \kappa/2)} \int \de^2 r \,  \frac{g_{\rm d} (r) g_{0} (r) \phi_i^*(r) \phi_j(r)}{\Delta_{\rm ad}(r) + i \Gamma/2} \right) c_i^\dagger c_j \ .
    \label{eq:Lindblad_op_photon_loss_trap_modes}
\end{equation}
The dependence on the speckle pattern implies that these jump operators have a random component. This is rather natural, as we expect the dissipation to depend on the local intensity of the light pattern (the speckle). We notice that the sum has a large diagonal component, which nonetheless is proportional to the number operator. Since the unitary dynamics commutes with the number operator, and if we initialize the state in a fixed sector of such operator, this diagonal component will not affect the subsequent dynamics. Therefore, schematically, the Lindblad jump operator is of the form
\begin{equation}
    L = \sum_{ij} K_{ij} c_i^\dagger c_j \ , \qquad \qquad \overline{|K_{ij}|^2} = \frac{K^2}{N^2} \ ,
    \label{eq:generic_quadratic_jump_ops}
\end{equation}
\textit{i.e.}, quadratic in the fermionic sites. The variance is set according to the fact that the $K_{ij}$ should be thought of as `the square root' of the SYK couplings, so that their variance scales as \eqref{eq:generic_quadratic_jump_ops}. Such jump operators have also been studied in \cite{Ryu_2022}, with the important difference that at finite $N$ and $R$, our jump operators are not independent from the SYK dynamics. Results in Section \ref{sec:Statistical_analysis_couplings} allow us to claim that in the large--$R$ regime, statistical independence holds, which suggests us to use the methods employed in \cite{Ryu_2022} for the present context.

We then take inspiration from \cite{Ryu_2022} to study how dissipation affects the dynamics of the system. The strategy employs coherent state path integrals, and the evolution around the infinite temperature state, which is a stable state of the system, and which can be written in terms of a Keldysh effective theory \cite{Sieberer:2015svu} as
\begin{equation}
    Z  = \int \mathcal D[c_\pm, c_\pm^\dagger] \, e^{iS[c_{\pm}, c_\pm^\dagger]} \ ,
\end{equation}
where the effective action is
\begin{multline}
    S[c_\pm, c_\pm^\dagger] = \int \de t \bigg[ \sum_{k = 1}^N i c_{k,+}^\dagger \dot c_{k,+} - \sum_{k = 1}^N i c_{k,-}^\dagger \dot c_{k,-} - H_+[c_+, c_+^\dagger] + H_-[c_-, c_-^\dagger] \\
    - i \sum_\alpha \left( 2 L_{\alpha, -}^\dagger L_{\alpha,+} - L_{\alpha,+}^\dagger L_{\alpha,+} - L_{\alpha, -}^\dagger L_{\alpha,-} \right) \bigg]  \ .
\end{multline}
It is no surprise that the resulting action is quadratic in the jump operators. A standard procedure to linearize it is to employ auxiliary fields $\theta_{\alpha, \pm}$ to perform a Hubbard--Stratonovich transformation, resulting in
\begin{multline}
    S[c_\pm, c_\pm^\dagger, \theta_{\alpha, \pm}, \bar \theta_{\alpha, \pm}] = \int \de t \left( \sum_{k = 1}^N i c_{k,+}^\dagger \dot c_{k,+} - \sum_{k = 1}^N i c_{k,-}^\dagger \dot c_{k,-} - H_+[c_+, c_+^\dagger] + H_-[c_-, c_-^\dagger] \right) + \\
    i \int \de t \sum_{\alpha} \left[ \begin{matrix}
    \bar \theta_{\alpha, +} (t) \\  \bar \theta_{\alpha, -} (t)
    \end{matrix} \right]^\mathsf{T} \left[ \begin{matrix}
    1 & 0 \\
    2 & 1
    \end{matrix} \right]
    \left[\begin{matrix}
    \theta_{\alpha, +} (t) \\
    \theta_{\alpha, -} (t)
    \end{matrix} \right] -i  \int \de t \sum_{\alpha} \Big( \bar \theta_{\alpha, +} (t) L_{\alpha, +} (t) +  \bar \theta_{\alpha, -} (t) L_{\alpha, -} (t) \, + \\
    L_{\alpha, +}^{\dagger} (t) \theta_{\alpha, +} (t) + L_{\alpha, -}^{\dagger} (t) \theta_{\alpha, -}(t) \Big)  \ .
    \label{SK_action}
\end{multline}
We now have an effective action which is quadratic in the fermionic sites. This allows us to perform a similar analysis as the usual one done for SYK, identifying the two--point function for the fermions $G(t) = \langle c_i^\dagger(t) c_i(0) \rangle$ and for the auxiliary fields $G_{ab}^\theta(t) = \langle \bar \theta_{\alpha, a} (t)  \theta_{\alpha, b} (0) \rangle$, and their corresponding self energies $\Sigma(t)$ and  $\Sigma^\theta_{ab}(t)$. From this, the Schwinger--Dyson equations for the auxiliary fields are simply
\begin{equation}
    \Sigma_{ab}^{\theta}(t) = - \frac{K^2}{4} G_{ab}^2(t) \ , \qquad \qquad \mathbf{G}^{\theta}(t) = \left[ \begin{pmatrix}
    1 & 0 \\
    -2 & 1\\
    \end{pmatrix} \delta(t) - \mathbf \Sigma^{\theta}(t)\right]^{-1} \ ,
\end{equation}
while the ones for the fermionic sites are
\begin{equation}
    \Sigma_{ab}(t) = \, - \frac{J^2 R}{N} \, s_{ab} G_{ab}^3 (t) + \frac{K^2 R}{N} \left(G_{ab}^\theta (t) + G_{ba}^\theta (-t) \right) G_{ab} (t) \ , \qquad \qquad \mathbf{G}(t) = \left[ \mathbf G_0^{-1}(t) - \mathbf \Sigma(t)\right]^{-1} \ .\label{Sigma_dissipation} 
\end{equation}
The subscripts $a$ and $b$ take values in the set $\{ +, - \}$ for the Schwinger-Keldysh contour, with the matrix $s_{ab}$ having components $s_{++} = s_{--} = 1$ and $s_{+-} = s_{-+} = -1$. The inverses refer to both the $\{a,b\}$ indices, as well as for the time--domain. Equation \eqref{Sigma_dissipation} makes it apparent that dissipation contributes at the same order as the unitary evolution in the large--$N$ limit. This implies that, at least for two--point functions,  the dissipation only depends on the ratio $\gamma = R /N$, provided we assume $R$ to scale linearly with $N$. 

We are now in the position to explain the effects we observed in the numerical simulations, and to extract lessons for the large--$N$ limit. From \eqref{Sigma_dissipation}, it is evident that, changing $\gamma$, both the unitary and the dissipative dynamics are affected, in particular resulting in timescales of the order
\begin{align}
    t_{\rm unitary} \, \sim & \; (\sqrt{\gamma} J)^{-1} \, \sim \,  \frac{\Delta_{\rm ad}^2 \Delta_{\rm cd}}{\sqrt{\gamma} \, \Omega_{\rm d}^2 \Omega_{\rm c}^2}  \ , \nonumber \\ t_{\rm dissipative} \, \sim & \, (\gamma K^2)^{-1} \, \sim \, \frac{(4 \Delta_{\rm cd}^2 + \kappa^2) \Delta_{\rm ad}^2}{\gamma \kappa \Omega_{\rm d}^2 \Omega_{\rm c}^2} \ .
\end{align}
The one on top reflects the fact that, summing over several realizations, the variance of the unitary SYK couplings are additive, and thus the timescale gets reduced by a factor of $\sqrt{\gamma}$. The one on the bottom results from the fact that the dissipative strength is additive in the dissipative modes, and thus such timescale gets reduced by a factor of $\gamma$, contrary to the unitary case. This confirms the numerical results obtained above (see Figure \ref{fig:dissipation_threepanel}), where fixing $N$, the higher the $R$, the more dominant the dissipation is. The same happens at large $N$, but with the crucial caveat that the relative strength between the coherent vs dissipative dynamics depends on $\gamma$ and not solely on $R$. This is reassuring news, as one hopes to reach both large $N$ and $R$ in an experimental realization. In order to have a dominant coherent dynamics, we have to compare the two timescales, which are related by
\begin{equation}
    t_{\rm dissipative} = \left(\frac{4 \Delta_{\rm cd}^2 + \kappa^2}{\sqrt{\gamma} \Delta_{\rm cd} \kappa} \right) \, t_{\rm unitary} \ .
\end{equation}
If we want the unitary dynamics to dominate over the dissipative one, we then must have
\begin{equation}
    \frac{4 \Delta_{\rm cd}^2 + \kappa^2}{\sqrt{\gamma} \Delta_{\rm cd} \kappa} \gg 1 \qquad \text{thus} \qquad \sqrt{\gamma} \ll \frac{4 \Delta_{\rm cd}^2 + \kappa^2}{\Delta_{\rm cd} \kappa} \ .
\end{equation}
Using the values in Appendix \ref{App:exp_parameters}, namely $\Delta_{\rm cd} = 2\pi \times 20$ MHz and $\kappa = 2 \pi \times 0.16$ MHz, we obtain the value $\sqrt{\gamma} \ll 500$, thus a very large ratio between trotter cycles and $N$. It is quite surprising to find such a large number, as the numerics we performed in Section \ref{sec:Lindbald_spectrum_numerical} for $\gamma = 1$ were suggesting dissipation to be smaller but somewhat comparable in size with the unitary dynamics. It should also be noted however that the numerics was performed for $N = 6$ (limited by the expensive nature of Lindblad numerical simulations), while in the present context we are considering large $N$.


\section{Discussion and future directions} \label{sec:discussion}

In this work, we have introduced a new Trotterized approach to simulate fully disordered all--to--all Hamiltonians, exemplified by the complex SYK$_4$ model within a single--mode cavity QED setup. The central idea is to `densify' the disorder by sequentially cycling through a set of sparse random interaction patterns (speckles), such that the time--averaged effect reproduces a dense, fully random Hamiltonian. This analog--digital strategy greatly expands the scope of models that can be realized in cavity platforms, enabling quantum simulations of disordered systems that were previously out of reach. We showed in detail how an ensemble of ultracold atoms inside a high--finesse optical cavity can implement the required time--dependent random couplings, with a concrete recipe for engineering the cSYK$_4$ interactions. A key outcome of our analysis is that the computational complexity of this scheme scales favorably with the system size $N$, improving over previous `sparsified' SYK proposals \cite{Xu:2020shn, Garcia_Garcia_2021, Granet2025}. 

We now expand on a few interesting directions that would be worthwhile to explore in future works.

\subsubsection*{Superradiance, self--organization, and disorder}

One of the most robust theoretical and experimental predictions of cQED platforms is the onset of Dicke superradiance \cite{Dicke_1954, Hepp_1973, Garraway_2011} (or self--organization) for $N$ two--level systems coupled to a photonic field. In the large--$N$ limit, the mean--field approach predicts a critical value of the coupling where the $N$ two--level systems spontaneously emit coherent photons, leading to a superradiant regime whose resulting photonic field's intensity scales as $N^2$, contrary to an incoherent sum of $N$ emitters whose photonic field would scale as $N$. Superradiant and self--organization transitions have been experimentally studied in numerous examples, such as \cite{Esslinger_2010, Brantut_2023}. As a clean experimental probe of light--matter interaction, it is extremely interesting to understand if and how such phase transitions are affected by the presence of disorder. In the case where a phase transition is still present with dense disorder, it is particularly tempting to understand possible connections with phase transitions in holographic models \cite{Hartnoll:2008vx, Hartnoll:2008kx, Gauntlett:2009dn, Gauntlett:2009bh}. To this aim, it would be fruitful perhaps to approach the theory without integrating out the photonic field, studying the so--called Yukawa--SYK model \cite{Esterlis_2019,Solis:2025clm}. In this direction, some works have studied the holographic dual of Yukawa--SYK and holographic superconductivity \cite{Schmalian_2022, Shankar:2025fok}.

\subsubsection*{Dissipative quantum chaos in cQED}

Another extremely interesting research direction would be to understand how the quantum--chaotic properties of such models are modified by the presence of dissipation. Some works in the literature \cite{Garcia-Garcia:2024tbd} have studied the behavior of the Lyapunov exponent of OTOCs for Majorana SYK coupled to a Markovian bath, modeled with a Lindblad evolution with linear (in the fermionic sites) jump operators. Such works noticed a decrease of the Lyapunov exponent when increasing the coupling $\mu$ with the bath, up to a critical $\mu_{\rm cr}$ where the Lyapunov exponent vanishes, signaling a transition from a chaotic to a non--chaotic system. A similar study could be conducted for the case of the dissipative evolutions studied in Section \ref{sec:Dissipation}, which would then bound the possible values of $\kappa$ and $\Gamma$ (and of the cooperativity) to have a positive Lyapunov exponent. A similar outcome to \cite{Garcia-Garcia:2024tbd} is foreseeable, perhaps tamed by the correlated disorder between system and environment, arising from the same physical effect. We expect these questions to be closely connected to ongoing studies of many-body quantum chaos in dissipative cavity setups~\cite{FerrariSavona_inprep}.

\subsubsection*{Quantum simulations of de Sitter}
Strikingly, recent works have found a connection between a double--scaled version of SYK \cite{Berkooz:2018jqr} at high temperature and the physics of two-- and three--dimensional de Sitter quantum gravity \cite{Narovlansky:2023lfz, Verlinde:2024znh, Verlinde:2024zrh}. While the model studied in the present work, and presumably any other model realized through analog simulators is far from the double--scaled limit (where the number of interacting fermions at each vertex is $p \to \infty$), it is an exciting possibility to extract some signatures of de Sitter quantum gravity in quantum simulators. Moreover, as remarked in Section \ref{App:connections_to_gravity}, the most natural states to prepare in such simulators are  very high--energy states for the effective models under study in this work, which puts de Sitter in a more favorable position than the AdS counterpart which appears only at low temperatures.

\subsubsection*{Prospects of experimental observations} 

We have focused in the analysis on the spectral form factors and OTOCs, which are key elements to demonstrate the convergence of the proposed simulation algorithm towards the SYK model, but those are unlikely to be directly experimentally accessible, even though protocols have been proposed in the cavity--QED context \cite{Swingle:2016var}, as
well as digital approaches \cite{Garcia-Alvarez:2016wem} and complementary protocols based on projected Loschmidt echoes and unitary-design formation  
\cite{Zhou:2025_renyientropy, Zhou:2025_kdesign}. Nevertheless, observables such as the mean occupation of the orbitals are accessible experimentally and can be used to track the dynamics in time and compare it with the predictions of the model. Techniques of fluorescence imaging have recently been demonstrated on few--fermion quantum systems, capable of detecting high--order correlations and entanglement generation \cite{PhysRevLett.125.180402, PhysRevLett.126.020401} and compatible with the next generation of cavity QED platforms with Fermi gases \cite{Bolognini:25}. This could then allow for the investigation of response functions sensitive to density--density correlations, which display a characteristic power--law dependence on frequency, sensitive to the rank of the SYK interactions \cite{Kim:2019lwh}.


\section*{Acknowledgements}
We thank Ehud Altman, Diego Barberena, Rohit Bhatt, Panagiotis Christodoulou, Léa Dubois, Mae Eichenberger, Ekaterina Fedotova, Yann Kiefer, Andrea Legramandi, Robin Löwenberg, David Pascual Solis, Nicola Reiter, Filippo Ferrari, Adrián Sánchez-Garrido, Nick Sauerwein, Alex Windey, Yi-Neng Zhou for insightful discussions.
This work was performed in part at Aspen Center for Physics, which is supported by National Science Foundation grant PHY-2210452. This work is supported by the Fonds National Suisse de la Recherche Scientifique through Project Grant $200021\_{}215300$, $200020E\_{}217124$ and the NCCR ``The Mathematics of Physics (SwissMAP)" (NCCR51NF40-141869.), the Swiss State Secretariat for Education, Research and Innovation (SERI) under contract number UeMO19-5.1 and $20\text{QU}-1\_{}215924$. This project has received funding from the Swiss State Secretariat for Education, Research and Innovation (SERI) under contract number UeMO19-5.1 and from from Fondazione Cassa di Risparmio di Trento e Rovereto (CARITRO) through the project SQuaSH - CUP E63C24002750007, from project DYNAMITE QUANTERA2\_00056 funded by the Ministry of University and Research through the ERANET COFUND QuantERA II – 2021 call and co-funded by the European Union (H2020, GA No 101017733). This project has received funding from the European Union’s Horizon Europe research and innovation programme under grant agreement No 101080086 NeQST.
Funded by the European Union. Views and opinions expressed are however those of the author(s) only and do not necessarily reflect those of the European Union or the European Commission. Neither the European Union nor the granting authority can be held responsible for them.
This work was supported by the Provincia Autonoma di Trento, and Q@TN, the joint lab between University of Trento, FBK—Fondazione Bruno Kessler, INFN—National Institute for Nuclear Physics, and CNR—National Research Council.

\bibliographystyle{utphys}
\bibliography{extendedrefs}

\appendix 

\section{Trotterization error}
\label{App:second_order}
In this Appendix, we want to give analytical bounds on the error between the exact time evolution and the Trotterized time evolution. We will be particularly concerned with understanding how it scales with $N$ and $R$ both assumed to be large, while we will be more cavalier on $\mathcal O(1)$ coefficients, which would nonetheless be hard to find exactly. From the Trotter formula \eqref{eq:U_T_steps}, we are thus interested in understanding the size of 
\begin{equation}
    \ve  = \frac{t^2}{2n} \sum_{\a < \b}^R [H_\a,H_\b]
\end{equation}
through a suitable matrix norm. Calling $D$ the dimension of the Hilbert space, as motivated at the beginning of Section \ref{sec:SparseToFullSYK4}, a convenient norm is the Frobenius norm, which for every matrix $A$, is defined as
\begin{equation}
	\| A \|^2 = \frac{1}{D} \, \Tr [ A^{\dagger} A ] = \frac{1}{D} \sum_{ij} |A_{ij}|^2 \ .
\end{equation}
This norm is convenient as any unitary matrix is a unit norm vector, and thus we can see the size $\| \ve \|$ as the tolerance we allow on the time evolution. Additionally, since our Hamiltonian $H$ and all the $H_{\alpha}$'s commute with the number operator
\begin{equation}
	\hat N = \sum_{i} c^{\dagger}_i c_i  \ ,
\end{equation}
we only consider a subspace of fixed occupation number. While we will keep $D$ generic for the moment, later on, we will focus on the half-filled sector, which is the biggest among all. 

To proceed, we notice the obvious fact that the precise size of the tolerance $\| \ve \|$ generally depends on the specific realization of the couplings $J_{ik}^\alpha$, and thus we are only interested in making average statements. While it would certainly be interesting to compute $\overline{\| \ve \|}$, we employ the fact that the square root is a concave function so that the inequality
\begin{equation}
    \overline{\| \ve \|} = \overline{ \sqrt{\| \ve \|^2} } \leq \sqrt{\overline{ \| \ve \|^2 }} 
	\label{eq:averaging}
\end{equation}
holds. Such inequality is important as we are interested in the LHS of \eqref{eq:averaging}, but the RHS is easier to compute and is able to produce an upper bound on the tolerance. With a clear abuse of notation, we will thus define
\begin{equation}
    \ve \equiv \sqrt{\overline{ \| \ve \|^2 }} 
	\label{eq:atolerance_definition}
\end{equation}
and focus on computing the RHS. The most difficult part is to obtain an estimate of
\begin{multline}
	\overline{ \sum_{\alpha < \beta , \gamma < \delta}  \Tr \Big( [H_\alpha, H_\beta] [H_\delta, H_\gamma] \Big) } = \sum_{\alpha < \beta , \gamma < \delta}  \overline{J^{\a}_{i_1 k_1} J^{\a}_{j_1 l_1} J^{\b}_{i_2 k_2} J^{\b}_{j_2 l_2} J^{\d}_{i_3 k_3} J^{\d}_{j_3 l_3} J^{\g}_{i_4 k_4} J^{\g}_{k_1 l_4}} \; \; \times \\ 
	\Tr \left( \Big[ c^{\dagger}_{i_1} c_{k_1} c^{\dagger}_{j_1} c_{l_1} , c^{\dagger}_{i_2} c_{k_2} c^{\dagger}_{j_2} c_{l_2} \Big] \Big[ c^{\dagger}_{i_3} c_{k_3} c^{\dagger}_{j_3} c_{l_3} , c^{\dagger}_{i_4} c_{k_4} c^{\dagger}_{j_4} c_{l_4} \Big] \right) 
    \label{eq:average_of_trace_of_comm_squared}
\end{multline}
which holds at large $R$ and $N$. We proceed in steps. First, for simplicity we take the couplings $J^{\a}_{i k}$ to be independent random variables, such that
\begin{equation}
    \overline{ J^{\a}_{i k} J^{\b}_{j l} } = \frac{\sqrt{2} J}{N^2} \, \delta^{\a \b} \delta_{il} \delta_{jk} \ .
\end{equation}
If we further assume them to be Gaussian, we can use Wick's theorem to find the various contractions contributing to the coupling average in the RHS of \eqref{eq:average_of_trace_of_comm_squared}. There are $(8-1)!!$ possible contractions, but not all of them contribute (or contribute equally) to the sum. For example, if at least one among the realizations $\{ \alpha, \beta, \gamma, \delta \}$ is contracted with itself, the commutators vanish, in formulae
\begin{equation}
	 \sum_{\alpha < \beta , \gamma < \delta}  \Tr \Big( \big[ \, \overline{ \big. H_\alpha } , \overline{ H_\beta \big] \big [H_\d, H_\g } \big] \Big) = 0 \ .
	 \label{avg_one_diff}
\end{equation}
The reason is that the mean Hamiltonian is proportional to the number operator
\begin{equation}
	\overline{ H_{\alpha} } = \sum_{ijkl} \overline{ J^{\alpha}_{ik} J^{\alpha}_{jl} } \, c^{\dagger}_i c_k c^{\dagger}_j  c_l = \frac{\sqrt{2} J}{N^2} \sum_{ij}  c^{\dagger}_i c_j c^{\dagger}_j c_i = \sqrt{2} J \left( \frac{\hat N}{N} - \frac{\hat N^2}{N^2} + \frac{\hat N}{N^2}  \right) \ ,
	\label{avg_H_a_N}
\end{equation}
and all the other terms in the commutators \eqref{avg_one_diff} commute with the number operator, so the result is always zero. This implies that, in the coupling average, each Greek index has to be equal to a different one. Additionally, the condition $\a < \b$ and $\g < \d$ implies that the only possibility is $\a = \g$ and $\b = \d$. Therefore
\begin{equation}
	\overline{ J^{\a}_{i_1 k_1} J^{\a}_{j_1 l_1} J^{\b}_{i_2 k_2} J^{\b}_{j_2 l_3} J^{\d}_{i_3 k_3} J^{\d}_{j_3 l_3} J^{\g}_{i_4 k_4} J^{\g}_{j_4 l_4} } = \overline{ J^{\a \textcolor{white}{\beta}}_{i_1 k_1} J^{\a}_{j_1 l_2} J^{\g}_{i_4 k_4} J^{\g}_{j_4 l_4} } \, \, \overline{ J^{\b}_{i_2 k_2} J^{\b}_{j_2 l_2} J^{\d}_{i_3 l_3} J^{\d}_{j_3 l_3} }
\end{equation}
Similarly
\begin{align}
	\overline{ J^{\a}_{i_1 k_1} J^{\a}_{j_1 l_1} J^{\g}_{i_4 k_4} J^{\g}_{j_4 l_4} } \, = & \;  \overline{ J^{\a}_{i_1 k_1} J^{\g}_{i_4 k_4} } \, \,  \overline{ J^{\a}_{j_1 l_1} J^{\g}_{j_4 l_4} } + \overline{ J^{\a}_{i_1 k_1} J^{\g}_{j_4 l_4} } \, \,  \overline{ J^{\a}_{j_1 l_1} J^{\g}_{i_4 k_4} } \nonumber \\
	= & \;  \frac{2 J^2}{N^4} \, \delta^{\a \g} \left( \delta_{i_1 k_4} \delta_{i_4 k_1} \delta_{j_1 l_4} \delta_{j_4 l_1} +   \delta_{i_1 l_4} \delta_{j_4 k_1} \delta_{j_1 k_4} \delta_{i_4 l_1} \right)  \ .
\end{align}
Moreover, performing the sum over $\alpha$ and $\beta$ in \eqref{eq:average_of_trace_of_comm_squared}, each term contributes equally, and we have a total of $R(R-1) / 2 \approx R^2/2$ terms due to the $\a < \b$ constraint. Putting everything together we have 
\begin{multline}
	\overline{ \sum_{\alpha < \beta , \gamma < \delta}  \Tr \Big( [H_\alpha, H_\beta] [H_\delta, H_\gamma] \Big) } = \\
    \frac{2 J^4 R^2}{N^8}  \, \Big [ \delta_{i_1 k_4} \delta_{i_4 k_1} \delta_{j_1 l_4} \delta_{j_4 l_1} \delta_{i_2 k_3} \delta_{i_3 k_2} \delta_{j_2 l_3} \delta_{j_3 l_2} + \delta_{i_1 k_4} \delta_{i_4 k_1} \delta_{j_1 l_4} \delta_{j_4 l_1} \delta_{i_2 l_3} \delta_{j_3 k_2} \delta_{j_2 k_3} \delta_{i_3 l_2} \qquad \quad \\
	+ \delta_{i_1 l_4} \delta_{j_4 k_1} \delta_{j_1 k_4} \delta_{i_4 l_1} \delta_{i_2 k_3} \delta_{i_3 k_2} \delta_{j_2 l_3} \delta_{j_3 l_2} + \delta_{i_1 l_4} \delta_{j_4 k_1} \delta_{j_1 k_4} \delta_{i_4 l_1} \delta_{i_2 l_3} \delta_{j_3 k_2} \delta_{j_2 k_3} \delta_{i_3 l_2} \Big ] \\
	\Tr \left( \Big[ c^{\dagger}_{i_1} c_{k_1} c^{\dagger}_{j_1} c_{l_1} , c^{\dagger}_{i_2} c_{k_2} c^{\dagger}_{j_2} c_{l_2} \Big] \Big[ c^{\dagger}_{i_3} c_{k_3} c^{\dagger}_{j_3} c_{l_3} , c^{\dagger}_{i_4} c_{k_4} c^{\dagger}_{j_4} c_{l_4} \Big] \right) \ ,
	\label{trace_expanded}
\end{multline}
where we have used the convention of summing over repeated indices. Expanding and omitting prefactors, we have
\begin{align}
	\text{RHS of \eqref{trace_expanded}} \, \propto \; & \Tr \left( \Big[ c^{\dagger}_{i_1} c_{i_4} c^{\dagger}_{j_1} c_{j_4} , c^{\dagger}_{i_2} c_{i_3} c^{\dagger}_{j_2} c_{j_3} \Big] \Big[ c^{\dagger}_{i_3} c_{i_2} c^{\dagger}_{j_3} c_{j_2}, c^{\dagger}_{i_4} c_{i_1} c^{\dagger}_{j_4} c_{j_1}  \Big] \right) + \nonumber \\
	 &\Tr \left( \Big[ c^{\dagger}_{i_1} c_{i_4} c^{\dagger}_{j_1} c_{j_4} , c^{\dagger}_{i_2} c_{j_3} c^{\dagger}_{j_2} c_{i_3} \Big] \Big[ c^{\dagger}_{i_3} c_{j_2} c^{\dagger}_{i_3} c_{j_2}, c^{\dagger}_{i_4} c_{i_1} c^{\dagger}_{j_4} c_{j_1}  \Big] \right) + \nonumber \\
	 &\Tr \left( \Big[ c^{\dagger}_{i_1} c_{j_4} c^{\dagger}_{j_1} c_{i_4} , c^{\dagger}_{i_2} c_{i_3} c^{\dagger}_{j_2} c_{j_3} \Big] \Big[ c^{\dagger}_{i_3} c_{i_2} c^{\dagger}_{j_3} c_{j_2}, c^{\dagger}_{i_4} c_{j_1} c^{\dagger}_{j_4} c_{i_1}  \Big] \right) + \nonumber \\
	 &\Tr \left( \Big[ c^{\dagger}_{i_1} c_{j_4} c^{\dagger}_{j_1} c_{i_4} , c^{\dagger}_{i_2} c_{j_3} c^{\dagger}_{j_2} c_{i_3} \Big] \Big[ c^{\dagger}_{i_3} c_{j_2} c^{\dagger}_{i_3} c_{j_2}, c^{\dagger}_{i_4} c_{j_1} c^{\dagger}_{j_4} c_{i_1}  \Big] \right) \ .
	\label{contraction}
\end{align}
We would like to make an estimate of this quantity. To simplify the analysis, we try to estimate the leading term at large-$N$. We have to consider all the possible combinations of indices in the expression above. Clearly, we have the largest number of terms when all the indices are different from each other, which would give
\begin{equation}
	\frac{N!}{(N-8)!}
\end{equation}
independent contributions. However, terms of this form do not contribute to the sum, since all the commutators independently vanish. Next in line are combinations in which exactly two indices are equal. We would like to argue that these terms also do not contribute to the sum. To see it, let's look at \eqref{contraction}. Without loss of generality, we can take $i_1$ to be equal to another index. However, $i_1$ cannot be equal to $j_1$, $i_4$, or $j_4$, because otherwise the commutators would again be zero. On the other hand, $i_1$ cannot be equal to $i_2$ or $j_2$, since in this case we would have two consecutive annihilation (or creation) operators, which on any vector of the Hilbert space gives zero (due to the fermionic nature of the sites). Therefore, the only possibilities are $i_1 = i_3$ or $i_1 = j_3$. Without loss of generality, let us consider the former possibility. To proceed, we notice that, analogously, also $j_1$ could be equal only to $i_3$ or $j_3$. We notice that if we consider the terms in which $j_3 = i_3$, these contribute exactly the same, but with an opposite sign, as the terms in which $i_1 = i_3$. Thus, they cancel each other, and the overall contribution from terms in which two indices are equal vanishes. Therefore, the leading term at large-$N$ must come from terms in which out of eight indices, only six are independent. This can be attained if either two pairs of indices are equal, or if three indices are all equal to each other. Both possibilities lead to non-zero results. To estimate the error, we notice that the commutators give a total of sixteen terms, and each term can be estimated as 
\begin{equation}
	\binom{8}{3} \frac{N!}{(N-6)!} \binom{N-6}{\frac{N-6}{2}} + \frac{1}{2} \binom{8}{2}^2 \frac{N!}{(N-6)!} \binom{N-6}{\frac{N-6}{2}} = 448 \, \frac{N!}{(N-6)!} \binom{N-6}{\frac{N-6}{2}}
\end{equation} 
We can use this result to obtain a numerical estimate of \eqref{trace_expanded}. In the half-filled sector, it is
\begin{equation}
	\overline{ \sum_{\alpha < \beta , \gamma < \delta}  \Tr \Big( [H_\alpha, H_\beta] [H_\delta, H_\gamma] \Big) } \, \lesssim \, \frac{14336 J^4 R^2}{N^8} \, \frac{N!}{(N-6)!} \frac{\binom{N-6}{(N-6)/2}}{\binom{N}{N/2}} \approx 2 \times 10^{2} \, \frac{J^4 R^2}{N^2}
    \label{eq:final_estimate}
\end{equation}
where in the last passage, we have used Stirling's approximation of a factorial and extracted the large-$N$ limit. We have obtained this result in the half-filled sector, but as long as we are not at the edge of the filling (thus almost empty or almost full), a similar result holds. All in all, the error is 
\begin{equation}
    \ve \approx 10 \, \frac{t^2 J^2 R}{n N} \ ,
\end{equation}
but we can also ask the `inverse question', namely, given a tolerance $\ve$ and a total simulation time $t$, how many Trotter steps should we perform in order to have the desired accuracy. The result is obviously
\begin{equation}
    n \gtrsim  10 \, \frac{t^2 J^2 R}{\ve N} \ ,
    \label{eq:number_trotter_steps_desired_accuracy}
\end{equation}
as was first reported in \cite{Baumgartner:2024ysk} without a proof. Perhaps the most remarkable feature of \eqref{eq:number_trotter_steps_desired_accuracy}, which originally comes from \eqref{eq:final_estimate}, is that for $R \sim N$ it is independent of the system size, which is quite promising for a quantum simulation involving possibly a large number of atoms. 

\section{Bound on Lipschitz functions}
\label{sec:lipscitz_bound}
Given metric spaces $X$ and $Y$, a function $f : X \to Y$ is called Lipschitz continuous is it exists a real positive constant $\kappa_{\rm L}$ for which 
\begin{equation}
    \| f(x_1) - f(x_2) \|_Y \leq \kappa_{\rm L}\, \| x_1 - x_2 \|_X  \ , \qquad \text{for all} \; \; x_1,x_2 \in X\ .
\end{equation}
The smallest constant $\kappa_{\rm L}$ is usually called (best) Lipschitz constant. In our case, we are interested in the SFF, which is a function that maps a unitary matrix to a real number. In formulae $\SFF : U(D) \to \mathbb R$ defined as
\begin{equation}
    \SFF(U) = \frac{1}{D^2} \Tr[U] \op{Tr}[U^\dagger] \ .
\end{equation}
We would like to find the (best) Lipschitz constant for the SFF. In order to it, we will use the following result, which is simple to obtain. 

\vspace{0.2cm}

\textbf{Theorem:} Let $f: X \to \mathbb R$ be a differentiable Lipschitz continuous function. Then, the Lipschitz constant $\kappa_{\rm L}$ of $f(x)$ is
\begin{equation}
	\kappa_{\rm L}= \op{sup}_{x \in X} \| \nabla f \| \ .
\end{equation} 

\vspace{0.2cm}
\noindent In our case, the metric is induced by the distance \eqref{eq:Frob_norm_definition_sec_3}, which results in
\begin{equation}
	\de s^2 = \frac{1}{D} \sum_{ij} \de a_{ij} \de \overline{a}_{ij} \ ,
\end{equation} 
which is flat. Moreover, as we saw previously, the SFF can be written in terms of eigenvalues of the unitary matrix, which are all phases. In particular, it is
\begin{equation}
	\SFF(\theta_i) = \frac{1}{D^2} \sum_{ij} e^{i (\theta_i - \theta_j)} = \frac{2}{D^2} \sum_{i \leq j} \cos(\theta_{i} - \theta_{j})  \ .
\end{equation}
In terms of these variables, the metric is 
\begin{equation}
	\de s^2 = \frac{1}{D} \sum_i \de \theta_i^2 + \dots \ .
\end{equation}
The dots indicate a sum over other components of the metric, which are flat directions for the SFF, and thus we omit them. Therefore, the only interesting derivatives of the SFF are
\begin{equation}
	\partial_{\theta_i}\SFF = - \frac{2}{D^2} \sum_{i \neq j} \sin(\theta_{i} - \theta_{j})  \ .
\end{equation}
To obtain a bound on the best Lipschitz constant, use that
\begin{equation}
	\kappa_{\rm L}= \op{sup}_{x \in X} \sqrt{\sum_{ij} g^{-1}_{ij}(x) \nabla_i f (x) \nabla_j f(x) } \leq  \sqrt{\sum_{ij} \op{sup}_{x \in X} g^{-1}_{ij}(x) \nabla_i f (x) \nabla_j f(x) } \ .
\end{equation} 
Then we are interested in $\op{sup} \left|  \partial_{\theta_i}\SFF  \right|$. An overestimate is 
\begin{equation}
    \op{sup} |\partial_{\theta_i}\SFF| = \frac{2 (D-1)}{D^2} \ .
\end{equation}
This implies in turn
\begin{equation}
	\kappa_{\rm L} \leq \frac{2 (D-1)}{D}  \ ,
 \label{overestimate_K}
\end{equation} 
which implies a large-$D$ behavior of $\kappa_{\rm L}\lesssim 2$. This is not a huge overestimate. Putting half of the phases at zero and the other half at $\pi/2$ gives an underestimate of
\begin{equation}
    \kappa_{\rm L} \geq 1 \ ,
\end{equation}
which has the same large-$D$ behavior of \eqref{overestimate_K}. In general, this shows that $\kappa_{\rm L}= \cO(1)$, and does not scale with $D$. This certainly is sufficient to set a bound. For instance, using this result, we can infer that
\begin{equation}
    \Delta \SFF \leq \kappa_{\rm L} \, \Delta U \sim \Delta U \ ,
\end{equation}
since $\kappa_{\rm L} \sim \cO(1)$. Given that our Trotterized protocol is able to bound $\Delta U$ independently of $D$, the same holds for $\Delta \SFF$. On the other hand, this bound is quite loose. For a generic unitary matrix, the SFF is expected to be of the order of $1 / D$. Therefore, an error on $\Delta \SFF$ that does not scale with $D$ means that for high $D$ there is a scale separation between the value of the SFF and its error, where the latter is much bigger than the former. 

\begin{figure}[t]
\centering
\includegraphics[width=0.96\textwidth]{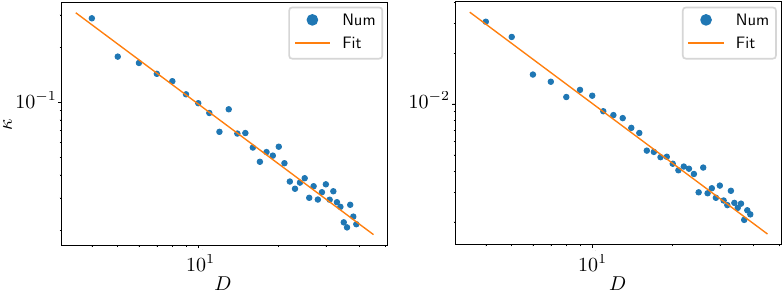}
\caption{Simulation of $\kappa_{\rm L}$ for random unitary matrices. Numerically, for each pair we extract $D$ eigenphases which are of the order $\Delta U$ distant, and then we extract another unitary matrix $P$ which represents the change of basis between the two. The matrix $P$ is sampled via the function \texttt{unitary\_group} of the Python library \texttt{scipy.stats}. For each $D$, we have repeated this process $n = 10^5$ times, keeping the largest value found for $\kappa_{\rm L}$. The dimensions of the unitary matrices in the plot range from $4$ to $40$. The values for $\kappa_{\rm L}$ found indicate a power-law behavior, which we check numerically with a fit of $\kappa_{\rm L}= A / D^B$. The fit has been done with the function \texttt{curve\_fit} of the library \texttt{scipy.optimize}. The value of $A$ should change with $n$ and $\Delta U$, while the value of $B$ should be universal. {\bf Left}: Simulation for $\Delta U = 0.1$. The optimal values found are $A = 1.2$ and $B = 1.1$. {\bf Right}: Simulation for $\Delta U = 0.01$. The optimal values found are $A = 0.15$ and $B = 1.1$.}
\label{fig:Simulation_Lip}
\end{figure}

The catch is that, while it is true that the largest value of $\kappa_{\rm L}$ does not scale with $D$, the set of points for which this happens is very small, especially if $D$ is very large, and typically we do not expect the unitary evolution to reach them. We have checked this numerically, showing that typically
\begin{equation}
    \kappa_{\rm L}\sim \frac{1}{D} \ .
\end{equation}
The numerics is implemented with the following protocol. We first fix the dimension of the unitary operators $D$. Then we draw a pair of unitaries $U_1$ and $U_2$, which have a distance of the order of $\Delta U$. Then we numerically compute the difference of SFF, and the relative $\kappa_{\rm L}$. We repeat this process $n$ times, estimating $\kappa_{\rm L}$ with the highest value obtained. We then repeat this process for different values of $D$ and see how $\kappa_{\rm L}$ scales with $D$. From the result obtained in Figure \ref{fig:Simulation_Lip}, we notice a power-law dependence between $\kappa_{\rm L}$ and $D$. Fitting
\begin{equation}
    \kappa_{\rm L} = \frac{A}{D^{B}} \ ,
\end{equation}
for different values of $\Delta U$, we found a constant $B \approx 1.1$, while $A$ varies both for different $n$ and $\Delta U$. However, for any of them, $A \sim \cO(1)$. For a more precise result, we refer to the caption of Figure~\ref{fig:Simulation_Lip}. This is an interesting result, since it means that in the large $D$ limit, the SFF is likely to be as precise as the time evolution.

\section{Distributions of couplings} \label{App:Statistical_analysis_couplings_extended}
In this Appendix we give additional details to the derivations that leads to the results presented in Section \ref{sec:Statistical_analysis_couplings}.

\subsubsection*{Single-coupling convergence}

We start from the distribution of a single coupling. Consider two independent random variables $X_{1,2} \sim \mathcal N(0, \sigma^2)$, and the product $Y = X_1 X_2$. We perform the change of variables
\begin{equation}
    \{X_1, X_2\} \; \mapsto \; \{Y, X_2\} \ ,
\end{equation} 
which can be implemented at the level of the Probability Distribution Function (PDF) as
\begin{equation}
    \frac{1}{2 \pi \sigma^2} \, e^{-\frac{X_1^2 + X_2^2}{2 \sigma^2}} \, \de X_1 \, \de X_2 \; \mapsto \; \frac{1}{2 \pi \sigma^2 |X_2|} \, e^{-\frac{Y^2}{2 \sigma^2 X_2^2} - \frac{X_2^2}{2 \sigma^2}} \,  \de Y \de X_2 \ ,
\end{equation}
where we have been careful in including how the measure changes. Integrating out the variable $X_2$ (thus taking the marginal), we obtain the PDF for $Y$
\begin{equation}
    P(Y) \, \de Y = \frac{1}{2 \pi \sigma^2} \, \de Y \int_{- \infty}^\infty  \frac{\de X_2}{|X_2|} \, e^{-\frac{Y^2}{2 \sigma^2 X_2^2} - \frac{X_2^2}{2 \sigma^2}} = \frac{1}{\pi \sigma^2} \, K_0\left(\frac{|Y|}{\sigma^2 } \right) \, \de Y \ ,
\end{equation}
which is the result presented in \eqref{P_Sigma_exact}. We now want to consider the distribution of the sum of different independent realizations of $Y$. To do it one has to compute the characteristic function
\begin{equation}
    \varphi_Y(s) = \mathbb E \Big[ e^{i s Y} \Big] = \left( 1 + \sigma^4 s^2 \right)^{- \frac{1}{2}}  \ ,
    \label{eq:characteristic_function_single_R} 
\end{equation}
or the Fourier transform of the probability distribution. Considering now the variable
\begin{equation}
    \mathcal Y = \frac{1}{\sqrt{R}} \sum_{\alpha = 1}^R Y_\alpha \ ,
\end{equation}
the characteristic function for $\mathcal Y$ is
\begin{equation}
    \varphi_{\mathcal Y}(s) = \left( 1 + \frac{\sigma^4 s^2}{R} \right)^{- \frac{R}{2}}  \ ,
    \label{eq:characteristic_function_different_R} 
\end{equation}
which is found appropriately rescaling \eqref{eq:characteristic_function_single_R}, and raising it to the power of $R$. Taking the inverse Fourier transform of \eqref{eq:characteristic_function_different_R} we obtain the probability distribution for $\mathcal Y$, which is 
\begin{equation}
    P(\mathcal Y) \, \de \mathcal Y = \frac{\sqrt{R}}{\sqrt{\pi} \, \sigma^2 \, \Gamma(R/2)} \left( \frac{\sqrt{R} \, |\mathcal Y|}{2 \sigma^2} \right)^{\frac{R-1}{2}} K_{\frac{1-R}{2}} \left( \frac{\sqrt{R}  \, |\mathcal Y|}{\sigma^2} \right) \, \de \mathcal Y \ ,
\end{equation}
as reported in \eqref{P_Sigma_exact}. To see the convergence to a Gaussian for large--$R$, it is convenient to expand the characteristic function \eqref{eq:characteristic_function_different_R} around $R \to \infty$ in $1/R$, and then take the inverse Fourier transform. The result is 
\begin{multline}
    P(\mathcal Y) = \frac{e^{-\frac{\mathcal Y^2}{2 \sigma ^4}}}{\sqrt{2 \pi \sigma^4}} 
    + \frac{1}{ R } \left( \frac{3}{4} - \frac{3 \mathcal Y^2}{2\sigma ^{4}} + \frac{\mathcal Y^4}{4\sigma^8} \right) \frac{e^{-\frac{\mathcal Y^2}{2 \sigma ^4}}}{\sqrt{2 \pi \sigma^4}} \\
    + \frac{1}{ R^2 } \left(\frac{25}{32} - \frac{45 \mathcal Y^2}{8 \sigma^{4}} + \frac{65 \mathcal Y^4}{16\sigma^8} - \frac{17 \mathcal Y^6}{24\sigma^{12}} + \frac{\mathcal Y^8}{32\sigma^{16} } \right) \frac{e^{-\frac{\mathcal Y^2}{2 \sigma ^4}}}{\sqrt{2 \pi \sigma^4}} + \mathcal O(R^{-3})  \ ,
\end{multline}
which is the result reported in \eqref{P_Sigma_expansion_second_order}.

\subsubsection*{Pairwise convergence and independence}

We can employ the same strategy to find the pairwise convergence of couplings. Take $X_{1,2,3} \sim \mathcal N(0, \sigma^2)$, and consider the change of variables 
\begin{equation}
    Y_1 = X_1 X_3 \ , \qquad \text{and} \qquad Y_2 = X_2 X_3 \ .
\end{equation}
In order to find the joint probability distribution for $Y_1$ and $Y_2$ we perform the change of variables 
\begin{equation}
    \{X_1, X_2, X_3\} \; \mapsto \; \{Y_1, Y_2, X_3\} \ ,
\end{equation} 
and then compute the marginal integrating out $X_3$. Explicitly, the calculation is 
\begin{equation}
    P(Y_1 , Y_2) \, \de Y_1 \de Y_2 = \frac{1}{(2 \pi \sigma^2)^{3/2}} \de Y_1 \de Y_2 \int_{- \infty}^{\infty}  \frac{\de X_3}{X_3^2} \, e^{- \frac{1}{2 \sigma^2} \left( \frac{Y_1^2 + Y_2^2}{X_3^2} + X_3^2 \right)} = \frac{e^{- \sqrt{Y_1^2 + Y_2^2}/\sigma^2}}{2 \pi \sigma^2 \sqrt{Y_1^2 + Y_2^2} } \, \de Y_1 \de Y_2 \ ,
\end{equation}
which is the result reported in \eqref{eq:two_couplings_probability_distribution}. As before, we would like to find the convergence when we consider sums of independent realizations of $Y_1$ and $Y_2$. We then compute the characteristic function associated to \eqref{eq:two_couplings_probability_distribution}, which is 
\begin{equation}
    \varphi_{Y_1, Y_2}(s_1, s_2) = \mathbb E_{Y_1, Y_2} \Big[ e^{i s_1 Y_1 + i s_2 Y_2} \Big] = \Big( 1 + \sigma^4 \big(s_1^2 + s_2^2 \big) \Big)^{-\frac{1}{2}} \ ,
\end{equation}
and we consider the variables
\begin{equation}
    \mathcal Y_1 = \frac{1}{\sqrt{R}} \sum_{\alpha = 1}^R Y_{1,\alpha} \ , \qquad \text{and} \qquad \mathcal Y_2 = \frac{1}{\sqrt{R}} \sum_{\alpha = 1}^R Y_{2,\alpha} \ .
\end{equation}
Then, the characteristic function for $\mathcal Y_1$ and $\mathcal Y_2$ is simply
\begin{equation}
    \varphi_{\mathcal Y_1, \mathcal Y_2}(s_1, s_2) = \left( 1 + \frac{ \sigma^4 \big(s_1^2 + s_2^2\big)}{R} \right)^{-\frac{R}{2}} \ .
    \label{eq:char_fn_two_coupl_indep}
\end{equation}
The inverse Fourier transform of \eqref{eq:char_fn_two_coupl_indep} is the joint probability distribution for $\mathcal Y_1$ and $\mathcal Y_2$, but unfortunately we have not found a way to express it in terms of elementary functions. We can then find the large--$R$ limit expanding \eqref{eq:char_fn_two_coupl_indep} around $R \to \infty$, and then performing an inverse Fourier transform. The result is 
\begin{equation}
    P(\mathcal Y_1, \mathcal Y_2) = \frac{e^{-\frac{\mathcal Y_1^2 + \mathcal Y_2^2}{2 \sigma^4}}}{2 \pi \sigma^4} + \frac{1}{R} \left(2 -  \frac{2( \mathcal Y_1^2 + \mathcal Y_2^2)}{\sigma^{4}} + \frac{(\mathcal Y_1^2 + \mathcal Y_2^2)^2}{4\sigma^{8}} \right) \frac{e^{-\frac{\mathcal Y_1^2 + \mathcal Y_2^2}{2 \sigma^4}}}{2 \pi \sigma^4} + \mathcal O(R^{-2}) \ ,
\end{equation}
which is the one reported in \eqref{eq.convergenceTrotterizedModel}.

\subsubsection*{Effective vs local independence}

We also present the detailed calculations leading to \eqref{eq:joint_distr_local_vs_eff}, which follows the same strategy presented above. Assume two random variables $X_{1,2} \sim \mathcal N(0, \sigma^2)$, and consider again the product $Y = X_1 X_2$. Using a simple change of coordinates, their joint probability distribution is 
\begin{equation}
    P(X_1,Y) \, \de X_1 \, \de Y = \frac{1}{2 \pi \sigma^2 |X_1|} \, e^{-\frac{Y^2}{2 \sigma^2 X_1^2} - \frac{X_1^2}{2 \sigma^2}} \, \de X_1 \, \de Y \ ,
\end{equation}
We can then compute the characteristic function associated to the above PDF, which reads
\begin{equation}
    \varphi_{X_1, Y}(u, s) = \mathbb E_{X_1, Y} \Big[ e^{i (u X_1 + s Y)}  \Big] = \frac{1}{\sqrt{1 + s^2 \sigma ^4}} \, \exp \left(-\frac{\sigma ^2 u^2}{2( 1 +  s^2 \sigma ^4)} \right)  \ .
    \label{eq:characteristic_function_single_vs_double}
\end{equation}
We then consider the sums over independent realizations of $X_1$ and $Y$, namely
\begin{equation}
    \mathcal X = \frac{1}{\sqrt{R}} \sum_{\alpha = 1}^R X_{1, \alpha} \ , \qquad \text{and} \qquad \mathcal Y = \frac{1}{\sqrt{R}} \sum_{\alpha = 1}^R Y_{\alpha} \ ,
\end{equation}
so that the characteristic function for $\mathcal X$ and $\mathcal Y$ is
\begin{equation}
    \varphi_{\mathcal X, \mathcal Y}(u,s) = \exp \left[-\frac{\sigma ^2 u^2}{2} \left( 1 +  \frac{s^2 \sigma ^4}{R} \right)^{-1} \right] \left( 1 + \frac{s^2 \sigma ^4}{R} \right)^{-\frac{R}{2}} \ .
    \label{eq:char_function_local_ind}
\end{equation}
We have not managed to express the inverse Fourier transform of \eqref{eq:char_function_local_ind} in terms of elementary functions. We then consider, as before, the large--$R$ expansion around $R \to \infty$, which is 
\begin{equation}
    \varphi_{\mathcal X, \mathcal Y}(u,s) = e^{- \frac{\sigma^2 u^2}{2} - \frac{\sigma^4 s^2}{2}}+ \frac{1}{R} \left(\frac{s^4 \sigma^8}{4} + \frac{u^2 s^2 \sigma^6}{2} \right)  e^{- \frac{\sigma^2 u^2}{2} - \frac{\sigma^4 s^2}{2}} + \mathcal O (R^{-2}) \ .
\end{equation}
Finally, an inverse Fourier transform gives the joint probability distribution
\begin{equation}
    P(\mathcal X, \mathcal Y) = \frac{e^{-\frac{\mathcal X^2}{2 \sigma^2} -\frac{\mathcal Y^2}{2 \sigma^4}}}{2 \pi \sigma^3} + \frac{1}{R}\left(\frac{5}{4} - \frac{\mathcal X^2}{2\sigma^2} - \frac{2 \mathcal Y^2}{\sigma^4} + \frac{\mathcal X^2 \mathcal Y^2}{2\sigma^6} + \frac{\mathcal Y^4 }{4\sigma^8 } \right) \frac{e^{-\frac{\mathcal X^2}{2 \sigma^2} -\frac{\mathcal Y^2}{2 \sigma^4}}}{2 \pi \sigma^3} + \mathcal O (R^{-2}) \ ,
\end{equation}
which is the result reported in \eqref{eq:joint_distr_local_vs_eff}.

\section{Numerical implementation of speckles}
\label{sec:App_on_speckles}

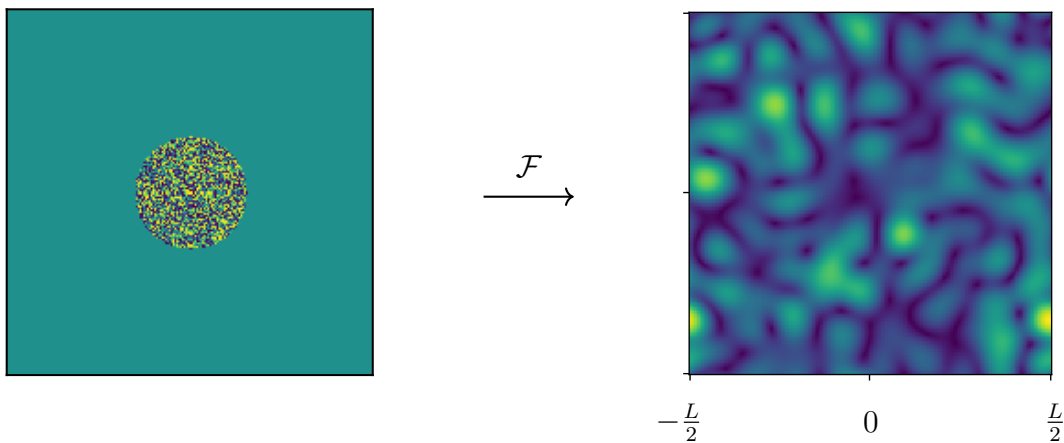
\begin{figure}[t]
\centering
\begin{tikzpicture}
    \node[] (img1) at (-4.5,0.05) {\includegraphics[width=0.3\textwidth]{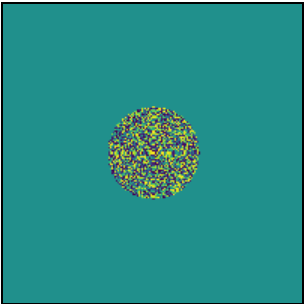}};

    \node[] at (0,0.4) {\large $\mathcal F$};
    \draw[->, thick] (-0.6,0) -- (0.6,0);

    \node[] at (2,-3) {\large $-\frac{L}{2}$};
    \node[] at (7,-3) {\large $\frac{L}{2}$};
    \node[] at (4.55,-3) {\large $0$};
    
    \node[] (img3) at (4.5,0) {\includegraphics[width=0.3\textwidth]{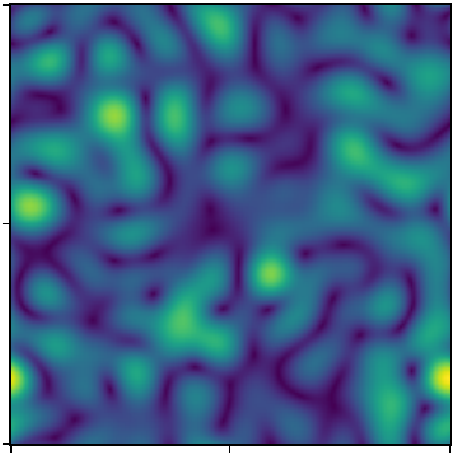}};
\end{tikzpicture}
\caption{\textbf{Left}: A mask used to generate speckle pattern. For ease of visualization, the radius of the mask is 30 pixels, while the length of the square is 200 pixels. In the text, and on the right speckle, we have mostly used masks with 6 pixels of radius. \textbf{Right}: A single realization of a speckle pattern coming from a mask with radius 6 pixels. The grid is composed of 200 pixels per axis, and it corresponds to 10 units in length of the harmonic trap in $H_{\rm kt}$.}
\label{fig:Mask_and_speckle}
\end{figure}

In this Section, we explain in detail how to implement numerically the spatially disordered detunings used in the main text. The main physical effect we want to simulate is the AC Stark shift happening when auxiliary states are coupled to the system with an additional blue laser. This imprint a spatially varying detuning as
\begin{equation}
    \Delta_{\rm ad}(r) = \Delta_{\rm ad} + \frac{|\Omega_{\rm b}(r)|^2}{4 \Delta_{\rm b}} \ ,
\end{equation}
where $\Delta_{\rm b}$ is the detuning between the ground states and the auxiliary states, and $\Omega_{\rm b}$ is the Rabi frequency of the transition, which is also proportional to the laser intensity. Thus, a spatially disordered detuning can be achieved when the additional blue laser imprints a randomized speckle of light intensity $|\Omega_{\rm b}(r)|^2$. This can be achieved by shining the laser through a spatial light modulator, whose resulting beam is then focused on the atomic cloud. To numerically reproduce this situation, we use the algorithm proposed in \cite{Duncan_Kirkpatrick}, which involves the Fourier transform of a collection of random phases cut out by a mask. 

Instead of explaining in detail the concepts behind \cite{Duncan_Kirkpatrick}, we lay out right away the numerical implementation we have used. We consider a square grid of $200 \times 200$ pixels (the entries of an array), filled with zero's except for a circular mask of radius $r$--pixels. We then fill each pixel of the mask (entry of the array) with a random complex phase. The resulting quantity is a $200 \times 200$ array, which is mostly filled with zeros except for the `central' part. This is shown on the left panel of Figure \ref{fig:Mask_and_speckle}. To create the speckle pattern, we then perform a Fast Fourier Transform on this array, and then consider only the absolute value. This generates a $200\times 200$ array with an intensity `varying' as on the right panel of Figure \ref{fig:Mask_and_speckle}. Furthermore, we rescale everything so that the spatial average of such an array is unity, and adding a further `one' in all entries, we have generated 
\begin{equation}
    \frac{\Delta_{\rm ad}(r)}{\Delta_{\rm ad}} = 1 + \frac{|\Omega_{\rm b}(r)|^2}{4 \Delta_{\rm b}\Delta_{\rm ad}} \ ,
\end{equation}
where on the RHS the spatial average of the disordered speckle is comparable with the deterministic side. Moreover, this implies that
\begin{equation}
    \overline{\Delta_{\rm ad}(r)} \approx 2 \Delta_{\rm ad} \ ,
    \label{eq:average_of_detuning_appendix}
\end{equation}
which is indirectly confirmed in Figure \ref{fig:probability_density_two_body_couplings} following Equation \eqref{eq:experimental_couplings_two_body_avg}. To implement the disordered couplings, we now only have to specify the `physical' relation between such speckle and the geometry of the cavity. In particular, we assume that this $200 \times 200$ grid equals $10$ harmonic trap lengths, which is the physical length entering the Hermite--Gauss functions. To find the couplings \eqref{eq:experimental_couplings_two_body_avg} we then numerically integrate over the grid. 

To finish this Section, we report that, while the mask in Figure \ref{fig:Mask_and_speckle} has a radius of $30$ pixels, in our numerical simulations, we have mostly used a radius of $r = 6$ pixels, as the former would give a much finer speckle pattern, which would self average to zero.

\section{Bulk spectrum and thermodynamics}
\label{App:connections_to_gravity}

\begin{figure}[t]
\centering
\begin{tabular}{ll}
\includegraphics[width=0.48\textwidth]{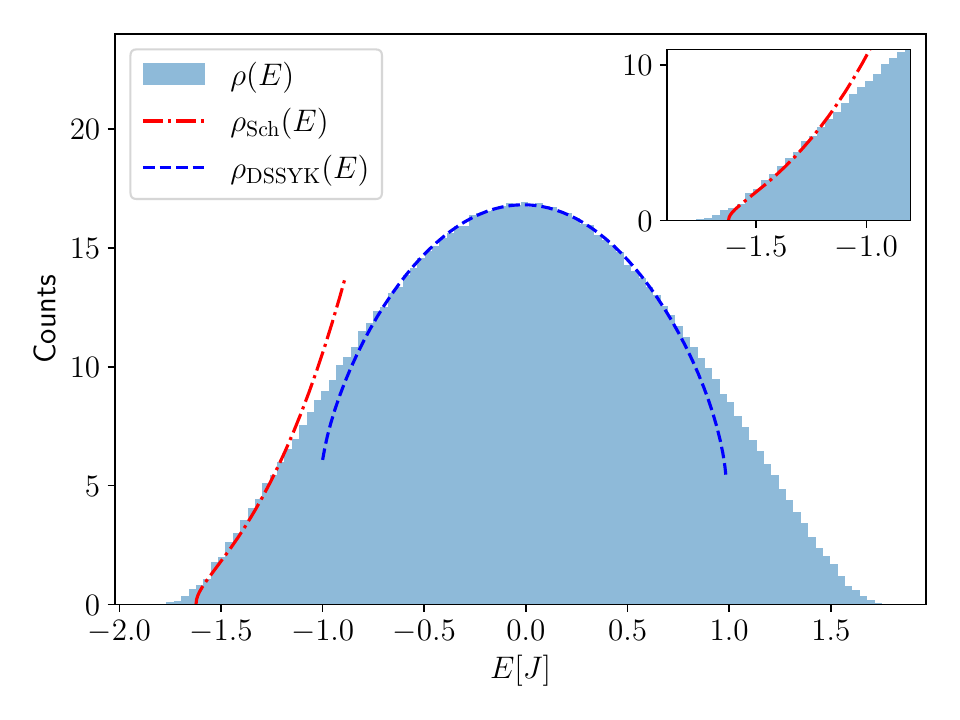}
&
\includegraphics[width=0.474\textwidth]{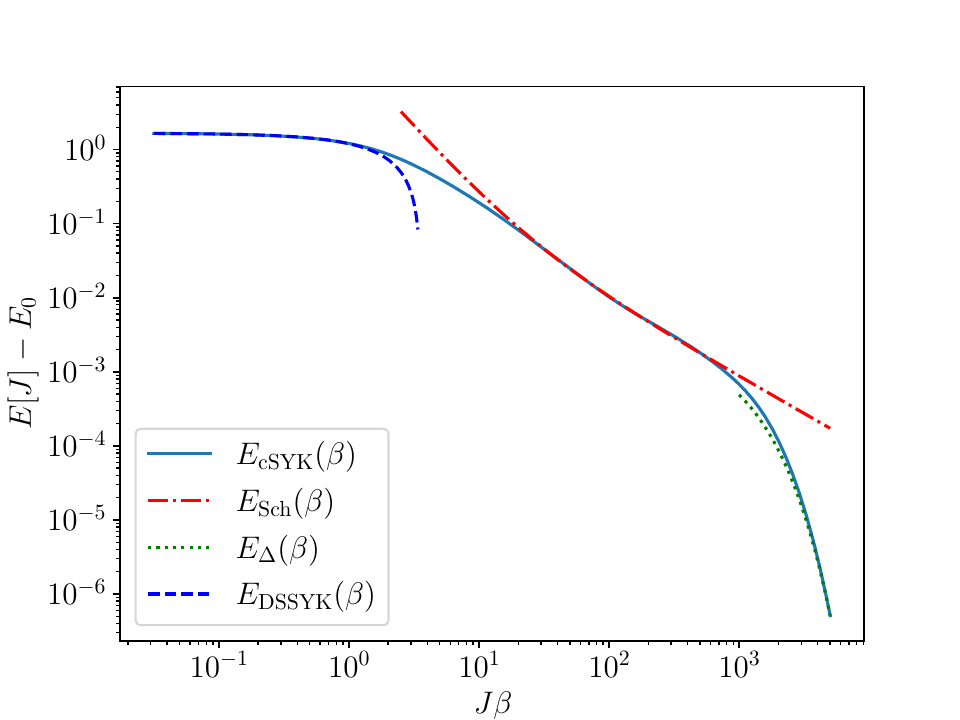}
\end{tabular}
\caption{{\bf Left}: Distribution of energy eigenvalues for $N = 12$ and $J= 1$, as in Figure \ref{fig:probability_density_spectrum}. The edge of the spectrum is fitted with the Schwarzian density of states \eqref{eq:Sch_density}, appropriately rescaled, while the bulk of the spectrum with the DSSYK density of states \eqref{eq:DSSYK_density}  {\bf Right}: Thermal energy $E(\beta)$ of complex SYK computed numerically with the spectrum shown on the left. The function fits well with the Schwarzian prediction \eqref{eq:Thermal_energy_Sch} for $J\beta \gtrsim 10$, and with the DSSYK prediction \eqref{eq:Thermal_energy_DSSYK} for $J \beta \lesssim 1$. At low temperatures, the finite-$N$ gap shows with an exponential behavior.}
\label{fig:SYK_Thermodynamics}
\end{figure}

In Section \ref{sec:product_states}, we have shown that product states in the `computational basis' have a relatively high temperature. This situation is not ideal to realize a system that is dual to JT-gravity, as this relation holds in the infrared limit, which thermally is realized when $J \beta \to \infty$. From the point of view of the energy spectrum, this implies that JT-gravity captures the lower edge of the spectrum, with the Schwarzian density of states
\begin{equation}
    \rho_{\rm Sch} (E) = \sinh(2 \pi \sqrt{\frac{E - E_0}{\mathcal E}}) \ ,
    \label{eq:Sch_density}
\end{equation}
which of course is meant to be understood in a `double--scaled' sense, namely taking a large Hilbert space and focusing only on the lower edge of the spectrum, so that the number of total eigenvalues becomes effectively infinite\footnote{Indeed, the integral over the density of states \eqref{eq:Sch_density} diverges. This is however not a problem, as we should not interpret the Schwarzian theory as a microscopic model.} In \eqref{eq:Sch_density} $E_{0}$ is the ground state energy ({\it i.e.} the offset of the spectral edge), while $\mathcal E$ is an energy scale required for dimensional reasons. Fitting \eqref{eq:Sch_density} to the numerical spectrum of the theory one finds a qualitative good agreement, as it is shown in Figure \ref{fig:SYK_Thermodynamics} on the left panel (and in particular in the inset), with $E_0 = -1.62J$ and $\mathcal E = 2.64J$. The small tail is a non-perturbative effect in $1/N$, and we neglect it. We can also try to compute the thermal energy of the system, 
\begin{equation}
    E(\beta) = \frac{1}{Z_{\beta}}\sum_n E_n e^{- \beta E_n} = - \frac{\de}{\de \beta} \, \log (Z_{\beta}) \ .
\end{equation}
and see if the physics of the Schwarzian sector is reproduced in the limit $J \beta$ big. The Schwarzian sector predicts a thermal energy of the form
\begin{equation}
    E_{\rm Sch}(\beta) = \frac{1}{Z_{\beta}}\int_{E_g}^\infty \de E \, \rho_{\rm Sch}(E) \, E \, e^{-\beta E} = E_0 + \frac{\pi^2}{\mathcal E \beta^2} + \frac{3}{2 \beta} \ .
    \label{eq:Thermal_energy_Sch}
\end{equation}
where the first term is the ground state energy, the second term comes from a saddlepoint approximation of \eqref{eq:Thermal_energy_Sch}, while the last one should be thought of as a one--loop contribution. In Figure \ref{fig:SYK_Thermodynamics} we compare this expectation with the cSYK thermal energy computed numerically. Fitting \eqref{eq:Thermal_energy_Sch}, we find $\mathcal E = 1.86 J$, which qualitatively agrees with the result previously found fitting the spectral density. On the other hand, fitting the coefficient of the last term in \eqref{eq:Thermal_energy_Sch} gives the value of $0.85$, which should be compared with the predicted $3/2$ and it is in a similar qualitative agreement as $\mathcal E$. At very low energy, the physics of the Schwarzian is cut--off by the fact that the Hilbert space is finite dimensional. In particular, for large enough $J \beta$, the energy starts resolving the gap between the ground state and the first excited state. This implies an exponential behavior of the form 
\begin{equation}
    E_{\Delta}(\beta) \approx E_0 + \tilde E_1 e^{-\beta \Delta} \ ,
    \label{eq:Thermal_energy_DSSYK}
\end{equation}
where $\Delta = E_1 - E_0$ is the gap, and $\tilde E _1$ is a constant close to $E_1$, but slightly different due to the effect of states with a higher energy. For the plot shown on the right of Figure \ref{fig:SYK_Thermodynamics}, we have used the spectrum on the left, which contains 100 single realizations of complex SYK\footnote{We have also removed some states at the lower edge of the spectrum, to compare with the predictions at $N \to \infty$. In particular, out of $92400$ energies ($100$ realizations of the $924$ energies of the half-filling sector for $N = 12$) we took out the lowest $40$ eigenvalues. Moreover, we notice that this numerical procedure is different from numerically computing $\langle E(\beta) \rangle$ over $100$ realizations.}. The presence of this gap limits a clean numerical derivation of the Schwarzian physics, and we believe that such limitation restricts the a good quantitative comparison between the two fits. 

On top of this, we can also study the high--temperature physics. This is motivated by the analysis of the previous Section, where we argued that an initial state is likely to be in the bulk of the spectrum. Motivated by \cite{Garcia-Garcia:2017pzl}, the density in such bulk can be approximated by
\begin{equation}
    \rho_{\rm DSSYK} (E) \propto \exp \left( - \arcsin^2 \left( \frac{E}{\tilde{\mathcal E}} \right) \right) \approx \exp \left( -  \frac{E^2}{\tilde{\mathcal E}^2} \right) \ ,
    \label{eq:DSSYK_density}
\end{equation}
which we have denoted with the subscript `DSSYK' since such functional form also appears in the semiclassical approximation of the Double Scaled SYK (DSSYK) model, even though the present situation is far from the double scaled regime. Moreover, as explained in \cite{Garcia-Garcia:2017pzl}, the ansatz \eqref{eq:DSSYK_density} is supposedly valid only at large $N$. However, its relative simplicity makes it convenient to use. We notice that \eqref{eq:DSSYK_density} fits rather well the bulk of the spectrum in Figure \ref{fig:SYK_Thermodynamics}, with fitting parameter $\tilde{\mathcal E} = 1.17 J$. On the other hand, to find the corresponding thermal energy, an expansion close to the middle of the spectrum ($E = 0$) of \eqref{eq:DSSYK_density} gives 
\begin{equation}
    E_{\rm DSSYK}(\beta) = - \frac{\tilde{\mathcal E}^2 \, \beta}{2} \ .
\end{equation}
The high temperature behavior of the numerical thermal energy fits well with a linear ansatz, and one obtains $\tilde{\mathcal E} = 0.95 J$, in good agreement with the value found from the spectrum.

This linear dependence on $\beta$ is a prediction of Gaussian densities of states, and is also found in DSSYK \cite{Okuyama:2023iwu}. Recent proposals have connected such high--temperature limit of DSSYK with Liouville gravity and three--dimensional quantum gravity in de Sitter \cite{Narovlansky:2023lfz, Verlinde:2024znh, Verlinde:2024zrh}. Given the good accuracy with which DSSYK seems to reproduce this high--temperature sector of cSYK, and given the intrinsically large energy (and temperature) of the initial states of the evolution, it is foreseeable that such evolution could be connected to de Sitter quantum gravity. We leave this intriguing speculation for future work.

\section{Realistic experimental parameters}
\label{App:exp_parameters}
This appendix summarizes the parameters used in the numerical simulations and computations presented in this work, which are chosen to be consistent with experimentally motivated values from recent cavity QED proposals.
\subsection*{Speckle Grid Parameters}
Parameters for numerical simulations of speckle patterns.

\begin{center}
\begin{tabular}{ | c | c| c | } 
  \hline
  Parameter & Value & Description \\ 
   \hline
   \hline
  \texttt{Dimgrid} & 10 & Size of grid (units of trap length) \\ 
  \hline
  \texttt{Ngrid} & 200 & Grid resolution \\ 
  \hline
  \texttt{dx} & 0.05 & \texttt{Dimgrid}/\texttt{Ngrid} \\ 
  \hline
  \texttt{radius} & 0.30 = (6 pixels) $\times$ \texttt{dx} &  Aperture of mask\\ 
  \hline
\end{tabular}
\end{center}

\subsection*{Experimental Cavity-QED Parameters}
Cavity parameters used in the numerical simulations which correspond to realistic experimental values. The cooperativity is normalized as $\eta = 4 g^2/(\kappa \Gamma)$.

\begin{center}
    \begin{tabular}{|c|c|c|}
    \hline
    Parameter & Value & Description \\
    \hline
    \hline 
        $\kappa$ & $2\pi \times 0.16 \, \text{MHz}$ & Cavity decay rate  \\
        \hline
        $g$ & $ 2\pi \times 2.05 \, \text{MHz}$ & Atom-cavity coupling strength \\
    \hline
        $\eta$  & 18 & Cooperativity \\
        \hline
        $\Gamma$  &  $ 2\pi \times 5.86 \, \text{MHz}$ & Effective spontaneous emission rate \\
        \hline
        $\Delta_{\rm cd}$  & $ 2\pi \times20 \, \text{MHz}$ & Cavity-drive detuning\\
        \hline
        $\Delta_{\rm ad}$  & $2\pi \times 80 \, \text{MHz}$ & Drive-atom detuning\\
        \hline
        $\Omega_{\rm d}$  & $1 \, \text{GHz}$ & Drive amplitude \\
        \hline
        $\Omega_{\rm c}$  & $g$ & Cavity mode Rabi frequency\\
    \hline
    \end{tabular}
\end{center}

\end{spacing}
\end{document}